\documentclass[preprint,12pt,authoryear]{elsarticle}

\usepackage{amsmath,amssymb,mathtools}
\usepackage{amsthm}
\usepackage{algorithm2e}
\LinesNumbered
\SetAlgoNlRelativeSize{0}
\SetNlSty{scriptsize}{}{}
\usepackage[section]{placeins}
\usepackage{booktabs}
\usepackage{array,tabularx}
\usepackage{graphicx}
\usepackage{xcolor}
\usepackage[hidelinks]{hyperref}
\usepackage[nameinlink,noabbrev]{cleveref}

\newtheorem{theorem}{Theorem}
\newtheorem{lemma}{Lemma}

\theoremstyle{definition}

\theoremstyle{remark}

\newtheorem{property}{Property}

\newcolumntype{L}{>{\raggedright\arraybackslash}X}

\journal{Computers \& Operations Research}

\begin{document}

\begin{frontmatter}

\title{Split Algorithm in Linear Time for the Vehicle Routing Problem with Simultaneous Pickup and Delivery and Time Windows}

\author[uw]{Ethan Gibbons\corref{cor1}}
\ead{ej2gibbons@uwaterloo.ca}

\author[purdue]{Mario Ventresca}

\author[brock]{Beatrice Ombuki-Berman}

\affiliation[uw]{
    organization={University of Waterloo},
    city={Waterloo},
    state={Ontario},
    country={Canada}
}

\affiliation[purdue]{
    organization={Purdue University},
    city={West Lafayette},
    state={Indiana},
    country={USA}
}

\affiliation[brock]{
    organization={Brock University},
    city={St. Catharines},
    state={Ontario},
    country={Canada}
}

\cortext[cor1]{Corresponding author}

\begin{abstract}
For many kinds of vehicle routing problems (VRPs), a popular heuristic approach involves constructing a Traveling Salesman Problem (TSP) solution, then partitioning segments of the solution into routes for different vehicles with respect to problem constraints.  Previously, a Split algorithm with a worst-case runtime of $O(n)$ was proposed for the capacitated VRP (CVRP) and its soft variant that finds the most cost-efficient partition of customers, given a TSP solution. This was an improvement over the previously fastest-known algorithm with a worst-case runtime of $O(n^2)$ that was based on Bellman's shortest path algorithm. While this linear Split has been an integral part of modern state-of-the-art CVRP approaches, little progress has been made in extending this algorithm to handle additional VRP variants, limiting the general applicability of the algorithm. 
In this work, we propose an extension of the linear Split that handles two cardinal VRP variants simultaneously: (i) simultaneous pickups and deliveries (VRPSPD) and (ii) time windows (VRPTW). The resulting $O(n)$ algorithm is guaranteed to be optimal, assuming travel times between nodes satisfy a weakened triangle inequality.
We generalize the properties of both variants that allow for the proposed approach. 
Additionally, we develop $O(n)$ Splits for the soft versions of VRPTW (with time-warp penalty), VRPSPD, and the combined VRPSPD with time windows (VRPSPDTW). The complications these variants introduce compound when combined, yet linearity remains obtainable. Computational experiments are performed to empirically validate the correctness and linearity of these new approaches.

\end{abstract}


\end{frontmatter}

\section{Introduction}

For over a decade, the Hybrid Genetic Search (HGS) metaheuristic has remained a dominant state-of-the-art heuristic approach for several variants of the vehicle routing problem (VRP), including the basic capacitated VRP (CVRP) \citep{toth2014vehicle,vidal2022hybrid}. The algorithmic framework for HGS has been iteratively improved over several works \citep{vidal2012hybrid,vidal2013hybrid,vidal2014unified} and has been shown to be easily extendable to many variants of the VRP due to its modular design. The open-source CVRP implementation of  \citet{vidal2022hybrid} has been extended and applied in various approaches \citep{kool2022hybrid,ghannam2023hybrid,wang2023decomposition,wouda2024pyvrp,zhao2025large}, including some that use machine learning techniques to improve search quality and efficiency \citep{santana2023neural,greenberg2025accelerating}.

Key to the design of HGS is the Split algorithm originally proposed by \citet{prins2004simple} for use in a genetic algorithm. In order to simplify the genetic crossover operator, VRP solutions are represented as a long tour, which is a permutation of the \(n\) customers that specifies their order of visitation but not the starts and ends of routes. After crossover and mutation, a Split algorithm based on Bellman's shortest-path algorithm finds the optimal starts and ends of the routes while respecting the problem constraints. It takes as input an implicit directed acyclic graph constructed from the permutation such that a shortest path represents a feasible VRP solution. The runtime is \(O(nB)\), where \(B\) is the average number of feasible routes in the tour that start at a given customer, with a worst case of \(O(n^2)\).

This long-tour and optimal-Split method, referred to as an ``order-first split-second'' approach \citep{prins2014order}, guarantees that if a long tour contains the optimal order in which customers are visited, Split will produce the optimal solution to the VRP. This is considered an attractive quality by practitioners \citep{prins2014order,yaddaden2022neural} and contrasts with greedy splits, which may be unable to obtain the optimal solution even if a long-tour representation of that solution is found during the search \citep{ombuki2006multi}. The HGS works in \citet{vidal2012hybrid,vidal2013hybrid,vidal2014unified} take advantage of this Split algorithm and note that many variant-specific constraints can be ignored during crossover and mutation because Split handles them after the genetic operators are applied.

A critical improvement to the Split algorithm, and therefore to HGS, was provided by \citet{vidal2016split}. That work proposed a Split with a worst-case runtime of \(O(n)\) specifically for the CVRP, together with an \(O(n)\) Split for a linear capacity penalty and an \(O(nK)\) Split for a limited set of \(K\) vehicles. However, a key limitation of the linear Split is that its rules work only for the CVRP with and without relaxed capacity constraints. In contrast, the Bellman Split was applied to eleven VRP variants in \citet{vidal2014unified} alone. To our knowledge, the only extension of linear Split to another VRP variant is the multi-trip time-dependent VRP, with a running time of \(O(nKW)\), where \(W\) is the number of breakpoints defining changes in travel time \citep{zhao_hybrid_2024}.

The primary contribution of this paper is to extend the linear Split to handle the additional constraints (penalties) that arise from two cardinal hard (soft) VRP variants: the VRP with simultaneous pickup and delivery (VRPSPD) and the VRP with time windows (VRPTW). For conciseness, this paper formulates the Split problem using the VRP with simultaneous pickup and delivery and time windows (VRPSPDTW) as the underlying problem, while treating VRPSPD and VRPTW as special cases. In an attempt to assist further research into Split algorithms, we capture and decouple the properties of the constraints (penalties) that make a linear Split possible for the variants at hand. The resulting Splits may be useful for other variants that exhibit the same properties or serve as a starting point for variants with similar properties.

We first consider hard constraints. We show that for variants with constraints such that (i) the prepending/appending of customers to the beginning/end of an infeasible route never results in feasibility, and (ii) the removal of customers from the beginning/end of a feasible route never results in infeasibility, it is enough to record the first customer index that cannot be reached from each predecessor. These indices are nondecreasing, and any fixed number of such constraints can be combined by taking their minimum. The resulting Split is optimal and runs in \(O(n)\) time whenever the indices can also be constructed in \(O(n)\) time. We provide such constructions for simultaneous pickup and delivery and time windows. Note that these properties hold for the VRPTW as long as singleton routes are feasible and travel times between customers and the depot satisfy a weakened triangle inequality. These inequalities are satisfied by instances whose travel times are defined by Euclidean distance, as is common for synthetically generated VRP instances \citep{solomon1987algorithms,augerat1995computational,gehring1999parallel,wang2012genetic,uchoa2017new,arnold2019efficiently}. They are also satisfied when travel times are shortest-path distances on a road network.

We next consider soft constraints. One clever technique employed in HGS is the use of dynamic penalty terms that allow the genetic search to consider solutions in the infeasible space, and \citet{vidal2022hybrid} employs the linear Split for the soft CVRP for this purpose. The soft VRPTW combines the CVRP capacity penalty with the popular time-warp penalty \citep{nagata2007effective,nagata2010penalty,vidal2013hybrid}. We identify a number of strong properties shared by these two penalties, including the property that two penalized predecessors increase in cost at the same rate, that allow the boundaries between penalized and unpenalized predecessors to be easily tracked and maintained. Under the previously stated triangle-inequality assumptions, this results in an $O(n)$ Split for the soft VRPTW with both capacity and time-warp penalties.

The capacity penalty for the soft VRPSPD does not share all such properties. Specifically, the point of maximum vehicle load can change as a route grows, and this can change the relative cost difference between penalized predecessors. Nevertheless, we establish weaker conditions specific to the VRPSPD that allow for an \(O(n)\) Split. Finally, we combine the VRPSPD penalty with time warp to obtain an \(O(n)\) Split for the soft VRPSPDTW. This last Split takes procedures from all of the above and additionally occasionally requires the use of a double-ended queue (deque) of deques, where the outer deque is maintained using a variation of the ``two stacks'' algorithm \citep{tangwongsan2017low}. While these last algorithms are highly specific to the VRPSPD and VRPSPDTW, we hope their design can serve as inspiration for other, more irregular VRP variants' Splits.

Computational experiments are performed to validate the algorithms' linear runtime and to measure the runtime overhead of tracking additional constraints and penalties in comparison to the simplest hard and soft CVRP Split. We alter the set of CVRP permutations used by \cite{vidal2016split} to be VRPSPDTW permutations and use these permutations, which include instances of up to 71000 customers, to observe asymptotic behavior. For an empirical correctness check, we employ randomized permutations of popular instances from the VRPTW, VRPSPD, and VRPSPDTW literature to compare the solutions of each new Split against its Bellman-based equivalent.

The remainder of this paper is organized as follows. Section \ref{sec:related-work} discusses related work. Section \ref{sec:split-problem} gives the common shortest-path problem and notation. Section \ref{sec:hard-split} generalizes the hard CVRP Split from \citet{vidal2016split} to work for any fixed number of constraints satisfying the conditions previously mentioned. Sections \ref{sec:generic-penalties}, \ref{sec:soft-vrpspd}, and \ref{sec:soft-vrpspdtw} give the algorithms for the soft VRPTW, VRPSPD, and VRPSPDTW Splits, respectively. Section \ref{sec:experiments} reports the computational experiments, and Section \ref{sec:conclusion} concludes the paper. Variant-specific preprocessing steps, implementation details, and longer proofs are deferred to the appendices.

\section{Related work}\label{sec:related-work}

While the aforementioned HGS by \citet{vidal2022hybrid} is perhaps the current most relevant order-first split-second metaheuristic, \citet{prins2014order} found more than 70 papers employing the approach. The following review does not enumerate all order-first split-second metaheuristics published since then. Instead, since the contributions of this work extend the linear CVRP Split by \citet{vidal2016split} to work for other VRP variants, we discuss works that employ the linear CVRP Split in contexts other than the CVRP.

For example, for the periodic VRP, \citet{zajac_adaptive_2017} employs the limited fleet Split on a long tour of the customers tentatively assigned to be serviced on a particular day. The Split serves as a local search tool, where split routes are shuffled, perturbed, and recombined into another long tour for another iteration of the Split procedure. Similarly, \citet{vadseth_iterative_2021} make creative use of the limited fleet Split repeatedly over a distribution of customer demands to find a tour partition that is robust to daily changes in customer demand for the inventory routing problem (IRP). \citet{skalnes_branch-and-cut_2023} apply the same Split approach as in \citet{vadseth_iterative_2021} for the same problem. Recently, \citet{zhao2025large} adapted HGS for the inventory routing problem (IRP), where the linear Split is used on all customers for each day after a perturbation occurs. For the capacitated electric VRP, \citet{feng_bilevel_2024} base their approach on HGS where the linear Split is used without alteration, and charging stops are strategically inserted when a resulting route is infeasible due to insufficient battery power. For the generalized VRP, where customers belong to clusters and only one customer in a cluster needs to be visited, \citet{latorre_hybrid_2025} applies Split on precomputed centroids of each cluster as part of their solution construction. 

A number of VRPTW methods also employ the Split algorithm, all ignoring time window constraints until after the Split terminates. For the single-objective VRPTW, \citet{kool2022hybrid} base their
implementation on HGS, evaluating candidate solutions after the CVRP
Split procedure using the time warp penalty \citep{nagata2010penalty}. As a direct follow-up for the dynamic VRPTW, \citet{ghannam2023hybrid} apply HGS and Split only on customers who were not yet assigned to a vehicle currently en route. The lexicographical bi-objective VRPTW was studied by \citet{wang2023decomposition}, who run the limited-fleet Split repeatedly with decrementing limits in an attempt to find solutions requiring fewer vehicles. The authors of these works do not empirically validate the algorithm design choice to favour a faster Split that ignores time windows over the known \(O(nB)\) VRPTW Split that can either optimize routes penalized by a time warp penalty or prune infeasible solutions \citep{vidal2014unified}. However, ignoring time windows removes the property typically desired by VRP practitioners that guarantees if a long tour representation of an optimal VRPTW is found during the search, the Split will produce the optimal solution.

As far as we are aware, only \citet{zhao_hybrid_2024} have fundamentally extended the linear Split. The authors developed a Split for the multi-trip time-dependent VRP (MT-TD-VRP) with a limited fleet, where the objective is to minimize total travel time and vehicle count. They employ up to $K|W|$ monotonically decreasing deques to maintain configurations of non-dominated breakpoints, where $K$ is the size of the fleet and $|W|$ is the maximum number of departure time breakpoints for any customer. Each breakpoint represents a possible duration to travel from the current best predecessor to the newly considered route-ending customer $i$. As a result, the worst-case complexity of their approach is \(O(nK|W|)\). 

The work in \citet{zhao_hybrid_2024} is an important showcase of the potential of the double-ended queue data structure in speeding up the Split. However, their resulting algorithm is not linear and does not handle penalty factors. In contrast, we develop linear Splits for the hard and soft VRPSPDTW.  As both the VRPSPD and VRPTW are well understood by the VRP community \citep{toth2014vehicle}, we suspect that the algorithmic strategies employed in the proposed Splits may provide hints toward developing faster Splits and evaluation routines for other VRP variants.
\section{VRPSPDTW Split shortest-path problems}
\label{sec:split-problem}\label{sec:vrpspdtw}

Let \(\pi\) be a fixed giant tour containing the \(n\) customers of a
VRPSPDTW instance. Following previous Split works
\citep{prins2004simple,vidal2016split}, customers are referred to by
their position in the tour, so that \(\pi=(1,\ldots,n)\). Customer
\(z\) has delivery and pickup quantities \(d_z,p_z\geq0\), time window
\([a_z,b_z]\), and service time \(s_z\geq0\). Vehicles have capacity
\(Q\), leave a common depot at time 0, and must return to the depot by
time \(T\). 

For a fixed giant tour, an induced route can use only travel from the
depot to a customer, between consecutive customers of the tour, and
from a customer back to the depot. The required travel costs are
therefore \(c_{0z}\) and \(c_{z0}\) for \(z\in\{1,\ldots,n\}\), and
\(c_{z,z+1}\) for \(z\in\{1,\ldots,n-1\}\). The corresponding travel
durations are denoted \(t_{0z}\), \(t_{z0}\), and \(t_{z,z+1}\). No
symmetry is assumed.

\begin{table}[t]
\centering
\caption{Principal notation for the Split problems.}
\label{tab:split-problem-notation}
\small
\begin{tabularx}{\linewidth}{@{}lL@{}}
\toprule
Symbol & Meaning \\
\midrule
\(R(i,x)\)
& Route \((0,i+1,\ldots,x,0)\) induced by predecessor \(i\) and final
customer \(x\). \\

\(c(i,x)\)
& Base, unpenalized travel cost of \(R(i,x)\), equal to
\(c_{0,i+1}+C[x]-C[i+1]+c_{x0}\). \\

\(D[x],P[x]\)
& Prefix deliveries and prefix pickups along the giant tour. \\

\(C[x],S[x]\)
& Prefix travel cost and prefix service-plus-travel duration along the
giant tour. \\

\(L(i,z,x),H(i,x)\)
& Load after position \(z\) and maximum load on \(R(i,x)\). \\

\(h(i,x)\)
& Latest position attaining $H(i,x)$ on \(R(i,x)\). \\

\(A_i(z),B_i(z)\)
& Service start before time warp and time warp incurred at customer
\(z\). \\

\(\operatorname{tw}(i,x)\)
& Total customer and depot-return time warp of \(R(i,x)\). \\

\addlinespace
\(\operatorname{pen}_{\alpha h}(i,x)\)
& VRPSPD capacity penalty
\(\alpha\max\{H(i,x)-Q,0\}\). \\

\(\operatorname{pen}_{\alpha d}(i,x)\)
& Delivery-only capacity penalty
\(\alpha\max\{D[x]-D[i]-Q,0\}\). \\

\(\operatorname{pen}_{\beta a}(i,x)\)
& Time-window penalty \(\beta\operatorname{tw}(i,x)\). \\

\addlinespace
\(c_{\alpha h}(i,x)\)
& Soft VRPSPD route cost
\(c(i,x)+\operatorname{pen}_{\alpha h}(i,x)\). \\

\(c_{\alpha d,\beta a}(i,x)\)
& Soft VRPTW route cost
\(c(i,x)+\operatorname{pen}_{\alpha d}(i,x)
+\operatorname{pen}_{\beta a}(i,x)\). \\

\(c_{\alpha h,\beta a}(i,x)\)
& Soft VRPSPDTW route cost
\(c(i,x)+\operatorname{pen}_{\alpha h}(i,x)
+\operatorname{pen}_{\beta a}(i,x)\). \\

\addlinespace
\(\operatorname{pot}(x),\operatorname{pred}(x)\)
& Shortest-path cost to Split vertex \(x\) and a predecessor attaining
it. \\
\bottomrule
\end{tabularx}
\end{table}

The following prefix values are used throughout the paper:
\begin{align}
D[x]&=\sum_{z=1}^{x}d_z,
& P[x]&=\sum_{z=1}^{x}p_z,                               \label{eq:prefix-loads}\\
C[x]&=\sum_{z=1}^{x-1}c_{z,z+1},
& S[x]&=\sum_{z=1}^{x-1}(s_z+t_{z,z+1}).                 \label{eq:prefix-tour}
\end{align}
Here, \(D[0]=P[0]=0\) and \(C[1]=S[1]=0\). The values
\(C[\cdot]\) give the accumulated travel costs along the giant tour,
while \(S[\cdot]\) gives the accumulated service and travel
durations. All prefix values require \(O(n)\) time to construct.

For \(0\leq i<x\leq n\), let \(R(i,x)=(0,i+1,\ldots,x,0)\) denote the
route serving the contiguous segment from customer \(i+1\) through
customer \(x\). Its base cost is \(c(i,x)\) (defined in
Table~\ref{tab:split-problem-notation}).

\subsection{Vehicle load}

Immediately after departing position \(z\in\{i,\ldots,x\}\), where
\(z=i\) denotes departure from the depot, the load on \(R(i,x)\) is
\begin{equation}\label{eq:route-load}
 L(i,z,x)=D[x]-D[z]+P[z]-P[i].
\end{equation}
At position \(z\), the vehicle carries the deliveries remaining for
customers \(z+1,\ldots,x\), together with the pickups collected from
customers \(i+1,\ldots,z\).

The maximum load incurred by the route is
\begin{equation}\label{eq:problem-spd-maximum}
 H(i,x)
 =\max_{i\leq z\leq x}L(i,z,x)
 =D[x]-P[i]+\max_{i\leq z\leq x}\{P[z]-D[z]\}.
\end{equation}
We denote the latest position attaining this maximum as
\begin{align}
    h(i,x)=\max\arg\max_{i\leq z\leq x}\{P[z]-D[z]\}\label{eq:highest-point},
\end{align} where the outer
maximum breaks ties in favor of the latest position. Thus,
\(H(i,x)=L(i,h(i,x),x)\).
The route satisfies the VRPSPD capacity constraint whenever
\(H(i,x)\leq Q\). Given a positive penalty multiplier \(\alpha\),
the soft VRPSPD capacity penalty is
\(\operatorname{pen}_{\alpha h}(i,x)\) (penalty functions are
defined in Table~\ref{tab:split-problem-notation}).

The delivery-only case assumes \(p_z=0\) for every customer. As in
\citet{vidal2016split} for the CVRP, the load incurred by
\(R(i,x)\) is then \(D[x]-D[i]\). This route-load calculation is also
used by the soft VRPTW Split. The capacity penalty for this case is denoted as
\(\operatorname{pen}_{\alpha d}(i,x)\).

\subsection{Service times and time warp}

For route \(R(i,x)\), let \(A_i(z)\) denote the service start time at
customer \(z\), before applying any time warp at that customer, and let
\(B_i(z)\) denote the warp incurred there. At the first customer,
\begin{equation}\label{eq:time-start-first}
 A_i(i+1)=\max\{t_{0,i+1},a_{i+1}\}.
\end{equation}
At each customer, \(B_i(z)=\max\{A_i(z)-b_z,0\}\). After this warp,
service is treated as having started at \(A_i(z)-B_i(z)\). Hence, for
\(z=i+2,\ldots,x\),
\begin{equation}\label{eq:time-start-recurrence}
 A_i(z)=\max\{A_i(z-1)-B_i(z-1)+s_{z-1}+t_{z-1,z},a_z\}.
\end{equation}
Thus, whenever a customer incurs time warp, the same amount is removed
from the time propagated to subsequent customers.

The total time warp, including a late return to the depot, is
\begin{equation}\label{eq:problem-time-warp}
 \operatorname{tw}(i,x)
 =\sum_{z=i+1}^{x}B_i(z)
 +\max\{A_i(x)-B_i(x)+s_x+t_{x0}-T,0\}.
\end{equation}
Hard time-window feasibility is equivalent to
\(\operatorname{tw}(i,x)=0\). Given a positive multiplier \(\beta\),
the time-warp penalty is denoted as
\(\operatorname{pen}_{\beta a}(i,x)\).

The time-window results assume the following weakened triangle
inequalities along the giant tour:
\begin{align}
 t_{0,z+1}
 &\leq \max\{t_{0z},a_z\}+s_z+t_{z,z+1},
 \label{eq:weak-triangle-out}\\
 b_z+s_z+t_{z0}
 &\leq \max\{b_z+s_z+t_{z,z+1},a_{z+1}\}
       +s_{z+1}+t_{z+1,0},
 \label{eq:weak-triangle-in}
\end{align}
for \(z=1,\ldots,n-1\). The first inequality states that traveling
directly from the depot to customer \(z+1\) cannot begin service there
later than traveling first to customer \(z\), waiting if necessary,
servicing it, and then continuing to \(z+1\). The second states that
returning directly to the depot after servicing customer \(z\) cannot
take longer than first visiting and servicing customer \(z+1\), with
any required waiting, and then returning to the depot.

Every singleton route \(R(z-1,z)\) is also assumed feasible with
respect to the time windows and depot return time, so
\(\operatorname{tw}(z-1,z)=0\) for \(z=1,\ldots,n\). These assumptions
apply only to the time-window results. Singleton feasibility is also
assumed for capacity constraints.

\subsection{Hard and soft Split graphs}\label{sec:cvrp}

Let \(V=\{0,\ldots,n\}\) be the vertices of the Split graph. Every pair
\(0\leq i<x\leq n\) corresponds to route \(R(i,x)\), and the set of
all possible Split arcs is
\(\mathcal A_{\mathrm{all}}=\{(i,x):0\leq i<x\leq n\}\).

For the hard VRPSPDTW, an arc is retained only when its route satisfies
both the capacity and time-window constraints:
\begin{equation}\label{eq:hard-arc-set}
 \mathcal A_{h,a}
 =\{(i,x)\in\mathcal A_{\mathrm{all}}:
      H(i,x)\leq Q,\ \operatorname{tw}(i,x)=0\}.
\end{equation}
The hard VRPSPDTW Split finds a shortest path in
\(\mathcal G_{h,a}=(V,\mathcal A_{h,a},c)\), where \(c\) is the
arc-weight function. All route-cost functions used below are defined in
Table~\ref{tab:split-problem-notation}. Setting all
pickups to zero gives the hard VRPTW, while omitting the time-window
condition gives the hard VRPSPD.

For each soft variant, every arc in \(\mathcal A_{\mathrm{all}}\) is
permitted. The soft VRPSPDTW Split graph is
\(\mathcal G_{\alpha h,\beta a}
=(V,\mathcal A_{\mathrm{all}},c_{\alpha h,\beta a})\). Its two
special cases considered below are the soft VRPSPD graph
\(\mathcal G_{\alpha h}
=(V,\mathcal A_{\mathrm{all}},c_{\alpha h})\) and the soft VRPTW graph
\(\mathcal G_{\alpha d,\beta a}
=(V,\mathcal A_{\mathrm{all}},c_{\alpha d,\beta a})\).

For any of these graphs, a path from 0 to \(n\) corresponds to a
partition of the giant tour, and its path weight is the corresponding
hard or penalized objective value. The label \(\operatorname{pot}(x)\)
denotes the shortest-path cost from 0 to \(x\), and 
\(\operatorname{pred}(x)\) denotes a predecessor attaining that cost.

\subsection{Constant cost differences}

The linear Split of \citet{vidal2016split} exploits the following
property of the base route cost \(c(i,x)\):

\begin{property}[Constant cost differences]
\label{prop:constant-cost-difference}
For any \(0\leq i<j<x\leq n\),
\[
\begin{aligned}
&\operatorname{pot}(i)+c(i,x)
 -\operatorname{pot}(j)-c(j,x)\\
&\quad=\operatorname{pot}(i)+c_{0,i+1}-C[i+1]
 -\operatorname{pot}(j)-c_{0,j+1}+C[j+1].
\end{aligned}
\]
Therefore, the difference between the candidate path costs through two
fixed predecessors is independent of their shared final customer.
\end{property}

As a result of this property, whenever \(i\) and \(j\) are both admissible predecessors of customer
\(x\), their candidate path-cost difference is unchanged
for every later customer for which both remain admissible.
Property~\ref{prop:constant-cost-difference} underlies the dominance
rules used by the linear CVRP Split. The hard VRPSPDTW Split preserves the
property directly, while the soft Splits require additional treatment
because their penalty differences need not remain constant.

\section{Generalized Hard Split in Linear Time}
\label{sec:hard-split}\label{sec:vrpspdtw-split}

The constraint-specific preprocessing for the VRPSPDTW is separated
from the generic shortest-path algorithm to make the extent of the
approach's generalizability explicit. Let \(\mathsf{feas}_q(i,x)\) indicate whether route
\(R(i,x)\) satisfies hard constraint \(q\). The proposed Split applies
to constraints satisfying
\begin{align}
\neg\mathsf{feas}_q(i,x)
&\Rightarrow \neg\mathsf{feas}_q(i,x+1),
&&0\leq i<x<n,
\tag{H1}\label{cond:H1}\\
\mathsf{feas}_q(i,x)
&\Rightarrow \mathsf{feas}_q(i+1,x),
&&0\leq i\leq x-2,\ x\leq n,
\tag{H2}\label{cond:H2}\\
\mathsf{feas}_q(i,i+1),
&&&0\leq i<n.
\tag{H3}\label{cond:H3}
\end{align}
In other words, appending customers cannot restore an infeasible route,
deleting the first customer cannot destroy feasibility, and every
customer can be served alone.

\begin{table}[t]
\centering
\caption{Notation for the generalized hard Split.}
\label{tab:hard-split-notation}
\small
\begin{tabularx}{\linewidth}{@{}lL@{}}
\toprule
Symbol & Meaning \\
\midrule
\(\mathsf{feas}_q(i,x)\) & Whether route \(R(i,x)\) satisfies hard constraint \(q\). \\
\(\rho_q(i)\) & First customer at which predecessor \(i\) violates constraint \(q\), or \(n+1\) if none exists. \\
\(\rho(i)\) & First customer violating any constraint in the collection \(\mathcal Q\). \\
\(\Lambda\) & Deque of nondominated predecessors, guaranteed to be ordered by index and candidate path cost after maintenance. \\
\(\operatorname{dominated}(i,j)\) & Test determining whether the older predecessor \(i\) can be removed in favor of \(j\). \\
\bottomrule
\end{tabularx}
\end{table}

Constraints satisfying these conditions have a well-defined
\emph{route-feasibility limit}
\begin{equation}\label{eq:hard-route-limit}
\rho_q(i)
=\min\left(
  \{x:i<x\leq n,\ \neg\mathsf{feas}_q(i,x)\}\cup\{n+1\}
 \right),
\qquad i=0,\ldots,n-1,
\end{equation}
where \(\rho_q(i)\) is the first customer at which route \(R(i,x)\)
violates constraint \(q\), or \(n+1\) when no violation occurs.
Conditions~\eqref{cond:H1} and~\eqref{cond:H3} imply that
\(R(i,x)\) satisfies constraint \(q\) as long as
\(i<x<\rho_q(i)\). Conditions~\eqref{cond:H2} and~\eqref{cond:H3}
imply \(\rho_q(i)\leq\rho_q(i+1)\).

For a fixed collection \(\mathcal Q\) of hard constraints, their
conjunction has the single nondecreasing limit
\(\rho(i)=\min_{q\in\mathcal Q}\rho_q(i)\). Route \(R(i,x)\) satisfies
every constraint in \(\mathcal Q\) as long as \(x<\rho(i)\).

For two predecessors \(i<j\), we denote the dominance indicator function
\begin{equation}\label{eq:dominated-hard}
\operatorname{dominated}(i,j)
\Longleftrightarrow
\operatorname{pot}(i)+c_{0,i+1}-C[i+1]
\geq
\operatorname{pot}(j)+c_{0,j+1}-C[j+1].
\end{equation}
The function determines whether the older predecessor
\(i\) is no cheaper than \(j\). If it does, predecessor \(j\) is
both weakly cheaper for every shared final customer by
Property~\ref{prop:constant-cost-difference} and feasible wherever
\(i\) is feasible by Condition~\eqref{cond:H2}. Thus, \(i\) is
\emph{dominated} by \(j\) and can be removed permanently. Ties are broken in favor of the newer predecessor.

When \(\neg\operatorname{dominated}(i,j)\), predecessor \(i\) is
cheaper than \(j\) for every customer \(x\) satisfying
\(j<x<\rho(i)\). However, predecessor \(j\) cannot be discarded because it may
remain feasible after \(i\) reaches its route-feasibility limit. This is the same logic on which the linear CVRP Split of
\citet{vidal2016split} relies.

Algorithm~\ref{alg:generic-hard-split} presents the Generalized Split
for any set of constraints \(\mathcal Q\) that each satisfy
Conditions~\eqref{cond:H1}--\eqref{cond:H3}. The deque \(\Lambda\)
stores nondominated predecessors in increasing order of index and
candidate path cost over final customers for which adjacent predecessors
are both feasible. The back-removal and insertion steps on
lines~\ref{line:hard-back-start}--\ref{line:hard-insert} use the
dominance indicator in Equation~\eqref{eq:dominated-hard}. The
front-removal loop on lines~\ref{line:hard-front-start}--\ref{line:hard-front-end}
discards predecessors whose route-feasibility limits have been reached.

\begin{algorithm}[t]
\caption{Generalized linear Split for hard constraints}
\label{alg:generic-hard-split}
\DontPrintSemicolon
\SetKwInOut{Input}{Input}\SetKwInOut{Output}{Output}
\Input{Tour length \(n\); route-feasibility limit
\(\rho\); travel costs \(c_{0z}\) and \(c_{z0}\), and prefix values
\(C[z]\), for \(z=1,\ldots,n\)}
\Output{Shortest-path costs
\(\operatorname{pot}(1),\ldots,\operatorname{pot}(n)\) and
predecessors
\(\operatorname{pred}(1),\ldots,\operatorname{pred}(n)\)}
\(\operatorname{pot}(0)\leftarrow0\); \(\Lambda\leftarrow()\)\nllabel{line:hard-init}\;
\For{\(x\leftarrow1\) \KwTo \(n\)}{
  \While{\(\Lambda\neq()\) \textnormal{and}
  \(\operatorname{dominated}(\Lambda.\mathrm{back},x-1)\)\nllabel{line:hard-back-start}}{
    \(\Lambda.\mathrm{removeBack}()\)\nllabel{line:hard-back-end}\;
  }
  \(\Lambda.\mathrm{pushBack}(x-1)\)\nllabel{line:hard-insert}\;
  \While{\(\rho(\Lambda.\mathrm{front})\leq x\)\nllabel{line:hard-front-start}}{
    \(\Lambda.\mathrm{removeFront}()\)\nllabel{line:hard-front-end}\;
  }
  \(\operatorname{pred}(x)\leftarrow\Lambda.\mathrm{front}\)\nllabel{line:hard-record-pred}\;
  \(\operatorname{pot}(x)\leftarrow
  \operatorname{pot}(\operatorname{pred}(x))
  +c(\operatorname{pred}(x),x)\)\nllabel{line:hard-record-pot}\;
}
\end{algorithm}

\begin{theorem}\label{thm:generic-hard-split}
Assume that the base route costs satisfy
Property~\ref{prop:constant-cost-difference}, each constraint in a
fixed collection \(\mathcal Q\) satisfies
Conditions~\eqref{cond:H1}--\eqref{cond:H3}, and all route-feasibility
limits \(\rho_q\) can be constructed in \(O(n)\) time. Then
Algorithm~\ref{alg:generic-hard-split} returns a minimum-cost
partition in \(O(n)\) time and \(O(n)\) space. Its predecessor labels
reconstruct an optimal partition.
\end{theorem}

The proof and constructions of the route-feasibility limits appear in
\ref{app:hard-reach}. In particular, the limits associated
with pickup-and-delivery capacity and hard time windows can each be constructed in
linear time by exploiting variant-specific arc-invariant properties.
The main Split consequently sees only the combined limit \(\rho\).

\section{Soft VRPTW Split in Linear Time}
\label{sec:generic-penalties}\label{sec:vrptw}\label{sec:soft-vrptw}

The soft CVRP Split of \citet{vidal2016split} accounts for one
capacity penalty. The soft VRPTW route cost \(c_{\alpha d,\beta a}\)
combines the CVRP capacity penalty
\(\operatorname{pen}_{\alpha d}\) and the time-warp penalty
\(\operatorname{pen}_{\beta a}\). The proposed Split tracks the two
penalty boundaries within the predecessor deque and relies on the fact
that both penalties satisfy five conditions defined below.

In the conditions below, \(q\in\{\alpha d,\beta a\}\) identifies the
penalty \(\operatorname{pen}_q\), whose route-feasibility limit is
\(\rho_q\). The candidate path cost through predecessor \(i\) is
\(\operatorname{pot}(i)+c_{\alpha d,\beta a}(i,x)\).

\begin{table}[t]
\centering
\caption{Additional notation for the soft VRPTW Split.}
\label{tab:soft-vrptw-notation}
\small
\begin{tabularx}{\linewidth}{@{}lL@{}}
\toprule
Symbol & Meaning \\
\midrule
\(c_{\alpha d,\beta a}(i,x)\)
& Penalized route cost for the soft VRPTW
(defined in Table~\ref{tab:split-problem-notation}). \\

\(\operatorname{pen}_{\alpha d}(i,x),
  \operatorname{pen}_{\beta a}(i,x)\)
& Delivery-capacity and time-warp penalties
(defined in Table~\ref{tab:split-problem-notation}). \\

\(q\in\{\alpha d,\beta a\}\)
& Penalty label: delivery capacity \(\alpha d\) or time warp \(\beta a\). \\

\(\rho_q(i)\)
& First customer at which predecessor \(i\) incurs a positive
\(\operatorname{pen}_q\), or \(n+1\) if none exists. \\

\(Y_x^1,Y_x^2,Y_x^3\)
& Predecessors penalized by both penalties, exactly one penalty, and
neither penalty, respectively. \\

\(\Lambda\)
& Deque of nondominated predecessors, guaranteed to be ordered by index
and candidate path cost after maintenance. \\

\(\Lambda.\mathrm{feas}\)
& Cursor to the oldest nondominated predecessor in \(Y_x^3\), or
\(\bot\) when \(Y_x^3\) is empty. \\

\(\Lambda.\mathrm{front2}\)
& Second predecessor in \(\Lambda\), when it exists. \\

\(\operatorname{dominated}_{\alpha d,\beta a}(i,j,x)\)
& Test determining whether the older predecessor \(i\) can be removed
in favor of \(j\) at customer \(x\). \\

\bottomrule
\end{tabularx}
\end{table}

First, the proposed Split requires that both penalty functions satisfy
the following monotonicity conditions:
\begin{align}
\operatorname{pen}_q(i,x)
&\leq \operatorname{pen}_q(i,x+1),
&&0\leq i<x<n,
\tag{P1}\label{cond:P1}\\
\operatorname{pen}_q(i+1,x)
&\leq \operatorname{pen}_q(i,x),
&&0\leq i\leq x-2,\ x\leq n,
\tag{P2}\label{cond:P2}\\
\operatorname{pen}_q(i,i+1)
&=0,
&&0\leq i<n.
\tag{P3}\label{cond:P3}
\end{align}
Thus, adding a customer to either end of a route cannot decrease its
penalty, and every singleton route is unpenalized.

Similarly to the hard-variant case, penalties satisfying these conditions
have a well-defined \emph{route-feasibility limit}, written as
\begin{equation}\label{eq:soft-route-limit}
\rho_q(i)
=
\min\bigl(
 \{x:i<x\leq n,\ \operatorname{pen}_q(i,x)>0\}
 \cup\{n+1\}
\bigr),
\qquad i=0,\ldots,n-1,
\end{equation}
where \(\rho_q(i)\) is the first customer at which route \(R(i,x)\)
incurs a positive value of \(\operatorname{pen}_q\), or \(n+1\) if the
penalty remains zero.
These limits coincide with the corresponding hard-feasibility limits:
\(\rho_{\alpha d}(i)=\rho_d(i)\) and
\(\rho_{\beta a}(i)=\rho_a(i)\).
Conditions~\eqref{cond:P1} and~\eqref{cond:P3} imply that predecessor
\(i\) is unpenalized when \(i<x<\rho_q(i)\), while
Conditions~\eqref{cond:P2} and~\eqref{cond:P3} imply that
\(\rho_q(i)\leq\rho_q(i+1)\).

Since \(\rho_q\) is nondecreasing, predecessors \(i<j\) are both
penalized at customer \(x\) whenever \(\rho_q(j)\leq x\). The proposed
Split assumes
\begin{align}
\operatorname{pen}_q(i,y)-\operatorname{pen}_q(j,y)
&=
\operatorname{pen}_q(i,x)-\operatorname{pen}_q(j,x),
&&
\substack{0\leq i<j<x\leq y\leq n,\\
          \rho_q(j)\leq x.}
\tag{P4}\label{cond:P4}
\end{align}
This condition states that once two predecessors are
penalized by \(\operatorname{pen}_q\), the difference between their penalties remains
constant as further customers are appended.

For \(i<j\) such that \(\rho_q(i)<\rho_q(j)\leq n\), the proposed
Split also assumes
\begin{align}
\operatorname{pen}_q(j,\rho_q(j))
&\leq
\operatorname{pen}_q(i,\rho_q(j))
-
\operatorname{pen}_q(i,\rho_q(j)-1).
\tag{P5}\label{cond:P5}
\end{align}
This condition states that when the later predecessor \(j\)
first becomes penalized, its initial penalty cannot exceed the increase
in penalty incurred by the already-penalized predecessor \(i\) at the
same customer.

\begin{lemma}\label{lem:penalty-difference-monotone}
Suppose the penalty function \(\operatorname{pen}_q\) satisfies
Conditions~\eqref{cond:P1}--\eqref{cond:P5}. For any predecessors
\(i<j\), the difference
\(\operatorname{pen}_q(i,x)-\operatorname{pen}_q(j,x)\) is
nondecreasing as customers are appended.
\end{lemma}

\begin{proof}
The route-feasibility limits satisfy
\(\rho_q(i)\leq\rho_q(j)\). In the case that both predecessors are unpenalized,
their penalty difference is zero. In the case that only \(i\) is penalized, the
difference equals \(\operatorname{pen}_q(i,x)\), which is
nondecreasing by Condition~\eqref{cond:P1}.

By Condition~\eqref{cond:P5}, the difference cannot decrease when \(j\)
first becomes penalized if \(\rho_q(i)<\rho_q(j)\leq n\). If
\(\rho_q(i)=\rho_q(j)\leq n\), repeated application of
Condition~\eqref{cond:P2} makes the new difference nonnegative. If
\(\rho_q(j)=n+1\), the later predecessor never becomes penalized and
the difference remains nondecreasing by Condition~\eqref{cond:P1}.
By Condition~\eqref{cond:P4}, the difference remains constant once both
predecessors are penalized.
\end{proof}

At customer \(x\), the two route-feasibility limits partition the
predecessors into three regions:
\begin{equation}\label{eq:penalty-regions}
\begin{aligned}
Y_x^1
&=
\left\{
 i:0\leq i<x,\
 \max\{\rho_{\alpha d}(i),\rho_{\beta a}(i)\}\leq x
\right\},\\
Y_x^2
&=
\left\{
 i:0\leq i<x,\
 \min\{\rho_{\alpha d}(i),\rho_{\beta a}(i)\}\leq x
 <
 \max\{\rho_{\alpha d}(i),\rho_{\beta a}(i)\}
\right\},\\
Y_x^3
&=
\left\{
 i:0\leq i<x,\
 x<\min\{\rho_{\alpha d}(i),\rho_{\beta a}(i)\}
\right\}.
\end{aligned}
\end{equation}
\(Y_x^1\) contains predecessors penalized by both delivery
capacity and time warp, \(Y_x^2\) contains predecessors penalized by one of the two penalties, and \(Y_x^3\) contains predecessors
penalized by neither.

Since \(\rho_{\alpha d}\) and \(\rho_{\beta a}\) are both
nondecreasing, the three regions are contiguous and occur in the order
\(Y_x^1,Y_x^2,Y_x^3\) from older to newer predecessors, although any
region may be empty. Depending on the relative positions of the two
penalty boundaries, \(Y_x^2\) contains either predecessors penalized
only by delivery-capacity excess or predecessors penalized only by
time warp.

By Property~\ref{prop:constant-cost-difference} and
Condition~\eqref{cond:P4}, adjacent predecessors that remain in the
same fixed region from iteration \(x-1\) to iteration \(x\) retain
their candidate-cost difference. Comparisons can change at region
boundaries or when predecessors change regions.

By Lemma~\ref{lem:penalty-difference-monotone}, the difference
\[
\operatorname{pot}(i)+c_{\alpha d,\beta a}(i,x)
-
\operatorname{pot}(j)-c_{\alpha d,\beta a}(j,x)
\]
is nondecreasing in \(x\) for any \(i<j\). Once an earlier predecessor
becomes no cheaper than a later predecessor, it can therefore be
removed permanently. We denote the corresponding dominance predicate
as
\begin{equation}\label{eq:dominated-penalized}
\begin{aligned}
\operatorname{dominated}_{\alpha d,\beta a}(i,j,x)
\Longleftrightarrow{}&
\operatorname{pot}(i)+c_{\alpha d,\beta a}(i,x)\\
&\geq
\operatorname{pot}(j)+c_{\alpha d,\beta a}(j,x),
\qquad i<j.
\end{aligned}
\end{equation}
Ties are broken in favor of the newer predecessor.

\subsection{Linear Split algorithm}

Algorithm~\ref{alg:stable-penalty-split} presents the linear Split for
the soft VRPTW. The algorithm maintains a deque \(\Lambda\) of
candidate predecessors in increasing index order. The nonempty regions among
\(Y_x^1,Y_x^2,Y_x^3\) form contiguous portions of \(\Lambda\). At the
end of each iteration, candidate path costs are guaranteed to strictly increase from the
front to the back of the deque, so \(\Lambda.\mathrm{front}\) is an
optimal predecessor of the current customer.

The cursor \(\Lambda.\mathrm{feas}\) points to the oldest nondominated
predecessor in \(Y_x^3\), thereby identifying the boundary between
\(Y_x^2\) and \(Y_x^3\). The value
\(\Lambda.\mathrm{feas}=\bot\) indicates that \(Y_x^3\) is empty.
The boundary between \(Y_x^1\) and \(Y_x^2\) is maintained from the
front of the deque. The algorithm retains at most one predecessor in
\(Y_x^1\), so this boundary, when present, lies between
\(\Lambda.\mathrm{front}\) and \(\Lambda.\mathrm{front2}\).

The operator \(\Lambda.\mathrm{feasPrev}\) accesses the predecessor
immediately before \(\Lambda.\mathrm{feas}\), while
\(\Lambda.\mathrm{removeFeasPrev}()\) removes that predecessor without
moving the cursor. The operator
\(\Lambda.\mathrm{moveFeasNext}()\) advances the cursor to its
successor, or sets it to \(\bot\) when no successor exists. Similarly,
\(\Lambda.\mathrm{front2}\) accesses the second predecessor in the
deque, and \(\Lambda.\mathrm{removeFront2}()\) removes it. These
 operations are detailed in
\ref{app:deque-operators}.

The route-feasibility limits and all data needed to evaluate
\(\operatorname{pen}_{\alpha d}\) and
\(\operatorname{pen}_{\beta a}\) in constant time are constructed
before the Split.
\ref{app:capacity-penalty-preprocessing} constructs
\(\rho_{\alpha d}\),
\ref{app:time-window-route-limits} constructs
\(\rho_{\beta a}\) and \(r_a\), and
\ref{app:time-warp-preprocessing} constructs the prefix-warp
data used for constant-time evaluation of
\(\operatorname{pen}_{\beta a}\).

\begin{algorithm}[t]
\caption{Linear Split for the soft VRPTW}
\label{alg:stable-penalty-split}
\footnotesize
\DontPrintSemicolon
\SetKwInOut{Input}{Input}
\SetKwInOut{Output}{Output}

\Input{Tour length \(n\); \(c_{0z}\), \(c_{z0}\), and \(C[z]\) for
\(z=1,\ldots,n\); route-feasibility limits
\(\rho_{\alpha d},\rho_{\beta a}\); data permitting \(O(1)\)
evaluation of \(\operatorname{pen}_{\alpha d}(i,x)\) and
\(\operatorname{pen}_{\beta a}(i,x)\)}
\Output{Shortest-path costs
\(\operatorname{pot}(1),\ldots,\operatorname{pot}(n)\) and
predecessors
\(\operatorname{pred}(1),\ldots,\operatorname{pred}(n)\)}

\(\operatorname{pot}(0)\leftarrow0\);
\(\Lambda\leftarrow()\);
\(\Lambda.\mathrm{feas}\leftarrow\bot\)\nllabel{line:vrptw-init}\;

\For{\(x\leftarrow1\) \KwTo \(n\)}{
  \While{\(\Lambda\neq()\) \textnormal{and}
         \(\operatorname{dominated}(\Lambda.\mathrm{back},x-1)\)\nllabel{line:vrptw-back-start}}{
    \If{\(\Lambda.\mathrm{feas}=\Lambda.\mathrm{back}\)}{
      \(\Lambda.\mathrm{feas}\leftarrow\bot\)\;
    }
    \(\Lambda.\mathrm{removeBack}()\)\nllabel{line:vrptw-back-end}\;
  }

  \(\Lambda.\mathrm{pushBack}(x-1)\)\nllabel{line:vrptw-insert}\;
  \If{\(\Lambda.\mathrm{feas}=\bot\)}{
    \(\Lambda.\mathrm{feas}\leftarrow\Lambda.\mathrm{back}\)\nllabel{line:vrptw-feas-init}\;
  }

  \While{\(\Lambda.\mathrm{feas}\neq\bot\) \textnormal{and}
         \(\bigl(
          \rho_{\alpha d}(\Lambda.\mathrm{feas})\leq x
          \textnormal{ or }
          \rho_{\beta a}(\Lambda.\mathrm{feas})\leq x
         \bigr)\)\nllabel{line:vrptw-feas-advance-start}}{
    \While{\(\Lambda.\mathrm{hasFeasPrev}\) \textnormal{and}
           \(\operatorname{dominated}_{\alpha d,\beta a}
             (\Lambda.\mathrm{feasPrev},
              \Lambda.\mathrm{feas},x)\)\nllabel{line:vrptw-cross-order-start}}{
      \(\Lambda.\mathrm{removeFeasPrev}()\)\nllabel{line:vrptw-cross-order-end}\;
    }
    \(\Lambda.\mathrm{moveFeasNext}()\)\nllabel{line:vrptw-feas-advance-end}\;
  }

  \While{\(\Lambda.\mathrm{feas}\neq\bot\) \textnormal{and}
         \(\Lambda.\mathrm{hasFeasPrev}\) \textnormal{and}
         \(\operatorname{dominated}_{\alpha d,\beta a}
           (\Lambda.\mathrm{feasPrev},
            \Lambda.\mathrm{feas},x)\)\nllabel{line:vrptw-current-boundary-start}}{
    \(\Lambda.\mathrm{removeFeasPrev}()\)\nllabel{line:vrptw-current-boundary-end}\;
  }

  \While{\(\Lambda.\mathrm{size}>1\) \textnormal{and}
         \(\Lambda.\mathrm{front2}\in Y_x^1\)\nllabel{line:vrptw-y1-start}}{
    \eIf{\(\operatorname{dominated}_{\alpha d,\beta a}
             (\Lambda.\mathrm{front},
              \Lambda.\mathrm{front2},x)\)}{
      \(\Lambda.\mathrm{removeFront}()\)\;
    }{
      \(\Lambda.\mathrm{removeFront2}()\)\nllabel{line:vrptw-y1-end}\;
    }
  }

  \If{\(\Lambda.\mathrm{size}>1\) \textnormal{and}
       \(\Lambda.\mathrm{front}\in Y_x^1\) \textnormal{and}
       \(\operatorname{dominated}_{\alpha d,\beta a}
         (\Lambda.\mathrm{front},
          \Lambda.\mathrm{front2},x)\)\nllabel{line:vrptw-first-boundary-start}}{
    \(\Lambda.\mathrm{removeFront}()\)\nllabel{line:vrptw-first-boundary-end}\;
  }

  \(\operatorname{pred}(x)\leftarrow\Lambda.\mathrm{front}\)\nllabel{line:vrptw-record-pred}\;
  \(\operatorname{pot}(x)\leftarrow
    \operatorname{pot}(\operatorname{pred}(x))
    +c_{\alpha d,\beta a}(\operatorname{pred}(x),x)\)\nllabel{line:vrptw-record-pot}\;
}
\end{algorithm}

The initial back-removal loop on
lines~\ref{line:vrptw-back-start}--\ref{line:vrptw-insert} performs the
same back insertion as the hard CVRP Split. The newly inserted
predecessor \(x-1\) belongs to \(Y_x^3\) by
Condition~\eqref{cond:P3}. An older predecessor already dominated in
base cost cannot become preferable after the nonnegative penalties are
included.

Lines~\ref{line:vrptw-feas-advance-start}--\ref{line:vrptw-feas-advance-end}
advance \(\Lambda.\mathrm{feas}\) past every predecessor that has left
\(Y_x^3\), simultaneously ensuring that predecessors entering
\(Y_x^2\) during the iteration remain in \(\Lambda\) only when they are
nondominated. Before each cursor move, the inner loop compares the
predecessor crossing the boundary with its immediate older neighbor and removes
that neighbor whenever it is dominated. Lines~\ref{line:vrptw-current-boundary-start}--\ref{line:vrptw-current-boundary-end}
check the current \(Y_x^2/Y_x^3\) boundary even when the cursor does
not move, since a predecessor in \(Y_x^2\) may continue accumulating
penalty while the oldest nondominated predecessor in \(Y_x^3\) remains
unpenalized.

Lines~\ref{line:vrptw-y1-start}--\ref{line:vrptw-y1-end} compare
predecessors in \(Y_x^1\) until at most one remains. By
Condition~\eqref{cond:P4}, every comparison within \(Y_x^1\) is
permanent. Lines~\ref{line:vrptw-first-boundary-start}--\ref{line:vrptw-first-boundary-end}
compare that remaining predecessor, when it exists, with the first
predecessor in the next nonempty region. After these operations,
candidate path costs are guaranteed to increase from the front to the
back of \(\Lambda\).

\begin{theorem}\label{thm:stable-penalty-split}
Assume that the base route costs satisfy
Property~\ref{prop:constant-cost-difference}, and that
\(\operatorname{pen}_{\alpha d}\) and
\(\operatorname{pen}_{\beta a}\) satisfy
Conditions~\eqref{cond:P1}--\eqref{cond:P5}. Assume further that
\(\rho_{\alpha d}\), \(\rho_{\beta a}\), and the data required to
evaluate both penalties in constant time can be constructed in
\(O(n)\) time. Then Algorithm~\ref{alg:stable-penalty-split} returns
an exact minimum-cost partition for
\(\mathcal G_{\alpha d,\beta a}\) in \(O(n)\) time and \(O(n)\)
space.
\end{theorem}

The proof in \ref{app:stable-penalties} establishes the
route-feasibility-limit, cursor, and deque-order invariants. Each
predecessor is inserted once and removed at most once.
The cursor \(\Lambda.\mathrm{feas}\) moves only toward newer
predecessors and crosses each predecessor at most once. Every
successful repetition of the remaining loops removes a predecessor,
and each loop performs at most one unsuccessful comparison per
customer. The total running time is therefore linear.

\ref{app:soft-vrptw-penalty-verification} and
\ref{app:time-warp-preprocessing} verify
Conditions~\eqref{cond:P1}--\eqref{cond:P5} for the two penalties.
The required linear preprocessing is given in
\ref{app:capacity-penalty-preprocessing},
\ref{app:time-window-route-limits}, and
\ref{app:time-warp-preprocessing}.

\section{Linear Split for the Soft VRPSPD}
\label{sec:vrpspd}\label{sec:soft-vrpspd}

Assuming capacity-feasible singleton routes, the simultaneous pickup and delivery penalty \(\operatorname{pen}_{\alpha h}\) satisfies
Conditions~\eqref{cond:P1}--\eqref{cond:P3}. These conditions make the
penalized and unpenalized predecessor regions \(Y_x^1\) and \(Y_x^2\)
contiguous for this variant. Condition~\eqref{cond:P4} does not hold
since two penalized predecessors may attain their maximum loads at
different positions, and these positions may change at different
customers. The proposed Split instead relies on
Conditions~\eqref{cond:S1}--\eqref{cond:S3}. Proofs that these
conditions hold for the soft VRPSPD are given in
\ref{app:soft-spd}.

For \(0\leq i<x\leq n\), recall from Equation~\eqref{eq:highest-point} that
    $h(i,x)$ denotes the latest position on route $R(i,x)$ that attains the highest load on the route.
 Then
\(H(i,x)=D[x]-P[i]+P[h(i,x)]-D[h(i,x)]\). Once \(h(i,x)\)
is known, \(\operatorname{pen}_{\alpha h}(i,x)\) is evaluated in
constant time as \(\alpha\max\{L(i,h(i,x),x)-Q,0\}\).

\begin{table}[t]
\centering
\caption{Additional notation for the soft VRPSPD Split.}
\label{tab:soft-spd-notation}
\small
\begin{tabularx}{\linewidth}{@{}lL@{}}
\toprule
Symbol & Meaning \\
\midrule
\(h(i,x)\) & Latest position attaining the maximum load on \(R(i,x)\). \\
\(Y_x^1,Y_x^2\) & penalized and unpenalized predecessors of
customer \(x\), respectively. \\
\(\Lambda\) & Deque of retained candidate predecessors. \\
\(\Lambda.\mathrm{best}\) & Cursor to the least-cost predecessor
currently retained in \(\Lambda\). \\
\(\Lambda.\mathrm{prev},\Lambda.\mathrm{next}\) & Entries immediately
before and after \(\Lambda.\mathrm{best}\). \\
\(\operatorname{dominated}_{\alpha p}(i,j)\) & Permanent comparison
used under the penalty-state and current-cost conditions described below. \\
\(\operatorname{cheaper}_{\alpha h}(i,h_i,j,h_j,x)\) & Direct comparison
of two candidate path costs at customer \(x\). \\
\bottomrule
\end{tabularx}
\end{table}

At customer \(x\), let \(Y_x^1\) contain the predecessors with positive penalty and let \(Y_x^2\) contain those with zero penalty. Let
\(\Delta_i(x)=\operatorname{pen}_{\alpha h}(i,x+1)
-\operatorname{pen}_{\alpha h}(i,x)\). The following substitute
conditions hold:
\begin{align}
h(i,x)=h(j,x)
&\Rightarrow h(i,y)=h(j,y),
&&0\leq i<j\leq x\leq y\leq n,
\tag{S1}\label{cond:S1}\\
h(j,x)=h(j,x+1)
&\Rightarrow h(i,x)=h(i,x+1),
&&0\leq i<j\leq x<n,
\tag{S2}\label{cond:S2}\\
\Delta_i(x)
&\leq \Delta_j(x),
&&i<j,\quad i,j\in Y_x^1.
\tag{S3}\label{cond:S3}
\end{align}
Equality holds in Condition~\eqref{cond:S3} whenever
\(h(j,x)=h(j,x+1)\). Condition~\eqref{cond:S1} states that two
predecessors sharing the same maximum-load position continue to share
it. Condition~\eqref{cond:S2} states that a maximum-load change for an
earlier predecessor can occur only when it also occurs for every later
predecessor under consideration. Condition~\eqref{cond:S3} states that
when the later predecessor changes maximum-load position, its penalty
grows at least as quickly as that of the earlier predecessor. Hence, if
an earlier penalized predecessor is already cheaper than a later one,
the later predecessor cannot overtake it. The reverse comparison is not
permanent since a currently cheaper later predecessor may lose its
advantage when its maximum-load position changes.

Condition~\eqref{cond:S1} implies that two penalized predecessors with
the same maximum-load position have a permanently fixed penalty
difference. We denote the resulting permanent dominance test by
\begin{equation}\label{eq:spd-permanent-comparison}
\begin{aligned}
\operatorname{dominated}_{\alpha p}(i,j)
\Longleftrightarrow{}&
\operatorname{pot}(i)+c_{0,i+1}-C[i+1]
+\alpha(P[j]-P[i])\\
&\geq\operatorname{pot}(j)+c_{0,j+1}-C[j+1],
\qquad i<j.
\end{aligned}
\end{equation}
For two currently penalized predecessors \(i<j\), the penalty
of \(j\) may grow faster than that of \(i\) as further customers are
appended, reducing the difference between their penalties. However,
this faster growth can continue only until the two routes share the
same maximum-load position. Their penalty difference then remains
fixed at \(\alpha(P[j]-P[i])\), which is also a lower bound on that
difference before their maximum-load positions coincide.
The test \(\operatorname{dominated}_{\alpha p}(i,j)\) determines
whether \(i\) is still no cheaper than \(j\) even with this smallest
possible penalty difference. If it holds, \(i\) can never become
cheaper than \(j\) and may therefore be removed permanently.

When the newer predecessor is an unpenalized cursor, the algorithm
additionally maintains that the older predecessor is currently no
cheaper. Its penalty cannot decrease while the cursor remains
unpenalized, and once both predecessors are penalized, the argument
above applies. These invariants are established in
\ref{app:soft-spd}.

Given \(h_i=h(i,x)\) and \(h_j=h(j,x)\), we evaluate the two
penalties from the corresponding loads \(L(i,h_i,x)\) and
\(L(j,h_j,x)\). We denote the direct comparison operator by
\begin{align}
&\operatorname{cheaper}_{\alpha h}(i,h_i,j,h_j,x) \notag\\
&\quad\Longleftrightarrow
\operatorname{pot}(i)+c(i,x)+\operatorname{pen}_{\alpha h}(i,x)
<\operatorname{pot}(j)+c(j,x)+\operatorname{pen}_{\alpha h}(j,x).
\label{eq:spd-current-comparison}
\end{align}
Ties are broken in favor of the newer predecessor.

\subsection{Linear Split Algorithm}

Algorithm~\ref{alg:soft-vrpspd-split} provides the linear Split for
the soft VRPSPD. The algorithm uses the predecessor deque \(\Lambda\)
and a cursor \(\Lambda.\mathrm{best}\). After maintenance, every entry
before the cursor belongs to \(Y_x^1\) and the retained predecessors are
ordered by nonincreasing penalized cost toward the cursor. Every entry
after the cursor belongs to \(Y_x^2\) and remains ordered by the
base-cost comparison. The cursor itself may belong to either region.
The older entries in \(Y_x^1\) are retained only when the permanent
comparison shows that they may become cheaper after their maximum-load
positions coincide with that of a later entry. The cursor is guaranteed
to be no more expensive than either neighbor. As a result,
\(\Lambda.\mathrm{best}\) is the minimum-cost predecessor.

A static rightmost range-maximum structure is constructed for the
sequence \(P[z]-D[z]\). It returns \(h(i,x)\) in constant time after
linear preprocessing. The construction is described in
\ref{app:soft-spd-rmq}. 

\begin{algorithm}[t]
\caption{Linear Split for the soft VRPSPD}
\label{alg:soft-vrpspd-split}\label{alg:soft-vrpspd}
\scriptsize
\DontPrintSemicolon
\SetKwInOut{Input}{Input}\SetKwInOut{Output}{Output}
\Input{Tour length \(n\); \(c_{0z}\), \(c_{z0}\), \(C[z]\),
\(D[z]\), and \(P[z]\) for \(z=1,\ldots,n\); capacity \(Q\),
multiplier \(\alpha\), and constant-time queries of \(h(i,x)\)}
\Output{Shortest-path costs \(\operatorname{pot}(1),\ldots,
\operatorname{pot}(n)\) and predecessors \(\operatorname{pred}(1),
\ldots,\operatorname{pred}(n)\)}
\(\operatorname{pot}(0)\leftarrow0\); \(\Lambda\leftarrow()\);
\(\Lambda.\mathrm{best}\leftarrow\bot\)\nllabel{line:spd-init}\;
\For{\(x\leftarrow1\) \KwTo \(n\)}{
  \While{\(\Lambda\neq()\) \textnormal{and}
         \(\operatorname{dominated}(\Lambda.\mathrm{back},x-1)\)\nllabel{line:spd-back-start}}{
    \If{\(\Lambda.\mathrm{best}=\Lambda.\mathrm{back}\)}{
      \(\Lambda.\mathrm{best}\leftarrow\bot\)\;
    }
    \(\Lambda.\mathrm{removeBack}()\)\nllabel{line:spd-back-end}\;
  }
  \(\Lambda.\mathrm{pushBack}(x-1)\)\nllabel{line:spd-insert}\;
  \If{\(\Lambda.\mathrm{best}=\bot\)\nllabel{line:spd-best-reset-start}}{
    \(\Lambda.\mathrm{best}\leftarrow\Lambda.\mathrm{back}\)\;
    \While{\(\Lambda.\mathrm{hasPrev}()\) \textnormal{and}
           \(\operatorname{dominated}_{\alpha p}
             (\Lambda.\mathrm{prev},\Lambda.\mathrm{best})\)}{
      \(\Lambda.\mathrm{removePrev}()\)\nllabel{line:spd-best-reset-end}\;
    }
  }

  \If{\(h(\Lambda.\mathrm{best},x)=x\)\nllabel{line:spd-max-update-start}}{
    \While{\(\Lambda.\mathrm{hasPrev}()\) \textnormal{and}
           \(h(\Lambda.\mathrm{prev},x)=x\)}{
      \(\Lambda.\mathrm{movePrev}()\)\;
      \(\Lambda.\mathrm{removeNext}()\)\;
    }
    \If{\(\Lambda.\mathrm{hasPrev}()\) \textnormal{and}
       \(\operatorname{cheaper}_{\alpha h}
       (\Lambda.\mathrm{prev},h(\Lambda.\mathrm{prev},x),
        \Lambda.\mathrm{best},h(\Lambda.\mathrm{best},x),x)\)}{
      \(\Lambda.\mathrm{movePrev}()\)\;
      \(\Lambda.\mathrm{removeNext}()\)\nllabel{line:spd-max-update-end}\;
    }
  }

  \While{\(\Lambda.\mathrm{hasNext}()\) \textnormal{and}
    \(\operatorname{pen}_{\alpha h}(\Lambda.\mathrm{next},x)>0\)\nllabel{line:spd-penalized-scan-start}}{
    \eIf{\(\operatorname{cheaper}_{\alpha h}
      (\Lambda.\mathrm{best},h(\Lambda.\mathrm{best},x),
       \Lambda.\mathrm{next},h(\Lambda.\mathrm{next},x),x)\)}{
      \(\Lambda.\mathrm{removeNext}()\)\;
    }{
      \(\Lambda.\mathrm{moveNext}()\)\;
      \While{\(\Lambda.\mathrm{hasPrev}()\) \textnormal{and}
             \(\operatorname{dominated}_{\alpha p}
               (\Lambda.\mathrm{prev},\Lambda.\mathrm{best})\)}{
        \(\Lambda.\mathrm{removePrev}()\)\;
      }
    }
  }

  \If{\(\Lambda.\mathrm{hasNext}()\) \textnormal{and not}
    \(\operatorname{cheaper}_{\alpha h}
      (\Lambda.\mathrm{best},h(\Lambda.\mathrm{best},x),
       \Lambda.\mathrm{next},h(\Lambda.\mathrm{next},x),x)\)\nllabel{line:spd-boundary-start}}{
    \(\Lambda.\mathrm{moveNext}()\)\;
    \While{\(\Lambda.\mathrm{hasPrev}()\) \textnormal{and}
           \(\operatorname{dominated}_{\alpha p}
             (\Lambda.\mathrm{prev},\Lambda.\mathrm{best})\)}{
      \(\Lambda.\mathrm{removePrev}()\)\nllabel{line:spd-boundary-end}\;
    }
  }

  \(\operatorname{pred}(x)\leftarrow\Lambda.\mathrm{best}\)\nllabel{line:spd-record-pred}\;
  \(\operatorname{pot}(x)\leftarrow
    \operatorname{pot}(\operatorname{pred}(x))
    +c(\operatorname{pred}(x),x)
    +\operatorname{pen}_{\alpha h}(\operatorname{pred}(x),x)\)\nllabel{line:spd-record-pot}\;
}
\end{algorithm}

Lines~\ref{line:spd-back-start}--\ref{line:spd-insert} perform the same
back insertion as the hard CVRP Split. When this removes the current
best, lines~\ref{line:spd-best-reset-start}--\ref{line:spd-best-reset-end}
place the cursor on the new back entry and delete permanently dominated
older neighbors.

Lines~\ref{line:spd-max-update-start}--\ref{line:spd-max-update-end}
process predecessors whose maximum-load position changes when customer
\(x\) is appended. By Conditions~\eqref{cond:S1} and
\eqref{cond:S2}, these predecessors are contiguous around the best
cursor. The scan beginning on
Line~\ref{line:spd-penalized-scan-start} moves the cursor whenever a
penalized successor is no more expensive.
Lines~\ref{line:spd-boundary-start}--\ref{line:spd-boundary-end} then
compare the current best with the first unpenalized successor.

\begin{theorem}\label{thm:soft-vrpspd-split}\label{thm:soft-vrpspd}
With nonnegative pickup and delivery quantities and capacity-feasible
singleton routes, Algorithm~\ref{alg:soft-vrpspd-split} returns an
exact minimum-cost partition for \(\mathcal G_{\alpha h}\) in
\(O(n)\) time and space.
\end{theorem}

\ref{app:soft-spd} proves the substitute conditions, the range-maximum
preprocessing, the best-cursor invariant, and the amortized bound. Each
predecessor is inserted once and removed at most once. Cursor movements
toward the front are paired with permanent deletions, while movements
toward the back cross each retained predecessor at most once, resulting in
linear total work.
\section{Soft VRPSPDTW Split in Linear Time}
\label{sec:soft-vrpspdtw}

The combination of pickup-and-delivery capacity and time-warp penalties introduces a new complication. Since both penalties satisfy Conditions~\eqref{cond:P1}--\eqref{cond:P3}, we again denote $Y^1_x$, $Y^2_x$, $Y^3_x$ as the predecessors to customer $x$ that are penalized by both capacity and time warp, penalized by capacity or time warp but not both, and not penalized, respectively. As before, the constant cost difference property keeps maintenance of nondominated predecessors among $Y^3_x$ simple. The constant penalty difference of the time-warp penalty implies that maintenance of nondominated predecessors among $Y^1_x$ will be no more complicated than managing penalized predecessors in the soft VRPSPD case. When $Y^2_x$ consists of predecessors penalized only by time warp, while comparing any predecessors $i<j$ among $Y^2_x$, if $i$ is more costly than $j$ to $x$ once time warp penalty is included, an eventual capacity penalty will not flip the cost relation due to capacity penalty monotonicity, so all such $i$ can be removed. The nondominated predecessors in $Y^2_x$ will be monotonically increasing in penalized cost from oldest to newest. Maintenance of $Y^2_x$ is therefore no more complicated than it was for the soft VRPTW case.

The only additional difficulty occurs when \(Y_x^2\) contains
predecessors penalized only by capacity. In this case, a predecessor that is not currently
the minimum of \(Y_x^2\) cannot necessarily be discarded. In the soft
VRPSPD, later already-penalized entries can be discarded when an earlier
penalized entry remains cheaper. A middle region penalized by capacity cannot use
that rule because later time-warp penalties may reverse the comparison.
Thus, unlike the penalized region in
Algorithm~\ref{alg:soft-vrpspd-split}, the minimum-cost retained
predecessor of \(Y_x^2\) will not necessarily be its newest retained entry. 

To handle this special case, the approach discussed here and presented in detail in \ref{app:soft-vrpspdtw} maintains three different deques $\Lambda^1, \Lambda^2$, and $\Lambda^3$ separately, one for each corresponding penalty region. Additionally, $\Lambda^2$ is used only when the penalty is time warp. If the penalty is capacity, a more complicated deque-of-deques structure $\Gamma^2$ is used that maintains predecessors in $Y^2_x$ by index order and keeps track of its minimum-cost retained predecessor (denoted as $\Gamma^2.\operatorname{minimum}$) without deleting any predecessors not proven to be dominated. All retained predecessors within $\Gamma^2$ that share the same maximum-load position $z$ are stored in a single inner deque $\Lambda_z$ guaranteed to be monotonically increasing in penalized cost ($\operatorname{pen}_{\alpha h }$ differences are permanent within each $\Lambda_z$, allowing for the safe deletions necessary to maintain monotonicity). As \(x\) advances, a suffix of the inner deques may acquire the same new maximum-load position and can then be merged, with permanent predecessor deletions restoring monotonicity. The outer deque $\Gamma^2$ maintains knowledge of the inner deque with the minimum-cost $\Lambda_z.\operatorname{front}$. The functionality of $\Gamma^2$ is also given in detail in \ref{app:soft-vrpspdtw}.

Since $\Lambda^1$ is maintained similarly to the penalized region of the soft VRPSPD case, its minimum-cost predecessor will be found at $\Lambda^1.\operatorname{back}$ after maintenance. If predecessors in $Y^2_x$  are penalized by time warp, $\Lambda^2$ will be maintained similarly to the middle region of the soft VRPTW case. Thus, its lowest-cost predecessor is $\Lambda^2.\operatorname{front}$ after maintenance. If $Y^2_x$ predecessors are penalized by capacity, then $\Gamma^2.\operatorname{minimum}$ will contain that structure's minimum-cost predecessor. Finally, $\Lambda^3$ handles the unpenalized predecessors the same as all previous variants, so $\Lambda^3.\operatorname{front}$ is its lowest-cost predecessor. The Split shown in \ref{app:soft-vrpspdtw} guarantees that at least one
minimum-cost predecessor to every later customer remains retained. Thus, a lowest-cost predecessor to each customer $x$ will be the minimizer among all deques' best predecessors.

\section{Computational Experiments} \label{sec:experiments}

To empirically evaluate the proposed algorithms, we compare the runtimes of the hard and soft linear VRPTW, VRPSPD, and VRPSPDTW Splits with their Bellman-based counterparts and with the existing linear hard and soft CVRP Splits. We also empirically verify correctness by comparing the \(\operatorname{pot}(x)\) and \(\operatorname{pred}(x)\) values produced by each proposed Split with those produced by its Bellman counterpart. We ran the correctness check on 1000 permutations of each of the VRPTW instances from \cite{solomon1987algorithms} and \cite{gehring1999parallel}; the VRPSPD instances from \cite{dethloff2001vehicle}, \cite{montane2006tabu}, and \cite{salhi1999cluster}; and the VRPSPDTW instances from \cite{wang2012genetic}, using the appropriate hard and soft Splits for each instance. For the soft Splits, we performed the checks using every applicable combination of \(\alpha\in\{0.1,1,5\}\) and \(\beta\in\{0.1,1,5\}\). All such comparisons passed.

Implementations for these Bellman and linear Splits as well as the framework used for the speed and correctness experiments can be found at \url{https://github.com/VRPSPDTW-Split/SPLIT-Library-VRPSPDTW}. Our experiments were run using a machine running macOS 14.6 Sonoma with an Apple M3, 8-core CPU. It should be noted that the Splits implemented in the source code are optimized by avoiding unnecessary preprocessing calculations. Route feasibility limits and highest load positions for penalized VRPSPD predecessors are maintained online. This avoids calculating values that are never used by the algorithm. For the soft VRPSPDTW, the buffering of predecessors entering new penalty regions is avoided. The correctness and speed experiments were all performed on these optimized implementations. The exhaustive preprocessing steps were presented in this paper for readability and generality.

We model our speed comparisons after the experiments run in \cite{vidal2016split}. We compare each new Split using the same 105 CVRP long tours employed in \cite{vidal2016split} by extending them to be VRPSPDTW instances/tours in the following manner. For each CVRP customer in the tour, the demand $d_i$ is split randomly between a new $d_i$ and a $p_i$ value, similar to \cite{salhi1999cluster}. A constant value was set for all service times $s_i$. We set \(t_{ij}=c_{ij}\), as is common for synthetic VRPTW instances \citep{solomon1987algorithms,wang2012genetic}. For the length of day $T$, we timed a vehicle traveling through the first 20 customers and returning to the depot as a suitable $T$ for testing purposes. We also confirmed that every singleton route was feasible with regard to $T$ and raised $T$ whenever this was not the case.  The code performing this extension is included in the available code.

We assigned time windows $[a_i,b_i]$ arbitrarily such that the opening time window would not prevent the singleton route from being feasible. It is noted that this arbitrary setting of time windows without changing the long tour permutation results in a very poor solution with many routes. This is acceptable for our purposes since we are not concerned with the quality of the overall VRP solution, but since the resulting solution has many short routes, the hard Bellman Splits may have better runtimes than one would expect during a heuristic search optimizing for efficient, tightly packed routes. We mitigate this phenomenon by multiplying closing times and $T$ by different factors (and by increasing vehicle capacity $Q$) to show how the runtimes of the hard Bellman Splits are affected by solutions requiring fewer and fewer routes. Note that the running times for Bellman Splits for the soft variants are always $O(n^2)$, since feasibility checks do not break the inner loop early. Thus, our poor solutions with many short routes will not asymptotically affect runtimes for these algorithms.

\begin{figure}
    \centering
    \includegraphics[width=\linewidth]{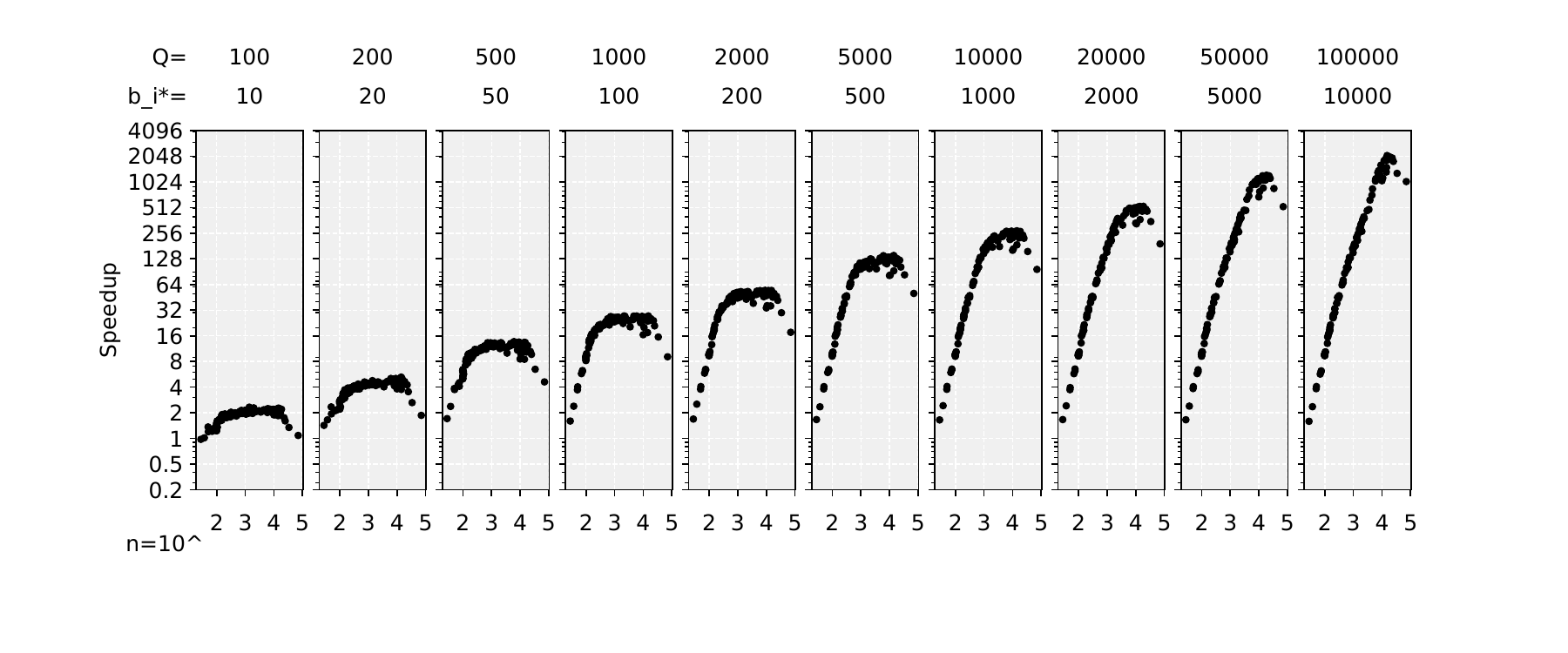}
    \caption{Speedup of the hard VRPTW linear Split over the Bellman-based algorithm for all 105 instances. Each graph corresponds to different $Q$ and closing time multiplier $b_i$ pairs. The X-axis indicates the instance size. The Y-axis gives the speedup relative to the Bellman Split. A logarithmic scale is used for both axes. }
    \label{fig:VRPTW_composite}
\end{figure}
\begin{figure}
    \centering
    \includegraphics[width=\linewidth]{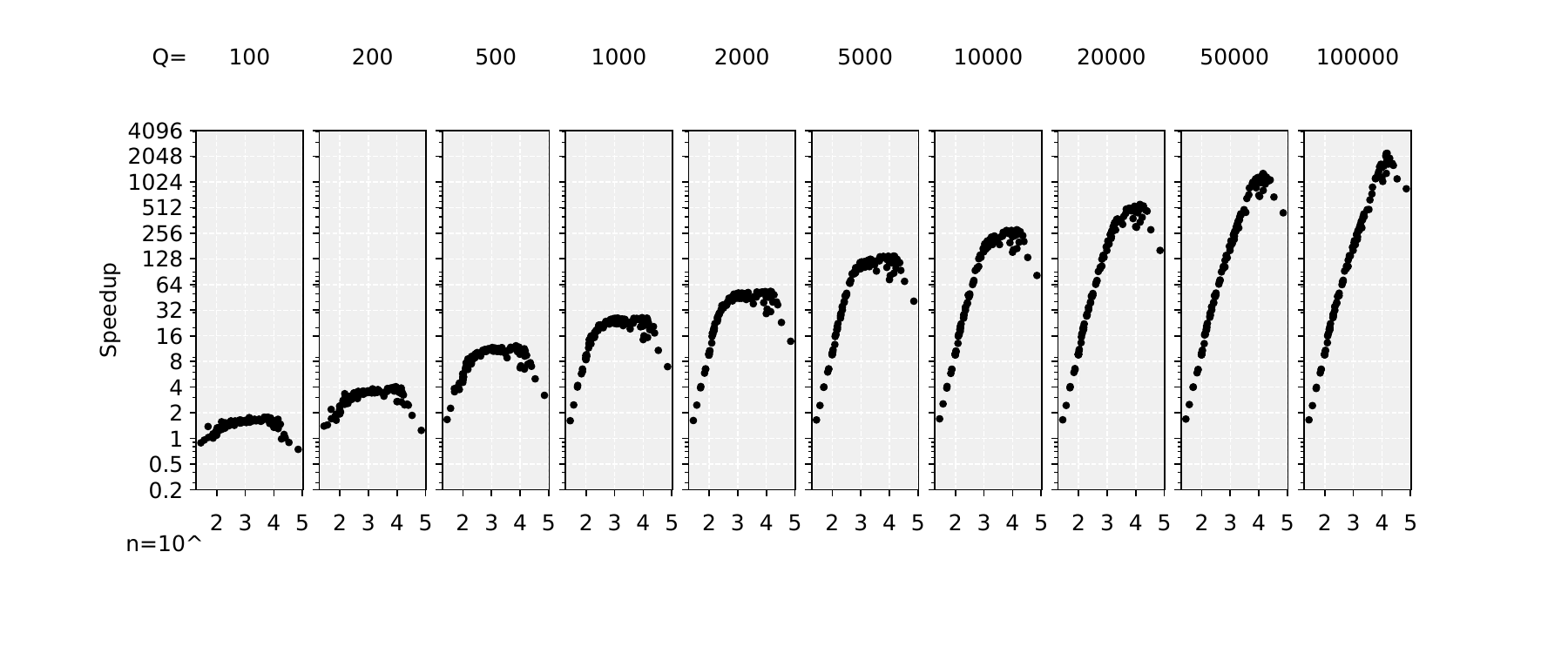}
    \caption{Speedup factors of the hard VRPSPD linear Split over the Bellman algorithm. }
    \label{fig:VRPSPD_composite}
\end{figure}
\begin{figure}
    \centering
    \includegraphics[width=\linewidth]{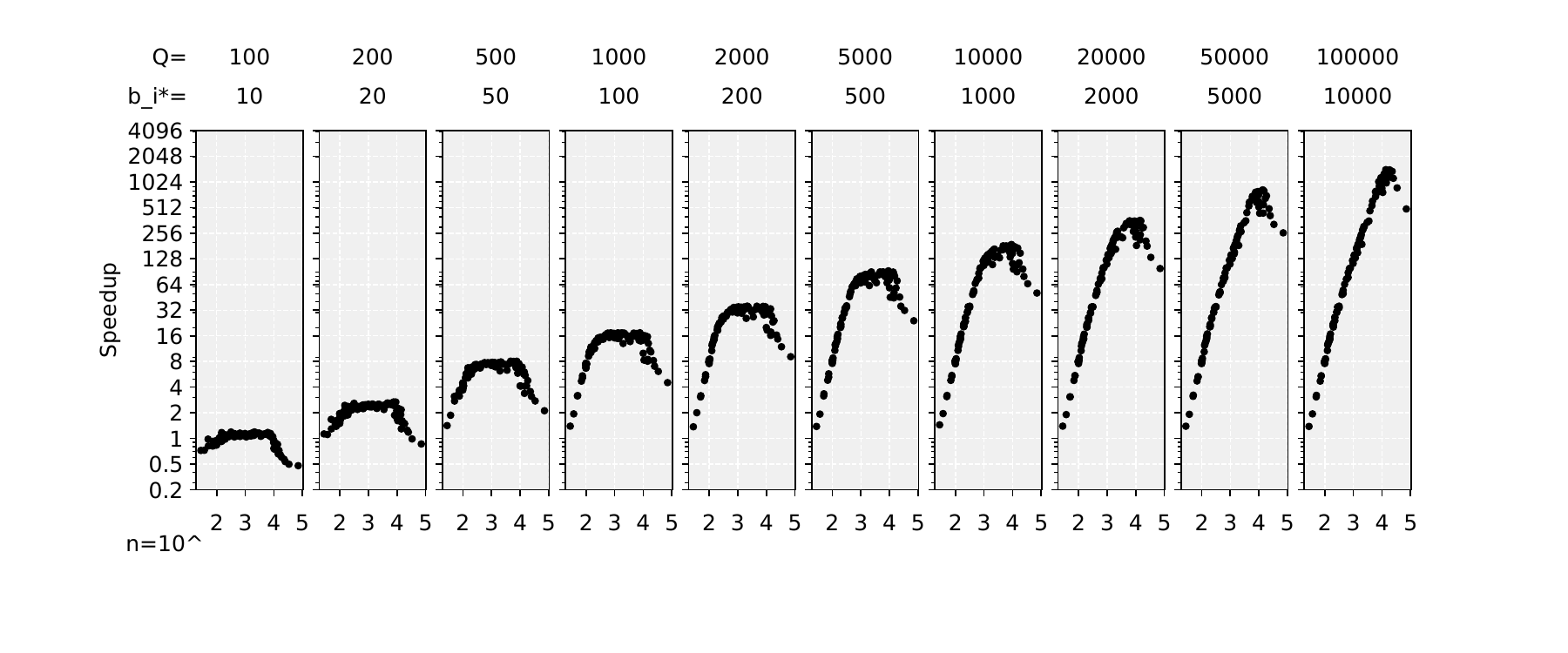}
    \caption{Speedup factors of the hard VRPSPDTW linear Split over the Bellman algorithm. }
    \label{fig:VRPSPDTW_composite}
\end{figure}

\begin{figure}
    \centering
    \includegraphics[width=\linewidth]{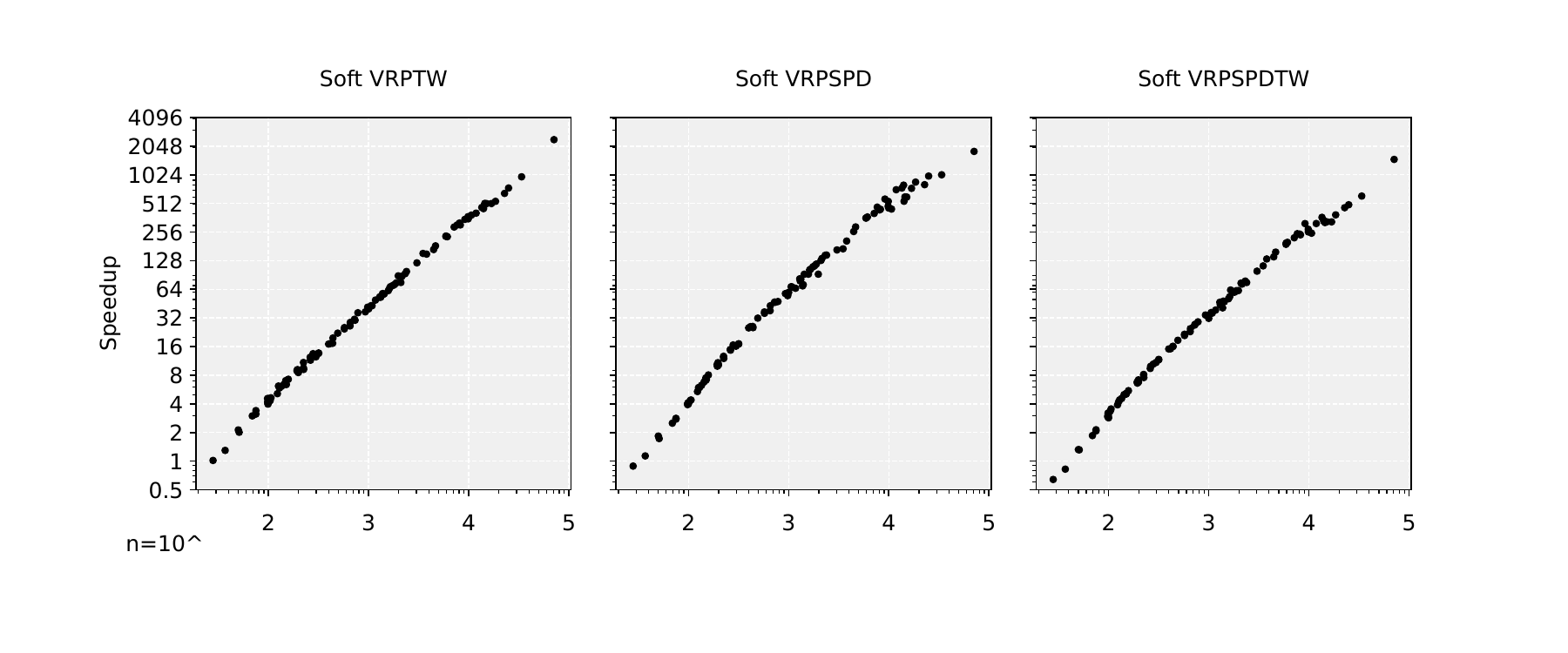}
    \caption{Speedup factors of the soft linear Splits over their Bellman counterparts. }
    \label{fig:soft_composite}
\end{figure}

\begin{table*}[t]
\centering
\small
\caption{CPU times (ms) of the hard VRPTW Bellman-based and linear Splits on a sample of extended instances originally from \cite{vidal2016split}, with increasing vehicle capacity and closing-window multiplier. The lower panel omits very small instances whose results no longer change due to nonbinding constraints. Faster times are bolded.}
\label{tab:hard-vrptw-timings}
\resizebox{\textwidth}{!}{%
\begin{tabular}{@{}lr*{5}{rr}@{}}
\toprule
Inst. & $n$ & \multicolumn{2}{c}{$Q=100$, $b_i^*=10$} & \multicolumn{2}{c}{$Q=200$, $b_i^*=20$} & \multicolumn{2}{c}{$Q=500$, $b_i^*=50$} & \multicolumn{2}{c}{$Q=1000$, $b_i^*=100$} & \multicolumn{2}{c}{$Q=2000$, $b_i^*=200$} \\
\cmidrule(lr){3-4} \cmidrule(lr){5-6} \cmidrule(lr){7-8} \cmidrule(lr){9-10} \cmidrule(lr){11-12}
 &  & $T_{\mathrm{Bellman}}$ & $T_{\mathrm{Linear}}$ & $T_{\mathrm{Bellman}}$ & $T_{\mathrm{Linear}}$ & $T_{\mathrm{Bellman}}$ & $T_{\mathrm{Linear}}$ & $T_{\mathrm{Bellman}}$ & $T_{\mathrm{Linear}}$ & $T_{\mathrm{Bellman}}$ & $T_{\mathrm{Linear}}$ \\
\midrule
wi29 & 28 & $\mathbf{7.90 \times 10^{-4}}$ & $8.09 \times 10^{-4}$ & $1.13 \times 10^{-3}$ & $\mathbf{7.96 \times 10^{-4}}$ & $1.34 \times 10^{-3}$ & $\mathbf{7.83 \times 10^{-4}}$ & $1.32 \times 10^{-3}$ & $\mathbf{8.25 \times 10^{-4}}$ & $1.32 \times 10^{-3}$ & $\mathbf{7.82 \times 10^{-4}}$ \\
eil51 & 50 & $1.82 \times 10^{-3}$ & $\mathbf{1.33 \times 10^{-3}}$ & $3.00 \times 10^{-3}$ & $\mathbf{1.28 \times 10^{-3}}$ & $4.99 \times 10^{-3}$ & $\mathbf{1.29 \times 10^{-3}}$ & $4.81 \times 10^{-3}$ & $\mathbf{1.29 \times 10^{-3}}$ & $4.75 \times 10^{-3}$ & $\mathbf{1.25 \times 10^{-3}}$ \\
rd100 & 99 & $2.38 \times 10^{-3}$ & $\mathbf{1.83 \times 10^{-3}}$ & $4.18 \times 10^{-3}$ & $\mathbf{1.88 \times 10^{-3}}$ & $9.38 \times 10^{-3}$ & $\mathbf{1.85 \times 10^{-3}}$ & $1.54 \times 10^{-2}$ & $\mathbf{1.85 \times 10^{-3}}$ & $1.74 \times 10^{-2}$ & $\mathbf{1.84 \times 10^{-3}}$ \\
d198 & 197 & $4.70 \times 10^{-3}$ & $\mathbf{2.53 \times 10^{-3}}$ & $8.87 \times 10^{-3}$ & $\mathbf{2.43 \times 10^{-3}}$ & $2.29 \times 10^{-2}$ & $\mathbf{2.31 \times 10^{-3}}$ & $4.31 \times 10^{-2}$ & $\mathbf{2.31 \times 10^{-3}}$ & $6.24 \times 10^{-2}$ & $\mathbf{2.31 \times 10^{-3}}$ \\
fl417 & 416 & $9.48 \times 10^{-3}$ & $\mathbf{5.08 \times 10^{-3}}$ & $1.89 \times 10^{-2}$ & $\mathbf{4.94 \times 10^{-3}}$ & $5.31 \times 10^{-2}$ & $\mathbf{4.86 \times 10^{-3}}$ & $0.105$ & $\mathbf{4.75 \times 10^{-3}}$ & $0.185$ & $\mathbf{4.74 \times 10^{-3}}$ \\
pr1002 & 1001 & $2.41 \times 10^{-2}$ & $\mathbf{1.19 \times 10^{-2}}$ & $4.99 \times 10^{-2}$ & $\mathbf{1.17 \times 10^{-2}}$ & $0.136$ & $\mathbf{1.13 \times 10^{-2}}$ & $0.273$ & $\mathbf{1.13 \times 10^{-2}}$ & $0.520$ & $\mathbf{1.09 \times 10^{-2}}$ \\
mu1979 & 1978 & $6.97 \times 10^{-2}$ & $\mathbf{3.05 \times 10^{-2}}$ & $0.127$ & $\mathbf{2.84 \times 10^{-2}}$ & $0.319$ & $\mathbf{2.82 \times 10^{-2}}$ & $0.619$ & $\mathbf{2.77 \times 10^{-2}}$ & $1.19$ & $\mathbf{2.77 \times 10^{-2}}$ \\
fnl4461 & 4460 & $0.122$ & $\mathbf{5.63 \times 10^{-2}}$ & $0.242$ & $\mathbf{5.30 \times 10^{-2}}$ & $0.658$ & $\mathbf{5.39 \times 10^{-2}}$ & $1.31$ & $\mathbf{5.29 \times 10^{-2}}$ & $2.56$ & $\mathbf{5.24 \times 10^{-2}}$ \\
kz9976 & 9975 & $0.371$ & $\mathbf{0.196}$ & $0.690$ & $\mathbf{0.181}$ & $1.60$ & $\mathbf{0.186}$ & $3.11$ & $\mathbf{0.189}$ & $6.06$ & $\mathbf{0.178}$ \\
d18512 & 18511 & $0.530$ & $\mathbf{0.241}$ & $1.06$ & $\mathbf{0.227}$ & $2.64$ & $\mathbf{0.215}$ & $5.35$ & $\mathbf{0.210}$ & $10.5$ & $\mathbf{0.210}$ \\
bm33708 & 33707 & $0.821$ & $\mathbf{0.610}$ & $1.54$ & $\mathbf{0.584}$ & $3.72$ & $\mathbf{0.573}$ & $7.55$ & $\mathbf{0.487}$ & $15.0$ & $\mathbf{0.501}$ \\
ch71009 & 71008 & $2.31$ & $\mathbf{2.13}$ & $3.83$ & $\mathbf{2.05}$ & $9.09$ & $\mathbf{1.97}$ & $18.5$ & $\mathbf{2.03}$ & $37.3$ & $\mathbf{2.12}$ \\
\bottomrule
\end{tabular}%
}
\vspace{0.8em}
\resizebox{\textwidth}{!}{%
\begin{tabular}{@{}lr*{5}{rr}@{}}
\toprule
Inst. & $n$ & \multicolumn{2}{c}{$Q=5000$, $b_i^*=500$} & \multicolumn{2}{c}{$Q=10000$, $b_i^*=1000$} & \multicolumn{2}{c}{$Q=20000$, $b_i^*=2000$} & \multicolumn{2}{c}{$Q=50000$, $b_i^*=5000$} & \multicolumn{2}{c}{$Q=100000$, $b_i^*=10000$} \\
\cmidrule(lr){3-4} \cmidrule(lr){5-6} \cmidrule(lr){7-8} \cmidrule(lr){9-10} \cmidrule(lr){11-12}
 &  & $T_{\mathrm{Bellman}}$ & $T_{\mathrm{Linear}}$ & $T_{\mathrm{Bellman}}$ & $T_{\mathrm{Linear}}$ & $T_{\mathrm{Bellman}}$ & $T_{\mathrm{Linear}}$ & $T_{\mathrm{Bellman}}$ & $T_{\mathrm{Linear}}$ & $T_{\mathrm{Bellman}}$ & $T_{\mathrm{Linear}}$ \\
\midrule
fl417 & 416 & $0.306$ & $\mathbf{4.69 \times 10^{-3}}$ & $0.313$ & $\mathbf{4.67 \times 10^{-3}}$ & $0.311$ & $\mathbf{4.61 \times 10^{-3}}$ & $0.309$ & $\mathbf{4.64 \times 10^{-3}}$ & $0.311$ & $\mathbf{4.66 \times 10^{-3}}$ \\
pr1002 & 1001 & $1.12$ & $\mathbf{1.10 \times 10^{-2}}$ & $1.75$ & $\mathbf{1.07 \times 10^{-2}}$ & $1.85$ & $\mathbf{1.09 \times 10^{-2}}$ & $1.85$ & $\mathbf{1.09 \times 10^{-2}}$ & $1.87$ & $\mathbf{1.11 \times 10^{-2}}$ \\
mu1979 & 1978 & $2.77$ & $\mathbf{2.82 \times 10^{-2}}$ & $4.95$ & $\mathbf{2.81 \times 10^{-2}}$ & $7.48$ & $\mathbf{2.85 \times 10^{-2}}$ & $7.82$ & $\mathbf{2.95 \times 10^{-2}}$ & $7.86$ & $\mathbf{2.92 \times 10^{-2}}$ \\
fnl4461 & 4460 & $6.24$ & $\mathbf{5.18 \times 10^{-2}}$ & $11.9$ & $\mathbf{5.09 \times 10^{-2}}$ & $21.4$ & $\mathbf{4.90 \times 10^{-2}}$ & $36.9$ & $\mathbf{5.25 \times 10^{-2}}$ & $37.2$ & $\mathbf{5.19 \times 10^{-2}}$ \\
kz9976 & 9975 & $14.8$ & $\mathbf{0.181}$ & $28.6$ & $\mathbf{0.177}$ & $55.9$ & $\mathbf{0.166}$ & $123$ & $\mathbf{0.180}$ & $189$ & $\mathbf{0.179}$ \\
d18512 & 18511 & $25.7$ & $\mathbf{0.200}$ & $51.3$ & $\mathbf{0.191}$ & $102$ & $\mathbf{0.192}$ & $241$ & $\mathbf{0.196}$ & $421$ & $\mathbf{0.209}$ \\
bm33708 & 33707 & $37.3$ & $\mathbf{0.449}$ & $74.1$ & $\mathbf{0.475}$ & $150$ & $\mathbf{0.428}$ & $362$ & $\mathbf{0.422}$ & $674$ & $\mathbf{0.522}$ \\
ch71009 & 71008 & $91.7$ & $\mathbf{1.82}$ & $181$ & $\mathbf{1.88}$ & $363$ & $\mathbf{1.89}$ & $921$ & $\mathbf{1.76}$ & $1.84 \times 10^{3}$ & $\mathbf{1.78}$ \\
\bottomrule
\end{tabular}%
}
\end{table*}

\begin{table*}[t]
\centering
\small
\caption{CPU times (ms) of the hard VRPSPD Bellman-based and linear Splits on the sampled instances.}
\label{tab:hard-vrpspd-timings}
\resizebox{\textwidth}{!}{%
\begin{tabular}{@{}lr*{5}{rr}@{}}
\toprule
Inst. & $n$ & \multicolumn{2}{c}{$Q=100$} & \multicolumn{2}{c}{$Q=200$} & \multicolumn{2}{c}{$Q=500$} & \multicolumn{2}{c}{$Q=1000$} & \multicolumn{2}{c}{$Q=2000$} \\
\cmidrule(lr){3-4} \cmidrule(lr){5-6} \cmidrule(lr){7-8} \cmidrule(lr){9-10} \cmidrule(lr){11-12}
 &  & $T_{\mathrm{Bellman}}$ & $T_{\mathrm{Linear}}$ & $T_{\mathrm{Bellman}}$ & $T_{\mathrm{Linear}}$ & $T_{\mathrm{Bellman}}$ & $T_{\mathrm{Linear}}$ & $T_{\mathrm{Bellman}}$ & $T_{\mathrm{Linear}}$ & $T_{\mathrm{Bellman}}$ & $T_{\mathrm{Linear}}$ \\
\midrule
wi29 & 28 & $\mathbf{7.25 \times 10^{-4}}$ & $8.17 \times 10^{-4}$ & $1.09 \times 10^{-3}$ & $\mathbf{7.82 \times 10^{-4}}$ & $1.29 \times 10^{-3}$ & $\mathbf{7.73 \times 10^{-4}}$ & $1.31 \times 10^{-3}$ & $\mathbf{8.10 \times 10^{-4}}$ & $1.27 \times 10^{-3}$ & $\mathbf{7.86 \times 10^{-4}}$ \\
eil51 & 50 & $1.73 \times 10^{-3}$ & $\mathbf{1.25 \times 10^{-3}}$ & $2.78 \times 10^{-3}$ & $\mathbf{1.27 \times 10^{-3}}$ & $4.87 \times 10^{-3}$ & $\mathbf{1.27 \times 10^{-3}}$ & $4.97 \times 10^{-3}$ & $\mathbf{1.24 \times 10^{-3}}$ & $4.87 \times 10^{-3}$ & $\mathbf{1.24 \times 10^{-3}}$ \\
rd100 & 99 & $2.16 \times 10^{-3}$ & $\mathbf{1.84 \times 10^{-3}}$ & $3.72 \times 10^{-3}$ & $\mathbf{1.81 \times 10^{-3}}$ & $9.22 \times 10^{-3}$ & $\mathbf{1.80 \times 10^{-3}}$ & $1.55 \times 10^{-2}$ & $\mathbf{1.78 \times 10^{-3}}$ & $1.76 \times 10^{-2}$ & $\mathbf{1.78 \times 10^{-3}}$ \\
d198 & 197 & $3.22 \times 10^{-3}$ & $\mathbf{2.40 \times 10^{-3}}$ & $6.76 \times 10^{-3}$ & $\mathbf{2.33 \times 10^{-3}}$ & $1.92 \times 10^{-2}$ & $\mathbf{2.39 \times 10^{-3}}$ & $3.83 \times 10^{-2}$ & $\mathbf{2.31 \times 10^{-3}}$ & $6.08 \times 10^{-2}$ & $\mathbf{2.32 \times 10^{-3}}$ \\
fl417 & 416 & $7.06 \times 10^{-3}$ & $\mathbf{4.95 \times 10^{-3}}$ & $1.44 \times 10^{-2}$ & $\mathbf{4.91 \times 10^{-3}}$ & $4.31 \times 10^{-2}$ & $\mathbf{4.65 \times 10^{-3}}$ & $9.34 \times 10^{-2}$ & $\mathbf{4.49 \times 10^{-3}}$ & $0.175$ & $\mathbf{4.69 \times 10^{-3}}$ \\
pr1002 & 1001 & $1.84 \times 10^{-2}$ & $\mathbf{1.16 \times 10^{-2}}$ & $4.02 \times 10^{-2}$ & $\mathbf{1.13 \times 10^{-2}}$ & $0.122$ & $\mathbf{1.07 \times 10^{-2}}$ & $0.255$ & $\mathbf{1.04 \times 10^{-2}}$ & $0.497$ & $\mathbf{1.05 \times 10^{-2}}$ \\
mu1979 & 1978 & $5.11 \times 10^{-2}$ & $\mathbf{3.16 \times 10^{-2}}$ & $0.113$ & $\mathbf{3.18 \times 10^{-2}}$ & $0.309$ & $\mathbf{2.94 \times 10^{-2}}$ & $0.634$ & $\mathbf{3.01 \times 10^{-2}}$ & $1.27$ & $\mathbf{2.96 \times 10^{-2}}$ \\
fnl4461 & 4460 & $9.36 \times 10^{-2}$ & $\mathbf{5.48 \times 10^{-2}}$ & $0.188$ & $\mathbf{5.31 \times 10^{-2}}$ & $0.556$ & $\mathbf{5.12 \times 10^{-2}}$ & $1.16$ & $\mathbf{5.15 \times 10^{-2}}$ & $2.35$ & $\mathbf{5.15 \times 10^{-2}}$ \\
kz9976 & 9975 & $0.274$ & $\mathbf{0.203}$ & $0.565$ & $\mathbf{0.209}$ & $1.39$ & $\mathbf{0.207}$ & $2.84$ & $\mathbf{0.197}$ & $5.72$ & $\mathbf{0.196}$ \\
d18512 & 18511 & $\mathbf{0.431}$ & $0.437$ & $0.984$ & $\mathbf{0.397}$ & $2.42$ & $\mathbf{0.326}$ & $5.09$ & $\mathbf{0.260}$ & $10.3$ & $\mathbf{0.212}$ \\
bm33708 & 33707 & $\mathbf{0.609}$ & $0.679$ & $1.23$ & $\mathbf{0.658}$ & $3.13$ & $\mathbf{0.625}$ & $7.21$ & $\mathbf{0.667}$ & $14.9$ & $\mathbf{0.647}$ \\
ch71009 & 71008 & $\mathbf{1.88}$ & $2.52$ & $3.15$ & $\mathbf{2.53}$ & $7.83$ & $\mathbf{2.44}$ & $16.9$ & $\mathbf{2.43}$ & $33.6$ & $\mathbf{2.44}$ \\
\bottomrule
\end{tabular}%
}
\vspace{0.8em}
\resizebox{\textwidth}{!}{%
\begin{tabular}{@{}lr*{5}{rr}@{}}
\toprule
Inst. & $n$ & \multicolumn{2}{c}{$Q=5000$} & \multicolumn{2}{c}{$Q=10000$} & \multicolumn{2}{c}{$Q=20000$} & \multicolumn{2}{c}{$Q=50000$} & \multicolumn{2}{c}{$Q=100000$} \\
\cmidrule(lr){3-4} \cmidrule(lr){5-6} \cmidrule(lr){7-8} \cmidrule(lr){9-10} \cmidrule(lr){11-12}
 &  & $T_{\mathrm{Bellman}}$ & $T_{\mathrm{Linear}}$ & $T_{\mathrm{Bellman}}$ & $T_{\mathrm{Linear}}$ & $T_{\mathrm{Bellman}}$ & $T_{\mathrm{Linear}}$ & $T_{\mathrm{Bellman}}$ & $T_{\mathrm{Linear}}$ & $T_{\mathrm{Bellman}}$ & $T_{\mathrm{Linear}}$ \\
\midrule
fl417 & 416 & $0.301$ & $\mathbf{4.55 \times 10^{-3}}$ & $0.305$ & $\mathbf{4.55 \times 10^{-3}}$ & $0.307$ & $\mathbf{4.50 \times 10^{-3}}$ & $0.309$ & $\mathbf{4.51 \times 10^{-3}}$ & $0.311$ & $\mathbf{4.50 \times 10^{-3}}$ \\
pr1002 & 1001 & $1.09$ & $\mathbf{1.06 \times 10^{-2}}$ & $1.70$ & $\mathbf{1.01 \times 10^{-2}}$ & $1.84$ & $\mathbf{1.01 \times 10^{-2}}$ & $1.83$ & $\mathbf{1.03 \times 10^{-2}}$ & $1.85$ & $\mathbf{1.04 \times 10^{-2}}$ \\
mu1979 & 1978 & $3.05$ & $\mathbf{2.94 \times 10^{-2}}$ & $5.54$ & $\mathbf{2.94 \times 10^{-2}}$ & $8.30$ & $\mathbf{2.97 \times 10^{-2}}$ & $8.76$ & $\mathbf{2.97 \times 10^{-2}}$ & $8.85$ & $\mathbf{2.99 \times 10^{-2}}$ \\
fnl4461 & 4460 & $5.88$ & $\mathbf{4.84 \times 10^{-2}}$ & $11.4$ & $\mathbf{4.82 \times 10^{-2}}$ & $21.1$ & $\mathbf{4.83 \times 10^{-2}}$ & $36.5$ & $\mathbf{5.04 \times 10^{-2}}$ & $37.3$ & $\mathbf{5.04 \times 10^{-2}}$ \\
kz9976 & 9975 & $14.4$ & $\mathbf{0.197}$ & $28.4$ & $\mathbf{0.185}$ & $54.0$ & $\mathbf{0.176}$ & $124$ & $\mathbf{0.174}$ & $191$ & $\mathbf{0.166}$ \\
d18512 & 18511 & $25.0$ & $\mathbf{0.198}$ & $50.0$ & $\mathbf{0.187}$ & $99.2$ & $\mathbf{0.185}$ & $236$ & $\mathbf{0.203}$ & $421$ & $\mathbf{0.217}$ \\
bm33708 & 33707 & $36.0$ & $\mathbf{0.517}$ & $72.9$ & $\mathbf{0.548}$ & $146$ & $\mathbf{0.521}$ & $356$ & $\mathbf{0.524}$ & $670$ & $\mathbf{0.603}$ \\
ch71009 & 71008 & $85.3$ & $\mathbf{2.09}$ & $173$ & $\mathbf{2.12}$ & $349$ & $\mathbf{2.17}$ & $892$ & $\mathbf{2.01}$ & $1.74 \times 10^{3}$ & $\mathbf{2.05}$ \\
\bottomrule
\end{tabular}%
}
\end{table*}

\begin{table*}[t]
\centering
\small
\caption{CPU times (ms) of the hard VRPSPDTW Bellman-based and linear Splits on sampled instances}
\label{tab:hard-vrpspdtw-timings}
\resizebox{\textwidth}{!}{%
\begin{tabular}{@{}lr*{5}{rr}@{}}
\toprule
Inst. & $n$ & \multicolumn{2}{c}{$Q=100$, $b_i^*=10$} & \multicolumn{2}{c}{$Q=200$, $b_i^*=20$} & \multicolumn{2}{c}{$Q=500$, $b_i^*=50$} & \multicolumn{2}{c}{$Q=1000$, $b_i^*=100$} & \multicolumn{2}{c}{$Q=2000$, $b_i^*=200$} \\
\cmidrule(lr){3-4} \cmidrule(lr){5-6} \cmidrule(lr){7-8} \cmidrule(lr){9-10} \cmidrule(lr){11-12}
 &  & $T_{\mathrm{Bellman}}$ & $T_{\mathrm{Linear}}$ & $T_{\mathrm{Bellman}}$ & $T_{\mathrm{Linear}}$ & $T_{\mathrm{Bellman}}$ & $T_{\mathrm{Linear}}$ & $T_{\mathrm{Bellman}}$ & $T_{\mathrm{Linear}}$ & $T_{\mathrm{Bellman}}$ & $T_{\mathrm{Linear}}$ \\
\midrule
wi29 & 28 & $\mathbf{6.97 \times 10^{-4}}$ & $9.64 \times 10^{-4}$ & $1.09 \times 10^{-3}$ & $\mathbf{9.70 \times 10^{-4}}$ & $1.34 \times 10^{-3}$ & $\mathbf{9.47 \times 10^{-4}}$ & $1.29 \times 10^{-3}$ & $\mathbf{9.28 \times 10^{-4}}$ & $1.32 \times 10^{-3}$ & $\mathbf{9.66 \times 10^{-4}}$ \\
eil51 & 50 & $\mathbf{1.59 \times 10^{-3}}$ & $1.62 \times 10^{-3}$ & $2.77 \times 10^{-3}$ & $\mathbf{1.65 \times 10^{-3}}$ & $4.87 \times 10^{-3}$ & $\mathbf{1.56 \times 10^{-3}}$ & $4.88 \times 10^{-3}$ & $\mathbf{1.55 \times 10^{-3}}$ & $4.88 \times 10^{-3}$ & $\mathbf{1.59 \times 10^{-3}}$ \\
rd100 & 99 & $\mathbf{2.16 \times 10^{-3}}$ & $2.32 \times 10^{-3}$ & $3.73 \times 10^{-3}$ & $\mathbf{2.28 \times 10^{-3}}$ & $9.10 \times 10^{-3}$ & $\mathbf{2.25 \times 10^{-3}}$ & $1.53 \times 10^{-2}$ & $\mathbf{2.14 \times 10^{-3}}$ & $1.76 \times 10^{-2}$ & $\mathbf{2.20 \times 10^{-3}}$ \\
d198 & 197 & $\mathbf{3.50 \times 10^{-3}}$ & $3.58 \times 10^{-3}$ & $6.95 \times 10^{-3}$ & $\mathbf{3.36 \times 10^{-3}}$ & $1.95 \times 10^{-2}$ & $\mathbf{3.07 \times 10^{-3}}$ & $3.81 \times 10^{-2}$ & $\mathbf{2.96 \times 10^{-3}}$ & $6.02 \times 10^{-2}$ & $\mathbf{2.99 \times 10^{-3}}$ \\
fl417 & 416 & $7.24 \times 10^{-3}$ & $\mathbf{6.95 \times 10^{-3}}$ & $1.48 \times 10^{-2}$ & $\mathbf{6.75 \times 10^{-3}}$ & $4.45 \times 10^{-2}$ & $\mathbf{6.50 \times 10^{-3}}$ & $9.55 \times 10^{-2}$ & $\mathbf{6.31 \times 10^{-3}}$ & $0.177$ & $\mathbf{6.17 \times 10^{-3}}$ \\
pr1002 & 1001 & $1.85 \times 10^{-2}$ & $\mathbf{1.70 \times 10^{-2}}$ & $4.11 \times 10^{-2}$ & $\mathbf{1.63 \times 10^{-2}}$ & $0.123$ & $\mathbf{1.58 \times 10^{-2}}$ & $0.260$ & $\mathbf{1.57 \times 10^{-2}}$ & $0.506$ & $\mathbf{1.57 \times 10^{-2}}$ \\
mu1979 & 1978 & $5.29 \times 10^{-2}$ & $\mathbf{4.89 \times 10^{-2}}$ & $0.104$ & $\mathbf{4.63 \times 10^{-2}}$ & $0.286$ & $\mathbf{4.61 \times 10^{-2}}$ & $0.575$ & $\mathbf{4.42 \times 10^{-2}}$ & $1.14$ & $\mathbf{4.43 \times 10^{-2}}$ \\
fnl4461 & 4460 & $8.86 \times 10^{-2}$ & $\mathbf{7.98 \times 10^{-2}}$ & $0.189$ & $\mathbf{7.63 \times 10^{-2}}$ & $0.561$ & $\mathbf{7.37 \times 10^{-2}}$ & $1.18$ & $\mathbf{7.49 \times 10^{-2}}$ & $2.35$ & $\mathbf{7.11 \times 10^{-2}}$ \\
kz9976 & 9975 & $\mathbf{0.278}$ & $0.365$ & $0.533$ & $\mathbf{0.309}$ & $1.37$ & $\mathbf{0.330}$ & $2.78$ & $\mathbf{0.279}$ & $5.61$ & $\mathbf{0.280}$ \\
d18512 & 18511 & $\mathbf{0.369}$ & $0.609$ & $0.803$ & $\mathbf{0.539}$ & $2.11$ & $\mathbf{0.442}$ & $4.56$ & $\mathbf{0.440}$ & $9.18$ & $\mathbf{0.381}$ \\
bm33708 & 33707 & $\mathbf{0.633}$ & $1.27$ & $\mathbf{1.23}$ & $1.25$ & $3.09$ & $\mathbf{1.12}$ & $6.61$ & $\mathbf{1.08}$ & $13.6$ & $\mathbf{1.14}$ \\
ch71009 & 71008 & $\mathbf{1.89}$ & $3.95$ & $\mathbf{3.43}$ & $3.98$ & $7.98$ & $\mathbf{3.77}$ & $16.8$ & $\mathbf{3.72}$ & $34.4$ & $\mathbf{3.78}$ \\
\bottomrule
\end{tabular}%
}
\vspace{0.8em}
\resizebox{\textwidth}{!}{%
\begin{tabular}{@{}lr*{5}{rr}@{}}
\toprule
Inst. & $n$ & \multicolumn{2}{c}{$Q=5000$, $b_i^*=500$} & \multicolumn{2}{c}{$Q=10000$, $b_i^*=1000$} & \multicolumn{2}{c}{$Q=20000$, $b_i^*=2000$} & \multicolumn{2}{c}{$Q=50000$, $b_i^*=5000$} & \multicolumn{2}{c}{$Q=100000$, $b_i^*=10000$} \\
\cmidrule(lr){3-4} \cmidrule(lr){5-6} \cmidrule(lr){7-8} \cmidrule(lr){9-10} \cmidrule(lr){11-12}
 &  & $T_{\mathrm{Bellman}}$ & $T_{\mathrm{Linear}}$ & $T_{\mathrm{Bellman}}$ & $T_{\mathrm{Linear}}$ & $T_{\mathrm{Bellman}}$ & $T_{\mathrm{Linear}}$ & $T_{\mathrm{Bellman}}$ & $T_{\mathrm{Linear}}$ & $T_{\mathrm{Bellman}}$ & $T_{\mathrm{Linear}}$ \\
\midrule
fl417 & 416 & $0.306$ & $\mathbf{6.23 \times 10^{-3}}$ & $0.315$ & $\mathbf{6.19 \times 10^{-3}}$ & $0.310$ & $\mathbf{6.21 \times 10^{-3}}$ & $0.311$ & $\mathbf{6.07 \times 10^{-3}}$ & $0.310$ & $\mathbf{6.05 \times 10^{-3}}$ \\
pr1002 & 1001 & $1.11$ & $\mathbf{1.55 \times 10^{-2}}$ & $1.73$ & $\mathbf{1.51 \times 10^{-2}}$ & $1.86$ & $\mathbf{1.49 \times 10^{-2}}$ & $1.87$ & $\mathbf{1.48 \times 10^{-2}}$ & $1.87$ & $\mathbf{1.51 \times 10^{-2}}$ \\
mu1979 & 1978 & $2.65$ & $\mathbf{4.27 \times 10^{-2}}$ & $4.81$ & $\mathbf{4.38 \times 10^{-2}}$ & $7.48$ & $\mathbf{4.51 \times 10^{-2}}$ & $7.81$ & $\mathbf{4.25 \times 10^{-2}}$ & $7.86$ & $\mathbf{4.14 \times 10^{-2}}$ \\
fnl4461 & 4460 & $5.90$ & $\mathbf{7.05 \times 10^{-2}}$ & $11.5$ & $\mathbf{7.04 \times 10^{-2}}$ & $21.5$ & $\mathbf{6.78 \times 10^{-2}}$ & $37.2$ & $\mathbf{6.96 \times 10^{-2}}$ & $38.2$ & $\mathbf{7.09 \times 10^{-2}}$ \\
kz9976 & 9975 & $14.3$ & $\mathbf{0.244}$ & $28.0$ & $\mathbf{0.248}$ & $54.9$ & $\mathbf{0.238}$ & $122$ & $\mathbf{0.235}$ & $191$ & $\mathbf{0.222}$ \\
d18512 & 18511 & $23.4$ & $\mathbf{0.331}$ & $47.3$ & $\mathbf{0.317}$ & $94.3$ & $\mathbf{0.316}$ & $226$ & $\mathbf{0.322}$ & $415$ & $\mathbf{0.293}$ \\
bm33708 & 33707 & $35.6$ & $\mathbf{1.12}$ & $71.3$ & $\mathbf{1.09}$ & $143$ & $\mathbf{1.07}$ & $348$ & $\mathbf{1.07}$ & $690$ & $\mathbf{0.791}$ \\
ch71009 & 71008 & $88.7$ & $\mathbf{3.69}$ & $178$ & $\mathbf{3.50}$ & $355$ & $\mathbf{3.60}$ & $962$ & $\mathbf{3.72}$ & $1.90 \times 10^{3}$ & $\mathbf{3.85}$ \\
\bottomrule
\end{tabular}%
}
\end{table*}

\subsection{Comparison against Bellman Split}
Tables \ref{tab:hard-vrptw-timings}, \ref{tab:hard-vrpspd-timings}, and \ref{tab:hard-vrpspdtw-timings} show a sample of run times of 12 of the 105 instances for the Bellman and linear Splits for the hard VRPTW, VRPSPD, and VRPSPDTW, respectively. Figures \ref{fig:VRPTW_composite}, \ref{fig:VRPSPD_composite}, and \ref{fig:VRPSPDTW_composite} graph the speedup attained by these linear Splits for all 105 instances. The capacity $Q$ and the multiplier of all $b_i$ are increased for every pair of columns in order to allow for longer routes within the tour. The full table of runtime speeds for all instances for all Splits is included with the available source code. 

We see from the graphs the linear speed-up for relatively unconstrained permutations. The speedup stalls at a particular instance size and then reliably lessens as instance size rises higher than around $n=10000$, resulting in a concave speedup shape. This same shape is seen in the results of \cite{vidal2016split}. The variable stall is due to the route constraints causing a large but constant average-route size $B$, rendering the Bellman Split effectively linear. The point of stalling is delayed for less constrained routes. The decrease in speedup is a result of the limits of caching the many preprocessed values for very large instances. It may be possible to implement memory-efficient versions of these Splits especially for use on very large instances.

Figure \ref{fig:soft_composite} shows the speed-up of the linear soft Splits against their quadratic counterparts. We see an almost exactly linear improvement as instance size increases, as was the case for the soft CVRP comparison in \cite{vidal2016split}. 

\subsection{Comparison against CVRP Split}

\begin{figure}
    \centering
    \includegraphics[width=\linewidth]{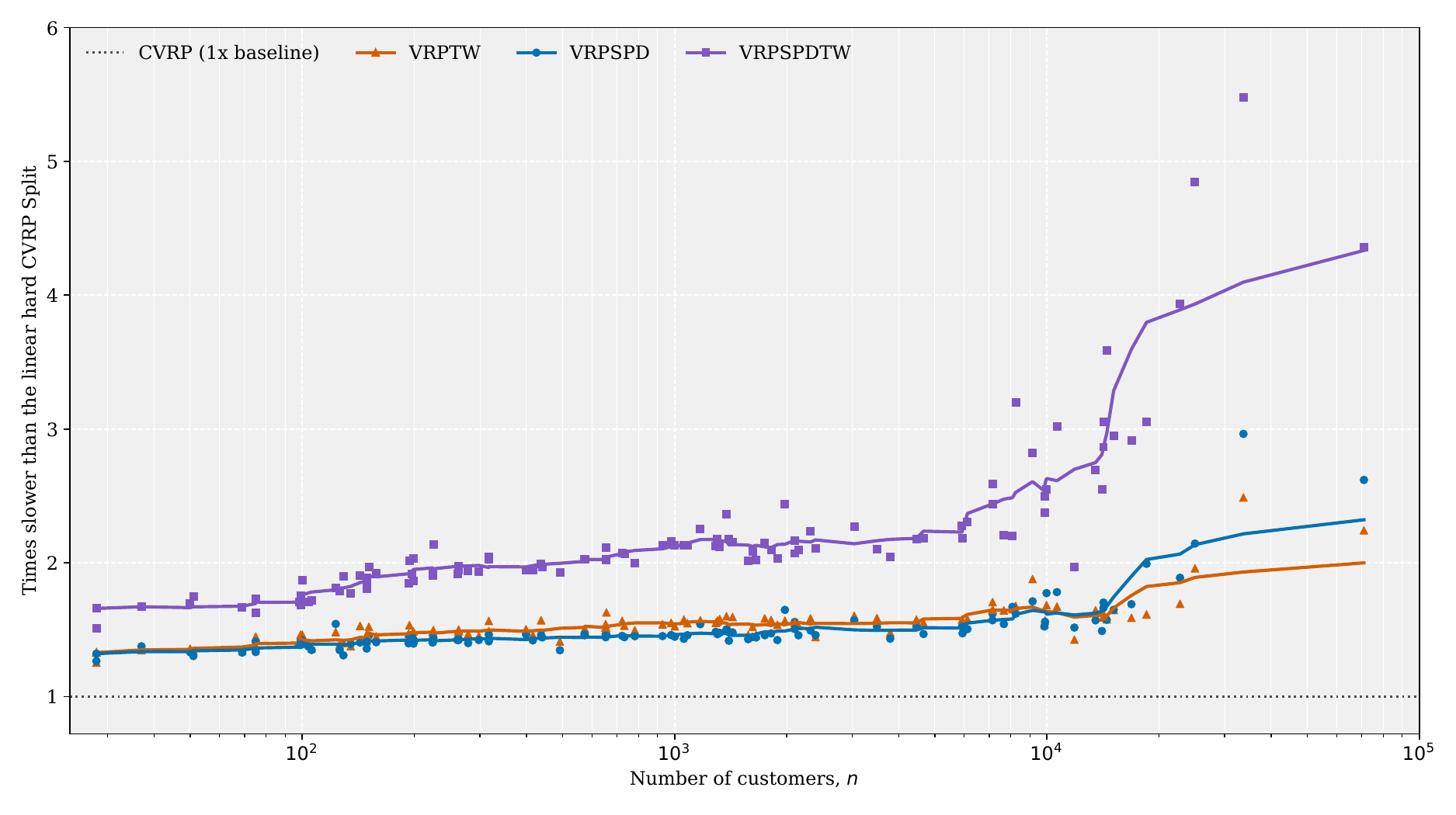}
    \caption{Slowdown factors of the hard VRPTW, VRPSPD, and VRPSPDTW linear Splits over the hard linear CVRP Split of \cite{vidal2016split}. The points are runtime speeds, while the same-colored lines are moving averages of the previous 9 instances. Only the X-axis is logarithmic. }
    \label{fig:hard_relative}
\end{figure}

\begin{figure}
    \centering
    \includegraphics[width=\linewidth]{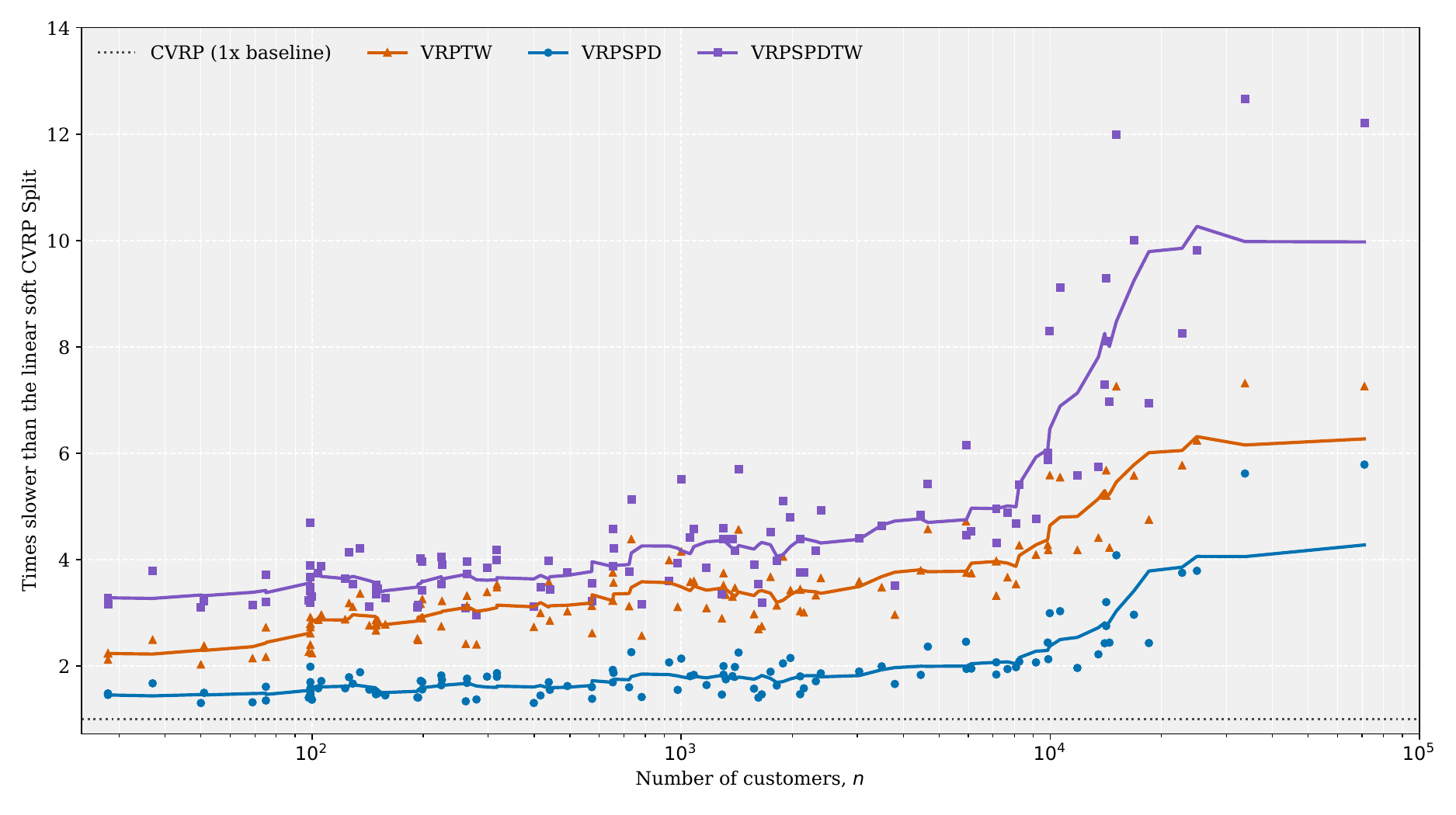}
    \caption{Slowdown factors of the soft VRPTW, VRPSPD, and VRPSPDTW linear Splits over the soft linear CVRP Split of \cite{vidal2016split}.}
    \label{fig:soft_relative}
\end{figure}

Figures \ref{fig:hard_relative} and \ref{fig:soft_relative} show comparisons of the hard and soft linear Splits introduced in this work against the existing hard and soft linear CVRP Splits. From these results, we see that the slowdown of each Split is constant up until around 10000 customers. This is presumably caused by caching limits, which would affect the linear CVRP Splits, but not to the same degree. The slowdown increase caused by this appears to be bounded, as expected from the asymptotic runtime guarantees.

For the hard variants, the variance in runtimes show less variation. The Splits for both the parent variants VRPTW and VRPSPD perform similarly, and both perform better than that for the combined variant. For the soft variants, the runtimes are more varied. The single-penalty VRPSPD is the fastest out of the three, revealing that the cost of maintaining two penalized regions, even if each penalty function's behavior is highly regular, is not negligible.  

\section{Conclusion}\label{sec:conclusion}

In this work, an important evaluation and search procedure for popular metaheuristics for the CVRP, the linear Split algorithm, was extended to handle simultaneous pickup-and-delivery constraints together with time-window constraints. Additionally, the Split was extended to handle the VRPSPD with a soft capacity penalty, the VRPTW with CVRP capacity and time warp penalties, and the soft VRPSPDTW. The assumptions and variant properties that ensure the algorithms' correctness were identified and decoupled as much as possible for the sake of future algorithmic development. These Splits were compared against their $O(nB)$ and $O(n^2)$ baseline algorithms and were shown theoretically and experimentally to be able to Split larger long tours asymptotically faster except in the case where route sizes are very small, where the $O(nB)$ Bellman Split also runs linearly. 

Several avenues of future research can be explored in relation to this work. Perhaps most obvious is the potential use of these Splits as an evaluation and/or search method within metaheuristics for the VRPSPDTW and its special-case variants, including the VRPSPD, the VRPTW, and the VRP with mixed pickup and delivery. The HGS by \cite{vidal2022hybrid} is currently the most popular approach which uses the Split. While it has recently been adapted a number of times to tackle the VRPTW, as discussed in Section \ref{sec:related-work}, these adaptations ignore time window constraints and penalties during the actual Split. For the VRPSPD, the last time an iteration of the HGS was applied to this problem was in \cite{vidal2014unified}. It would be worth retrying the HGS on this variant with the well-validated improvements applied in the open-source code while making use of the soft VRPSPD Split described in this work. Further experimentation is required to determine which VRP variants and instance types benefit from an ``order-first split-second'' subroutine. 

While the Bellman Split has been relatively easy to adapt to many variants \citep{vidal2014unified}, we see in this work that adapting the CVRP linear Split to some of the most conceptually simple VRP variants requires a great deal of careful algorithmic design. We hope this work provides developers interested in extending the Split to other problems with some initial theoretical scaffolding to make further algorithmic developments less burdensome.

\appendix
\section{Deque and Cursor Operations}
\label{app:deque-operators}\label{appendix-operators}

Except for the inner deques \(\Lambda_z\) of \(\Gamma^2\), all
\(\Lambda\) deques that do not require cursors are implemented as
array-based deques. The \(\Lambda\) deque used by the soft VRPSPD Split
and the \(\Lambda\) deque used by the soft VRPTW Split require cursors $\Lambda.\mathrm{best}$ and $\Lambda.\mathrm{feas}$, respectively, 
and are therefore implemented as doubly linked lists with direct handles to
cursor positions for constant-time access. In either representation, accessing the front or back,
pushing or removing a back entry, and removing the front take constant time.
For a linked-list entry \(i\), the operations for checking existence of, reading, and deleting elements adjacent to the deque's cursor also take
constant time. The sentinel \(\bot\) represents a missing predecessor
or successor.

The soft VRPSPDTW Split also uses the auxiliary deque-of-deques structure \(\Gamma^2\).
Its outer deque uses two array-based aggregation stacks with
rebalancing, while each inner deque stores its retained predecessors in a
doubly linked list, allowing for the ends of two adjacent deques to merge in constant time. The full operations and amortized analysis of $\Gamma^2 $ are
given in \ref{app:soft-vrpspdtw}.

\section{Generalized Hard Split: Proof and Route-Limit Constructions}
\label{app:hard-reach}\label{appendix-hard}\label{appendix:generalized-proof}

\subsection{Correctness of Algorithm~\ref{alg:generic-hard-split}}

\begin{lemma}\label{lem:hard-safe-removals}
Every predecessor deleted by
Algorithm~\ref{alg:generic-hard-split} can be discarded without
excluding an optimal predecessor at the current or any later customer.
\end{lemma}

\begin{proof}
Suppose an older predecessor \(i\) is removed from the back on
lines~\ref{line:hard-back-start}--\ref{line:hard-back-end} when a newer
predecessor \(j\) is inserted. By
Property~\ref{prop:constant-cost-difference}, the dominance indicator
gives \(\operatorname{pot}(i)+c(i,y)\geq
\operatorname{pot}(j)+c(j,y)\) for every common final customer \(y\).
By Condition~\eqref{cond:H2},
\(\mathsf{feas}_q(i,y)\Rightarrow\mathsf{feas}_q(j,y)\) for every
constraint \(q\). Thus, whenever \(i\) could be used, \(j\) is both
feasible and no more expensive.

A predecessor \(i\) is removed from the front on
lines~\ref{line:hard-front-start}--\ref{line:hard-front-end} only when
\(\rho(i)\leq x\). It is then infeasible for customer \(x\), and by
Condition~\eqref{cond:H1} it is infeasible for every later customer.
Hence neither type of removal can delete the unique strictly better
feasible predecessor of a future customer.
\end{proof}

\begin{lemma}\label{lem:hard-deque-invariant}
After iteration \(x\), every predecessor in \(\Lambda\) is feasible for
customer \(x\), the predecessors occur in increasing index order, and
their candidate path costs for customer \(x\) strictly increase from
the front to the back.
\end{lemma}

\begin{proof}
The claim is immediate for \(x=1\), because the singleton predecessor
0 is feasible by Condition~\eqref{cond:H3}. Assume it holds after
iteration \(x-1\). By
Property~\ref{prop:constant-cost-difference}, all candidate-cost
differences among retained predecessors are preserved when the final
customer changes from \(x-1\) to \(x\). The loop on
lines~\ref{line:hard-back-start}--\ref{line:hard-back-end} removes
only the suffix predecessors that are no cheaper than the new predecessor
\(x-1\), so the insertion on line~\ref{line:hard-insert} preserves
strict cost order. The loop on
lines~\ref{line:hard-front-start}--\ref{line:hard-front-end} removes
all predecessors whose limits have been reached. Since \(\rho\) is
nondecreasing and indices increase along the deque, all remaining
predecessors are feasible. Removing entries cannot disturb the order
of the survivors.
\end{proof}

\begin{proof}[Proof of Theorem~\ref{thm:generic-hard-split}]
By Lemma~\ref{lem:hard-safe-removals}, no predecessor capable of
strictly improving a future label is lost. By
Lemma~\ref{lem:hard-deque-invariant}, the front of \(\Lambda\) is the
least-cost feasible retained predecessor at every customer. Therefore, the predecessor and potential written on
lines~\ref{line:hard-record-pred}--\ref{line:hard-record-pot} are indeed
the predecessor attaining the cheapest-cost path to \(x\) and the cost of
that path, respectively.

Each predecessor is inserted on line~\ref{line:hard-insert} once,
removed from the back at most once, and removed from the front at most
once. All deque operations and cost comparisons take constant time, so
the Split scan is \(O(n)\). The arrays and deque contain \(O(n)\)
entries.
\end{proof}

\subsection{Preprocessing VRPSPD route limits}

Algorithm~\ref{alg:hard-spd-limits} computes the first customer
\(\rho_h(i)\) that makes the route starting after predecessor \(i\)
infeasible with respect to the VRPSPD constraint:
\[
\rho_h(i)
=
\min\{x\in\{i+1,\ldots,n\}:H(i,x)>Q\},
\]
with \(\rho_h(i)=n+1\) if no such customer exists. Hence,
\(R(i,x)\) satisfies the capacity constraint if and only if
\(x<\rho_h(i)\).
The corresponding penalty limit is
\(\rho_{\alpha h}(i)=\rho_h(i)\).

Recall from Equation~\eqref{eq:route-load} that the load $L$ on route $R(i,x)$ just after leaving position $z$ is
 $L(i,z,x)=D[x]-D[z]+P[z]-P[i].$
From Equation~\eqref{eq:problem-spd-maximum} the highest load attained on that route is 
\(
H(i,x)
=
D[x]-P[i]
+
\max_{i\leq z\leq x}\{P[z]-D[z]\}.
\)
From Equation~\eqref{eq:highest-point}, the label
$
h(i,x)
=
\max\arg\max_{i\leq z\leq x}\{P[z]-D[z]\}
$
denotes the latest position attaining this maximum.

These quantities satisfy the monotonicity conditions required by the
generalized Split. Appending customer \(x+1\) increases \(D[x]\) by the
nonnegative delivery \(d_{x+1}\) and adds one candidate to the
maximization in Equation~\eqref{eq:problem-spd-maximum}. Therefore,
\[
H(i,x+1)\geq H(i,x),
\]
so an infeasible route remains infeasible after extension, establishing
Condition~\eqref{cond:H1}. Conversely, advancing the predecessor from
\(i\) to \(i+1\) increases \(P[i]\) to \(P[i+1]\) and removes position
\(i\) from the maximization. Hence,
\[
H(i+1,x)\leq H(i,x),
\]
establishing Condition~\eqref{cond:H2}. Condition~\eqref{cond:H3} is
the assumed feasibility of every singleton route. These inequalities also establish Conditions~\eqref{cond:P1}
and~\eqref{cond:P2} for \(\operatorname{pen}_{\alpha h}\).

It remains to maintain \(h(i,x)\) efficiently as the route endpoints
advance. For positions \(u<v\), define
\begin{equation}\label{eq:dominated-h}
\operatorname{dominated}_h(u,v)
\Longleftrightarrow
P[u]-D[u]\leq P[v]-D[v].
\end{equation}
Whenever \(\operatorname{dominated}_h(u,v)\) holds, position \(u\)
cannot equal \(h(i,x)\) for any subsequent route containing both
positions since \(v\) has an at least as large value of
\(P[\cdot]-D[\cdot]\) and remains
eligible at least as long as \(u\). Equality is included so that the
latest maximizing position is retained.

Algorithm~\ref{alg:hard-spd-limits} maintains candidates of $h(\cdot,\cdot)$ in a
monotone deque \(\Lambda_h\). Its indices increase from front to back,
while their values \(P[z]-D[z]\) strictly decrease. Consequently,
\[
\Lambda_h.\mathrm{front}=h(i,x)
\]
for the current route \(R(i,x)\), so \(H(i,x)\) can be evaluated in
constant time using Equation~\eqref{eq:route-load}. For empty routes, the algorithm assumes
\(L(i,i,i)=H(i,i)=0\) and \(h(i,i)=i\).

\begin{algorithm}[t]
\caption{VRPSPD route-feasibility limits}
\label{alg:hard-spd-limits}
\DontPrintSemicolon
\SetKwInOut{Input}{Input}\SetKwInOut{Output}{Output}
\Input{Tour length \(n\); prefixes \(D,P\); capacity \(Q\)}
\Output{First-infeasible customers
\(\rho_h(0),\ldots,\rho_h(n-1)\)}

\(x\leftarrow0\); \(\Lambda_h\leftarrow(0)\)\;

\For{\(i\leftarrow0\) \KwTo \(n-1\)}{
  \While{\(x<n\) \textnormal{ and }
    \(L(i,\Lambda_h.\mathrm{front},x)\leq Q\)}{
    \(x\leftarrow x+1\)\;
    \While{\(\Lambda_h\neq()\) \textnormal{ and }
      \(\operatorname{dominated}_h(\Lambda_h.\mathrm{back},x)\)}{
      \(\Lambda_h.\mathrm{removeBack}()\)\;
    }
    \(\Lambda_h.\mathrm{pushBack}(x)\)\;
  }

  \eIf{\(L(i,\Lambda_h.\mathrm{front},x)>Q\)}{
    \(\rho_h(i)\leftarrow x\)\;
  }{
    \(\rho_h(i)\leftarrow n+1\)\;
  }

  \If{\(\Lambda_h.\mathrm{front}=i\)}{
    \(\Lambda_h.\mathrm{removeFront}()\)\;
  }
}
\end{algorithm}

At the beginning of iteration \(i\), the route endpoint \(x\) need not
be moved backwards since every endpoint before \(x\) was feasible for the
previous predecessor, and remains feasible after deleting the first
customer by Condition~\eqref{cond:H2}. The first while loop therefore
continues the scan from \(x\), extending the route while
\(
L(i,\Lambda_h.\mathrm{front},x)
=
H(i,x)
\leq Q.
\)
Whenever \(x\) advances, the back-removal loop deletes
candidates that can no longer equal \(h(i,x)\) due to domination by a later position with at least as large a load contribution, preserving
\(\Lambda_h.\mathrm{front}=h(i,x)\).

If the extension loop stops because
\(L(i,\Lambda_h.\mathrm{front},x)>Q\), then \(R(i,x)\) is infeasible.
All smaller endpoints are feasible, and Condition~\eqref{cond:H1}
guarantees that every larger endpoint is also infeasible. Thus
\(x=\rho_h(i)\). If the scan instead reaches \(x=n\) while
\(H(i,n)\leq Q\), no infeasible endpoint exists and
\(\rho_h(i)=n+1\).

When the predecessor advances from \(i\) to \(i+1\), position \(i\) 
is not included in any remaining route. If it has remained in \(\Lambda_h\) up until this point, then it has not been dominated. In this case it is the front element and is now removed from further consideration as a highest point. Otherwise it was already
discarded from the back in favor of a later candidate. The deque
invariant therefore holds for the next iteration.

The pointer \(x\) advances at most \(n\) times. Each position is
inserted into \(\Lambda_h\) once and removed at most once, while all
other operations take constant time. Therefore,
Algorithm~\ref{alg:hard-spd-limits} computes all route limits in
\(O(n)\) time and uses \(O(n)\) space.
\subsection{Time-window route limits and initial time warp}
\label{app:time-window-route-limits}
Algorithm~\ref{alg:hard-tw-limits} computes the first customer
\(\rho_a(i)\) that makes the route starting after predecessor \(i\)
infeasible with respect to the time-window constraints, together with
the initial time warp \(r_a(i)\) incurred at that customer.

Define the effective latest service start
\[
\bar b_z=\min\{b_z,T-s_z-t_{z0}\}.
\]
Under Inequality~\eqref{eq:weak-triangle-in}, route \(R(i,x)\) is
feasible with respect to the customer time windows and depot return
time if and only if
$
A_i(z)\leq \bar b_z
$
for every $z\in\{i+1,\ldots,x\}$.
Thus, the first infeasible customer is
\[
\rho_a(i)
=
\min\{x\in\{i+1,\ldots,n\}:A_i(x)>\bar b_x\},
\]
with \(\rho_a(i)=n+1\) if no such customer exists.
The corresponding penalty limit is denoted
\(\rho_{\beta a}(i)=\rho_a(i)\).
When \(\rho_a(i)\leq n\), its initial time warp is
\[
r_a(i)
=
A_i(\rho_a(i))-\bar b_{\rho_a(i)}.
\]

Using the effective-bound representation, appending a customer leaves
every earlier service start unchanged. Any violation of
\(A_i(z)\leq\bar b_z\) therefore persists after extension, establishing
Condition~\eqref{cond:H1}.

To establish Condition~\eqref{cond:H2}, suppose that \(R(i,x)\) is
feasible and \(x\geq i+2\). No customer on this route incurs time
warp. By Inequality~\eqref{eq:weak-triangle-out}, deleting its first
customer gives
\[
\begin{aligned}
A_{i+1}(i+2)
&=\max\{t_{0,i+2},a_{i+2}\}\\
&\leq
\max\{A_i(i+1)+s_{i+1}+t_{i+1,i+2},a_{i+2}\}\\
&=A_i(i+2).
\end{aligned}
\]
Thus, the shorter route begins service at its first customer no later
than the original route, satisfies that customer's effective closing
time, and incurs no time warp there. Applying
Equation~\eqref{eq:time-start-recurrence} inductively gives
\[
A_{i+1}(z)\leq A_i(z)\leq\bar b_z,
\qquad z=i+2,\ldots,x.
\]
At each step, the shorter route starts service no later than
the feasible original route and therefore also incurs no time warp.
It consequently satisfies every customer window and the depot-return
deadline, proving Condition~\eqref{cond:H2}.
Condition~\eqref{cond:H3} is the assumed feasibility of every
singleton route.

The quantitative monotonicity conditions
\eqref{cond:P1} and~\eqref{cond:P2} for
\(\operatorname{pen}_{\beta a}\) are addressed separately in
\ref{app:time-warp-preprocessing}.

We now show how to evaluate \(A_i(x)\) efficiently as the route endpoints
advance. Before the first violation, no preceding customer incurs time
warp, so unrolling Equations~\eqref{eq:time-start-first} and
\eqref{eq:time-start-recurrence} using prefix values $S[\cdot]$ from Equation~\eqref{eq:prefix-tour} gives
\[
A_i(x)
=
\max\left\{
t_{0,i+1}+S[x]-S[i+1],
S[x]+
\max_{i<z\leq x}
\{a_z-S[z]\}
\right\},
\]
The left term calculates the arrival time to customer $x$ in the case where the vehicle travels from customer $i+1$ to customer $x$ without being required to wait at any customer. The right term chooses the customer with the most obstructive opening time window, with the $-S[z]$ term accounting for the increasing time required to reach higher indexed candidates. This customer, denoted
\[
w(i,x)=\max\arg\max_{i<z\leq x}\{a_z-S[z]\}
\]
is the last customer the vehicle waits at on $R(i,x)$, if the vehicle waits at all. If the vehicle does wait, the opening window of $w(i,x)$ determines the arrival time at $x$. If $w(i,x)$ is known for a given $i,x$, then $A_i(x)$ is calculable in constant time.
For positions \(u<v\), let
\begin{equation}\label{eq:dominated-a}
\operatorname{dominated}_a(u,v)
\Longleftrightarrow
a_u-S[u]\leq a_v-S[v].
\end{equation}
Whenever \(\operatorname{dominated}_a(u,v)\) holds, customer \(u\)
cannot be $w(i,x)$ for any
subsequent route $R(i,x)$ containing $v$, since \(v\) is at least as obstructive and remains in the current route for at least as
long. Equality is included so that ties are retained in favor of the
later position.

Algorithm~\ref{alg:hard-tw-limits} maintains candidates of $w(\cdot,\cdot)$ in a
monotone deque \(\Lambda_a\). Its indices increase from front to back,
while their values \(a_z-S[z]\) strictly decrease. As a result, for
the current route \(R(i,x)\), $w(i,x)=\Lambda_a.\operatorname{front}$ after maintenance, and
\[
A_i(x)
=
\max\left\{
t_{0,i+1}+S[x]-S[i+1],
a_{\Lambda_a.\mathrm{front}}
+S[x]-S[\Lambda_a.\mathrm{front}]
\right\},
\]
so the service start at the current endpoint can be evaluated in
constant time.

\begin{algorithm}[t]
\caption{Time-window route-feasibility limits and initial time warp}
\label{alg:hard-tw-limits}
\DontPrintSemicolon
\SetKwInOut{Input}{Input}\SetKwInOut{Output}{Output}
\Input{Tour length \(n\); \(t_{0z}\), \(S[z]\), \(a_z\), and
\(\bar b_z\), for \(z=1,\ldots,n\)}
\Output{First-infeasible customers
\(\rho_a(0),\ldots,\rho_a(n-1)\) and initial warps
\(r_a(0),\ldots,r_a(n-1)\)}

\(x\leftarrow1\); \(\Lambda_a\leftarrow(1)\)\;

\For{\(i\leftarrow0\) \KwTo \(n-1\)
\nllabel{line:hard-tw-pre-loop}}{

  \While{\(x<n\) \textnormal{ and } \(A_i(x)\leq\bar b_x\)
  \nllabel{line:hard-tw-pre-extend}}{
    \(x\leftarrow x+1\)\;

    \While{\(\Lambda_a\neq()\) \textnormal{ and }
      \(\operatorname{dominated}_a(\Lambda_a.\mathrm{back},x)\)}{
      \(\Lambda_a.\mathrm{removeBack}()\)\;
    }

    \(\Lambda_a.\mathrm{pushBack}(x)\)\;
  }

  \eIf{\(A_i(x)>\bar b_x\)}{
    \(\rho_a(i)\leftarrow x\)
    \nllabel{line:hard-tw-pre-assign}\;
    \(r_a(i)\leftarrow A_i(x)-\bar b_x\)
    \nllabel{line:hard-tw-pre-warp}\;
  }{
    \(\rho_a(i)\leftarrow n+1\)
    \nllabel{line:hard-tw-pre-sentinel}\;
    \(r_a(i)\leftarrow0\)\;
  }

  \If{\(\Lambda_a.\mathrm{front}=i+1\)}{
    \(\Lambda_a.\mathrm{removeFront}()\)
    \nllabel{line:hard-tw-pre-drop}\;
  }
}
\end{algorithm}

At the beginning of iteration \(i\), the route endpoint \(x\) need not
be moved backwards since every endpoint before \(x\) was feasible for
the previous predecessor and remains feasible after deleting the first
customer by Condition~\eqref{cond:H2}. The first while loop therefore
continues the scan from \(x\), extending the route while
\(A_i(x)\leq\bar b_x\). Whenever \(x\) advances, the back-removal loop
discards candidates that can no longer be $w(i,x)$ due to dominance by another customer, preserving the
constant-time evaluation of \(A_i(x)\) using
\(\Lambda_a.\mathrm{front}\).

If the extension loop stops because \(A_i(x)>\bar b_x\), all smaller
endpoints are feasible and \(x\) is therefore the first infeasible
customer, so \(x=\rho_a(i)\). The excess
\(A_i(x)-\bar b_x\) is the initial time warp \(r_a(i)\). If the
scan instead reaches \(x=n\) while \(A_i(n)\leq\bar b_n\), no infeasible
endpoint exists and \(\rho_a(i)=n+1\).

When the predecessor advances from \(i\) to \(i+1\), customer \(i+1\)
is not included for any remaining routes. If it has remained in
\(\Lambda_a\), then it was never dominated and is therefore the front element, and is thus removed
from further consideration. Otherwise it was already discarded from
the back in favor of a later candidate. The deque invariant therefore
holds for the next iteration.

The pointer \(x\) advances at most \(n\) times. Each customer is
inserted into \(\Lambda_a\) once and removed at most once, while all
other operations take constant time. Therefore,
Algorithm~\ref{alg:hard-tw-limits} computes all route limits and initial
time warps in \(O(n)\) time and uses \(O(n)\) space.
\section{Details for the Soft VRPTW Split}
\label{app:stable-penalties}

For Algorithm~\ref{alg:stable-penalty-split}, we say an adjacent pair \(i<j\) in \(\Lambda\) is \emph{ordered} at customer
\(x\) when
\[
\operatorname{pot}(i)+c_{\alpha d,\beta a}(i,x)
<
\operatorname{pot}(j)+c_{\alpha d,\beta a}(j,x).
\]
Equality is not retained because the dominance indicators favor the
newer predecessor. The linked-deque operators used by
Algorithm~\ref{alg:stable-penalty-split} are described in
\ref{app:deque-operators}.

\begin{lemma}\label{lem:stable-order-invariant}
Immediately before
lines~\ref{line:vrptw-record-pred}--\ref{line:vrptw-record-pot}
of every iteration \(x\), the predecessors in \(\Lambda\) occur in
increasing index order and every adjacent pair is ordered at customer
\(x\). Moreover, \(\Lambda.\mathrm{feas}\) points to the oldest
retained predecessor in \(Y_x^3\), or equals \(\bot\) when
\(Y_x^3\) is empty, and at most one predecessor from \(Y_x^1\)
remains.
\end{lemma}

\begin{proof}
The claim is immediate after the first insertion. Assume that it holds
at the end of iteration \(x-1\).

The back-removal loop on
lines~\ref{line:vrptw-back-start}--\ref{line:vrptw-back-end} removes
the suffix dominated in base cost by the new predecessor \(x-1\),
which is then appended on line~\ref{line:vrptw-insert}. By
Condition~\eqref{cond:P3}, this new predecessor belongs to \(Y_x^3\).
If the previous cursor was removed, it is first set to \(\bot\), and
line~\ref{line:vrptw-feas-init} then makes the new entry the cursor.
Thus, after insertion, \(\Lambda.\mathrm{feas}\) still identifies the
oldest retained candidate that may belong to \(Y_x^3\).

Since both route-feasibility limits are nondecreasing,
\(Y_x^1,Y_x^2,Y_x^3\) occur as contiguous regions from front to back.
By Property~\ref{prop:constant-cost-difference} and
Condition~\eqref{cond:P4} holding for both penalty functions (see \ref{app:soft-vrptw-penalty-verification} and
\ref{app:time-warp-preprocessing}), the relative candidate cost of two
predecessors is fixed while they remain in the same region, and can
therefore change only when they belong to different regions or when one or both predecessors change regions.
Consequently, every pair of adjacent predecessors that both remained in the same region
from iteration \(x-1\) to iteration \(x\) remains ordered.

The loop beginning on
line~\ref{line:vrptw-feas-advance-start} advances
\(\Lambda.\mathrm{feas}\) across all retained predecessors
that have left \(Y_x^3\). Before each advance, the nested loop on
lines~\ref{line:vrptw-cross-order-start}--\ref{line:vrptw-cross-order-end}
removes dominated older neighbors until the newly exposed adjacency is
ordered. The cursor then advances on
line~\ref{line:vrptw-feas-advance-end}. Thus, all adjacencies created
as predecessors enter \(Y_x^2\) are ordered, and when this process
terminates the cursor points to the oldest retained predecessor in
\(Y_x^3\), or to \(\bot\) if none remains. The loop on
lines~\ref{line:vrptw-current-boundary-start}--\ref{line:vrptw-current-boundary-end}
similarly restores the current \(Y_x^2/Y_x^3\) boundary when the cursor
does not move.

The algorithm next maintains the front region. Since \(Y_x^1\) is a prefix
of \(\Lambda\), the loop on
lines~\ref{line:vrptw-y1-start}--\ref{line:vrptw-y1-end} repeatedly
compares its first two entries and removes the more expensive one,
resolving equality in favor of the newer predecessor. Both entries are
penalized under both penalties, so by Condition~\eqref{cond:P4} and
Property~\ref{prop:constant-cost-difference} their relative candidate
cost is fixed. The loop therefore performs a tournament over
\(Y_x^1\), leaving at most its cheapest retained predecessor at the
front. Finally,
lines~\ref{line:vrptw-first-boundary-start}--\ref{line:vrptw-first-boundary-end}
order this remaining predecessor, when it exists, with the first
predecessor of the next nonempty region.

All adjacencies are therefore ordered at the end of maintenance, and
the stated region invariants hold.
\end{proof}

\begin{lemma}\label{lem:stable-safe-deletion}
Every predecessor deleted by
Algorithm~\ref{alg:stable-penalty-split} can be discarded without
excluding an optimal predecessor at the current or any later customer.
\end{lemma}

\begin{proof}
We consider all cases where a predecessor is deleted from $\Lambda$. Consider a deletion in which an older predecessor \(i\) is
removed because
\[
\operatorname{dominated}_{\alpha d,\beta a}(i,j,x)
\]
holds for a newer retained predecessor \(j\). This includes the deletions at the
\(Y_x^2/Y_x^3\) boundary, the deletion of the older entry during the
\(Y_x^1\) tournament, and the deletion at the \(Y_x^1\) boundary.
By Lemma~\ref{lem:penalty-difference-monotone}, the difference
contributed by each penalty between \(i\) and \(j\) is nondecreasing as
the final customer advances. By
Property~\ref{prop:constant-cost-difference}, their base candidate-cost
difference is constant. Since \(i\) is already no cheaper than \(j\),
it can never become cheaper at a later customer.

The \(Y_x^1\) tournament may instead delete
\(j=\Lambda.\mathrm{front2}\), the newer predecessor, in favor of
\(i=\Lambda.\mathrm{front}\), the older predecessor. In this case both predecessors are
already penalized under both penalties. Condition~\eqref{cond:P4} fixes
both penalty differences for all later customers, while
Property~\ref{prop:constant-cost-difference} fixes their base-cost
difference. Their complete candidate-cost difference is therefore
permanent, so the deleted newer predecessor can never overtake the
retained older one.

Finally, a back deletion on
lines~\ref{line:vrptw-back-start}--\ref{line:vrptw-back-end} removes an
older predecessor \(i\) in favor of the newly inserted predecessor
\(j=x-1\). The base candidate cost through \(i\) is already no smaller
than that through \(j\). By Condition~\eqref{cond:P3}, \(j\) is
unpenalized at customer \(x\), whereas the penalties of \(i\) are
nonnegative, so the complete candidate cost through \(i\) is also no
smaller. Lemma~\ref{lem:penalty-difference-monotone} prevents this
ordering from reversing later.

Thus every deletion leaves another predecessor that is no more
expensive wherever the deleted predecessor could subsequently be used.
\end{proof}

\begin{proof}[Proof of Theorem~\ref{thm:stable-penalty-split}]
By Lemma~\ref{lem:stable-order-invariant}, candidate path costs
strictly increase from the front to the back of \(\Lambda\) after
maintenance in every iteration. By
Lemma~\ref{lem:stable-safe-deletion}, no predecessor removed in the
current or an earlier iteration can yield a strictly cheaper path.
Therefore, \(\Lambda.\mathrm{front}\) is an optimal predecessor of
customer \(x\), and
lines~\ref{line:vrptw-record-pred}--\ref{line:vrptw-record-pot}
record a predecessor attaining the cheapest-cost path to \(x\) and
the cost of that path, respectively. Hence the returned partition is
exact.

Each predecessor is inserted on line~\ref{line:vrptw-insert} once and
removed at most once. The cursor \(\Lambda.\mathrm{feas}\) advances
only on line~\ref{line:vrptw-feas-advance-end} and crosses each
predecessor at most once. Every successful repetition of the remaining
maintenance loops removes a predecessor, and each loop performs at
most one unsuccessful terminating comparison per iteration. All
penalty evaluations and deque operations take constant time after
preprocessing. The total running time is therefore \(O(n)\), with
\(O(n)\) storage.
\end{proof}

\subsection{Verification of the penalty conditions}
\label{app:soft-vrptw-penalty-verification}

The correctness of the soft VRPTW Split requires that Conditions~\eqref{cond:P1}--\eqref{cond:P5} hold for both penalties. The delivery-capacity conditions are verified in
\ref{app:capacity-penalty-preprocessing}, while the time-warp conditions are verified in \ref{app:time-warp-preprocessing}.

\subsubsection{Delivery-capacity excess}
\label{app:capacity-penalty-preprocessing}

The CVRP penalty
\(\operatorname{pen}_{\alpha d}\), defined in Table \ref{tab:split-problem-notation}, satisfies
Conditions~\eqref{cond:P1}--\eqref{cond:P3} whenever \(d_z\leq Q\)
for every customer. Its route-feasibility limit is
\begin{equation}\label{eq:cvrp-soft-route-limit}
\rho_{\alpha d}(i)
=
\min\left(
 \{x:i<x\leq n,\ D[x]-D[i]>Q\}
 \cup\{n+1\}
\right).
\end{equation}

Once two predecessors \(i<j\) are both penalized,
\begin{equation}\label{eq:cvrp-penalty-difference}
\operatorname{pen}_{\alpha d}(i,x)
-
\operatorname{pen}_{\alpha d}(j,x)
=
\alpha(D[j]-D[i]),
\end{equation}
which is independent of \(x\). Thus,
Condition~\eqref{cond:P4} holds.

To verify Condition~\eqref{cond:P5}, suppose
\(\rho_{\alpha d}(i)<\rho_{\alpha d}(j)\leq n\), and let
\(x^\star=\rho_{\alpha d}(j)\). Since predecessor \(j\) is
unpenalized at customer \(x^\star-1\),
\[
D[x^\star-1]-D[j]\leq Q,
\]
and therefore
\[
\operatorname{pen}_{\alpha d}(j,x^\star)
=
\alpha(D[x^\star]-D[j]-Q)
\leq
\alpha d_{x^\star}.
\]
Predecessor \(i\) is already penalized at customer \(x^\star-1\), so
\[
\operatorname{pen}_{\alpha d}(i,x^\star)
-
\operatorname{pen}_{\alpha d}(i,x^\star-1)
=
\alpha d_{x^\star}.
\]
Hence Condition~\eqref{cond:P5} also holds.

The penalty itself is evaluated in constant time from \(D[\cdot]\).
Since \(D\) is nondecreasing, all values
\(\rho_{\alpha d}(i)\) are obtained in \(O(n)\) time by scanning their
first violating endpoints with a single nondecreasing right pointer.

\subsubsection{Time warp}

For the soft VRPTW, the first customer at which time warp becomes
positive is
\begin{equation}\label{eq:time-warp-route-limit}
\rho_{\beta a}(i)
=
\min\left(
 \{x:i<x\leq n,\ \operatorname{tw}(i,x)>0\}
 \cup\{n+1\}
\right).
\end{equation}
Algorithm~\ref{alg:hard-tw-limits} supplies these limits together with
the initial warp \(r_a(i)\) whenever \(\rho_{\beta a}(i)\leq n\).

For \(\operatorname{pen}_{\beta a}\), singleton feasibility establishes
Condition~\eqref{cond:P3}. The proofs of Conditions~\eqref{cond:P1},
\eqref{cond:P2}, \eqref{cond:P4}, and~\eqref{cond:P5}, together with
the linear-time preprocessing for constant-time penalty evaluation,
are given below.

\subsection{Time-warp preprocessing and remaining penalty properties}
\label{app:time-warp-preprocessing}

As in the time-window route-limit construction, define the effective
closing time
\[
\bar b_z=\min\{b_z,T-s_z-t_{z0}\}.
\]
Using \(\bar b_z\) instead of $b_z$ for time warp calculations allows return-depot lateness that is already
unavoidable to be charged as time warp at customer \(z\), rather than
waiting until a later customer or the final return to the depot.

To see why this does not change the total time warp, suppose that,
after applying any warp required by \(b_z\), service at customer \(z\)
still begins \(\delta>0\) time units after
\(T-s_z-t_{z0}\). If the route ended at \(z\), the vehicle would
therefore return to the depot exactly \(\delta\) time units late.
These \(\delta\) units of warp are thus already unavoidable.

We must also show that charging this warp immediately at \(z\) cannot
cause part of it to be counted more than once as a result of additional waiting later in the
route that would not have happened if $\delta$ warp had not been applied at $z$. When the depot closing window is active at \(z\),
\(\bar b_z=T-s_z-t_{z0}<b_z\). By
Inequality~\eqref{eq:weak-triangle-in} and singleton feasibility at
customer \(z+1\),
\[
\bar b_z+s_z+t_{z,z+1}
\geq
T-s_{z+1}-t_{z+1,0}
\geq
a_{z+1}.
\]
Thus, after moving the service start at \(z\) back to \(\bar b_z\),
the vehicle still reaches \(z+1\) no earlier than its opening time, so
the additional warp cannot create waiting there. The same argument
applies successively while any of the \(\delta\)-unit shift remains.
Therefore, each unit of warp charged at \(z\) replaces one unit of
warp that would otherwise be incurred at a later customer or on the
return to the depot. Using \(\bar b_z\) therefore changes only where
the warp is recorded, not its total amount.

We therefore use the effective closing times \(\bar b_z\) throughout
the preprocessing to calculate
\(\operatorname{tw}(i,x)\).

We first calculate the time warp of the prefix routes
\(R(0,x)\), \(1\leq x\leq n\). Define
\begin{equation}\label{eq:prefix-time-warp}
W[x]=\operatorname{tw}(0,x),
\qquad W[0]=0.
\end{equation}
The values \(W[x]\) can be obtained in one forward traversal of the
giant tour. Let \(e_z\) be the service-start time after applying warp
at customer \(z\), with \(e_0=0\) and \(s_0=0\). For
\(z=1,\ldots,n\), calculate
\[
\widehat A_z
=
\max\{e_{z-1}+s_{z-1}+t_{z-1,z},a_z\},
\]
\[
w_z=\max\{\widehat A_z-\bar b_z,0\},
\qquad
e_z=\widehat A_z-w_z,
\]
and
\[
W[z]=W[z-1]+w_z.
\]
Thus, \(W[x]-W[y]\) is the amount of time warp incurred by the
prefix route after customer \(y\) through customer \(x\).

We next note two monotonicity properties of time warp. First,
appending another customer to a route does not affect warp that has
already occurred and can only introduce additional warp. Hence
\[
\operatorname{pen}_{\beta a}(i,x)
\leq
\operatorname{pen}_{\beta a}(i,x+1),
\]
establishing Condition~\eqref{cond:P1}.

Second, removing the first customer from a route cannot increase its
time warp. Consider the two routes \(R(i-1,x)\) and \(R(i,x)\).
Their first shared customer is \(i+1\). By
Inequality~\eqref{eq:weak-triangle-out} and singleton feasibility,
\(R(i-1,x)\) does not warp at customer $i$ and arrives at customer \(i+1\) no earlier than
\(R(i,x)\). If both $R(i-1,x)$ and $R(i,x)$ wait at a common customer $u$, then their arrival times are identical from then on, and warps will be applied equally. If $R(i,x)$ experiences warp for the first time at some customer $\rho_{\beta a}(i)$, then $R(i-1,x)$ arrives at $\rho_{\beta a}(i)$ at least as late, so $R(i-1)$ has at least as much warp by the time both arrive at $\rho_{\beta a}(i)$, and their penalty updates are equivalent from then on. Thus,
$
\operatorname{pen}_{\beta a}(i,x)
\leq
\operatorname{pen}_{\beta a}(i-1,x),
$
establishing Condition~\eqref{cond:P2}.

Repeated application of the fact employed above (i.e., that \(R(i-1,x)\) arrives at customer \(i+1\) no earlier than
\(R(i,x)\)) immediately results in the fact that for any predecessors $i<j$ reaching customer $x$, the arrival time for $R(i,x)$ to customer $j+1$ is no earlier than it is for $R(j,x)$. This fact is necessary to show the remaining conditions and allows for the following constant-time evaluation.

Suppose route \(R(i,x)\) incurs its first warp
at customer \(\rho_{\beta a}(i)\). The prefix route
\(R(0,x)\), which arrives at this customer no earlier, must also warp at
\(\rho_{\beta a}(i)\). After warping, both routes begin service at \(\bar b_{\rho_{\beta a}(i)}\). From the following customer onward,
both routes arrive at every customer at the same time. Thus, the warp incurred at every customer is exactly the same.

As a result, the amount of warp incurred after this first warped customer can be read directly from \(W\), where
\[
W[x]-W[\rho_{\beta a}(i)]
\]
is the amount of warp incurred by \(R(i,x)\) after customer
\(\rho_{\beta a}(i)\). Adding its initial warp \(r_a(i)\) gives
\begin{equation}\label{eq:time-warp-growth}
\operatorname{tw}(i,x)
=
r_a(i)+W[x]-W[\rho_{\beta a}(i)],
\qquad x\geq\rho_{\beta a}(i).
\end{equation}
Consequently,
\[
\operatorname{pen}_{\beta a}(i,x)
=
\begin{cases}
0,
& x<\rho_{\beta a}(i),\\[1mm]
\beta\bigl(
r_a(i)+W[x]-W[\rho_{\beta a}(i)]
\bigr),
& x\geq\rho_{\beta a}(i),
\end{cases}
\]
which is calculable in constant time.

Equation~\eqref{eq:time-warp-growth} also establishes
Condition~\eqref{cond:P4}. Once two predecessors \(i<j\) have both
incurred time warp,
\[
\begin{aligned}
\operatorname{tw}(i,x)-\operatorname{tw}(j,x)
={}&
r_a(i)-W[\rho_{\beta a}(i)]\\
&-
r_a(j)+W[\rho_{\beta a}(j)],
\end{aligned}
\]
which does not depend on \(x\). Thus, once both time-warp penalties
are active, their difference remains fixed as further customers are
appended.

Finally, suppose
\(\rho_{\beta a}(i)<\rho_{\beta a}(j)\leq n\).
Predecessor \(j\) incurs its first warp at
\(\rho_{\beta a}(j)\).
The prefix route reaches this customer no earlier than
\(R\bigl(j,\rho_{\beta a}(j)\bigr)\), so the initial warp of $R(j,\rho_{\beta a}(j))$ is
\[
r_a(j)
\leq
W[\rho_{\beta a}(j)]
-
W[\rho_{\beta a}(j)-1].
\]
Since predecessor \(i\) was already warped before this customer,
Equation~\eqref{eq:time-warp-growth} gives
\[
W[\rho_{\beta a}(j)]
-
W[\rho_{\beta a}(j)-1]
=
\operatorname{tw}\bigl(i,\rho_{\beta a}(j)\bigr)
-
\operatorname{tw}\bigl(i,\rho_{\beta a}(j)-1\bigr).
\]
Therefore,
\begin{equation}\label{eq:time-warp-boundary-increase}
\operatorname{tw}\bigl(j,\rho_{\beta a}(j)\bigr)
\leq
\operatorname{tw}\bigl(i,\rho_{\beta a}(j)\bigr)
-
\operatorname{tw}\bigl(i,\rho_{\beta a}(j)-1\bigr),
\end{equation}
establishing
Condition~\eqref{cond:P5} for
\(\operatorname{pen}_{\beta a}\).

\section{Soft VRPSPD Split}
\label{app:soft-spd}

\subsection{Maximum-load properties}

Recall that \(h(i,x)\) is the latest position attaining
\(\max_{i\leq z\leq x}\{P[z]-D[z]\}\).

\begin{proof}[Proof of Condition~\eqref{cond:S1}]
Suppose \(h(i,x)=h(j,x)=r\). Then \(r\) is the highest-load
position in terms of \(P[z]-D[z]\) among both
\(\{i,\ldots,x\}\) and \(\{j,\ldots,x\}\), with ties broken in favor
of the later position.

When customer \(x+1\) is appended, it is the only new candidate for
the highest-load position of either route. If
$
P[x+1]-D[x+1] < P[r]-D[r],
$
then \(r\) remains the highest-load position for both routes. Otherwise,
\(x+1\) becomes the latest position attaining the maximum for both routes.
Hence
$
h(i,x+1)=h(j,x+1).
$
Repeating this argument proves the claim for every \(y\geq x\).
\end{proof}

\begin{proof}[Proof of Condition~\eqref{cond:S2}]
Since ties are resolved in favor of the latest position,
\(h(j,x)=h(j,x+1)\) implies
\(P[x+1]-D[x+1]<P[h(j,x)]-D[h(j,x)]\). The interval defining
predecessor \(i<j\) contains the interval defining \(j\), so
\(P[h(i,x)]-D[h(i,x)]\geq P[h(j,x)]-D[h(j,x)]\). Therefore,
\(P[x+1]-D[x+1]<P[h(i,x)]-D[h(i,x)]\), and the maximizer of
predecessor \(i\) cannot change either.
\end{proof}

\begin{proof}[Proof of Condition~\eqref{cond:S3}]
For a predecessor already penalized at customer \(x\), appending
customer \(x+1\) increases its penalty by
\begin{align*}
\Delta_i(x)
=\alpha\bigl(&d_{x+1}
 +\max\{P[h(i,x)]-D[h(i,x)],P[x+1]-D[x+1]\}\\
 &-(P[h(i,x)]-D[h(i,x)])\bigr).
\end{align*}
For \(i<j\), the interval of \(i\) contains that of \(j\), so
\(P[h(i,x)]-D[h(i,x)]\geq P[h(j,x)]-D[h(j,x)]\). The function
\(u\mapsto\max\{u,P[x+1]-D[x+1]\}-u\) is nonincreasing; hence
\(\Delta_i(x)\leq\Delta_j(x)\). If the maximizer of \(j\) does not
change, Condition~\eqref{cond:S2} shows that neither maximizer changes,
and both increments equal \(\alpha d_{x+1}\).
\end{proof}

When two penalized predecessors share a maximizer, their penalty
difference is \(\alpha(P[j]-P[i])\), independent of the final
customer. More generally, if both predecessors are penalized and
\(\operatorname{dominated}_{\alpha p}(i,j)\) holds, then the older
predecessor \(i\) is permanently dominated by \(j\), even before their
maximum-load positions coincide. Since \(i<j\), the interval defining
\(h(i,x)\) contains the interval defining \(h(j,x)\), so the 
penalty difference is bounded below by \(\alpha(P[j]-P[i])\), exactly
the difference used by the permanent comparison.

When an already penalized predecessor \(i\) is tested against an
unpenalized cursor \(j\), the algorithm additionally maintains that
\(i\) is currently no cheaper than \(j\). Until \(j\) becomes
penalized, this cost difference cannot decrease by
Condition~\eqref{cond:P1}; after that, the permanent comparison above
applies. The test is therefore safe in this case as well. Within the
already penalized prefix, comparisons need rechecking only when the
later predecessor's maximum-load position changes. Penalty activations
are handled separately by the successor and boundary scans.

\subsection{Rightmost maximum-load queries}
\label{app:soft-spd-rmq}

Algorithm~\ref{alg:soft-vrpspd-split} repeatedly queries the latest
position maximizing \(P[z]-D[z]\) over an interval \([i,x]\). These are
static range-maximum queries on the sequence
\[
\bigl(P[z]-D[z],z\bigr),\qquad z=0,\ldots,n,
\]
ordered lexicographically so that the higher position wins a tie.
A max Cartesian tree for this sequence can be constructed in \(O(n)\)
time. The tree preserves the tour positions under an inorder traversal
and satisfies the max-heap property with respect to the lexicographic
keys. Consequently, for any interval \([i,x]\), the lowest common
ancestor of nodes \(i\) and \(x\) is the node having the
largest lexicographic key in that interval, and is therefore \(h(i,x)\).

After \(O(n)\)-time and \(O(n)\)-space preprocessing of the Cartesian
tree for lowest-common-ancestor queries, each such ancestor can be
retrieved in \(O(1)\) time \citep{bender2000lca}. Thus, every
occurrence of \(h(i,x)\) in Algorithm~\ref{alg:soft-vrpspd-split} is
evaluated in constant time after a single linear preprocessing step.
\subsection{Best-cursor invariant}

For Algorithm~\ref{alg:soft-vrpspd-split} at some iteration $x$, the retained predecessor deque can be denoted as
\[
\Lambda=(i_1,\ldots,i_{\ell-1},i_\ell,i_{\ell+1},\ldots,i_r),
\qquad i_\ell=\Lambda.\mathrm{best}.
\]
Before recording $\operatorname{pred}(x)$ and $\operatorname{pot}(x)$, the algorithm maintains the following
invariant:
\begin{enumerate}
\item the predecessor indices in $\Lambda$ increase from front to back;
\item every predecessor after \(\Lambda.\mathrm{best}\) belongs to
\(Y_x^2\) after maintenance, and their base candidate costs strictly
increase toward the back;
\item every predecessor before \(\Lambda.\mathrm{best}\) belongs to
\(Y_x^1\) after maintenance, and their penalized candidate costs
are nonincreasing toward the cursor;
\item \(\Lambda.\mathrm{best}\) may belong to either region and is
no more expensive than either neighbor;
\item for all retained predecessors \(i,j\in\Lambda\),
\[
i<j\leq\Lambda.\mathrm{best}\quad\Longrightarrow\quad
\operatorname{dominated}_{\alpha p}(i,j)=\mathrm{false}.
\]
\end{enumerate}
An absent neighbor makes the corresponding statement vacuous.

The back insertion on
lines~\ref{line:spd-back-start}--\ref{line:spd-insert} preserves the
first two conditions. If an older predecessor is no cheaper in base
cost than the newly inserted predecessor, Condition~\eqref{cond:P2}
ensures that its capacity penalty is also no smaller at every customer, so
its deletion is permanent. If this deletion removes the best cursor,
lines~\ref{line:spd-best-reset-start}--\ref{line:spd-best-reset-end}
place the cursor at the new back and remove older predecessors satisfying
\(\operatorname{dominated}_{\alpha p}\).

By Condition~\eqref{cond:S2}, if appending customer \(x\) leaves the
maximum-load position of \(\Lambda.\mathrm{best}\) unchanged, then it
also leaves every earlier maximum unchanged. Otherwise,
lines~\ref{line:spd-max-update-start}--\ref{line:spd-max-update-end}
visit exactly the contiguous predecessors whose maximizer becomes
\(x\). By Condition~\eqref{cond:S1}, predecessors whose maximizers
coincide thereafter retain the same maximizer, so the permanent
comparison in Equation~\eqref{eq:spd-permanent-comparison} justifies the
associated deletions.

The scan beginning on Line~\ref{line:spd-penalized-scan-start}
processes successors as they become penalized. If the current best, which is the earlier predecessor, remains cheaper, Condition~\eqref{cond:S3}
ensures that the successor cannot overtake it and the successor is
removed. Otherwise, the cursor moves to the no-more-expensive successor. Equation~\eqref{eq:spd-permanent-comparison}
then removes older predecessors that are permanently dominated under
the conditions above, retaining those not ruled out by this test.
Lines~\ref{line:spd-boundary-start}--\ref{line:spd-boundary-end} perform
the same comparison with the first unpenalized successor and restore the
penalized--unpenalized boundary.

\begin{lemma}\label{lem:spd-best-invariant}
At the end of every iteration of Algorithm~\ref{alg:soft-vrpspd-split},
the invariant above holds and \(\Lambda.\mathrm{best}\) is a
minimum-cost predecessor for the current customer.
\end{lemma}

\begin{proof}
The claim is immediate at customer 1. Assume the invariant after
customer \(x-1\). By Condition~\eqref{cond:P2}, the back deletions on
lines~\ref{line:spd-back-start}--\ref{line:spd-back-end} are permanent,
and line~\ref{line:spd-insert} preserves the order of the unpenalized
suffix. By Conditions~\eqref{cond:S1} and~\eqref{cond:S2}, the
comparisons within the already penalized prefix that can change at
customer \(x\) are checked on
lines~\ref{line:spd-max-update-start}--\ref{line:spd-max-update-end}.
By Condition~\eqref{cond:S3}, the scan beginning on
line~\ref{line:spd-penalized-scan-start} can remove a more expensive
later predecessor permanently. Under the penalty-state and current-cost
conditions above, Equation~\eqref{eq:spd-permanent-comparison}
identifies earlier predecessors that can be discarded permanently;
those not ruled out by this test are retained.

After these lines terminate, neither neighbor is cheaper than the
cursor, the entries on each side have the required monotone ordering,
and every removed entry is permanently dominated by a retained entry.
The cursor is therefore a global minimum among all predecessors.
\end{proof}

\begin{proof}[Proof of Theorem~\ref{thm:soft-vrpspd-split}]
By Lemma~\ref{lem:spd-best-invariant}, the predecessor and potential
recorded on lines~\ref{line:spd-record-pred}--\ref{line:spd-record-pot}
identify the predecessor giving a lowest-cost path to $x$ in \(\mathcal G_{\alpha h}\) and the cost associated with that path, respectively. The range-maximum structure is constructed in linear time and
answers each query in constant time. Each predecessor is inserted once
and removed at most once. Any move of \(\Lambda.\mathrm{best}\) toward
the front is paired with a permanent deletion, while moves toward the
back cross a predecessor at most once before that predecessor is
deleted or becomes the cursor. The total number of loop iterations is
therefore \(O(n)\), and all stored arrays, range-query data, and deque
entries use \(O(n)\) space.
\end{proof}

\section{Soft VRPSPDTW Split}
\label{app:soft-vrpspdtw}

As discussed in Section~\ref{sec:soft-vrpspdtw}, maintenance of the
fully penalized and unpenalized regions \(Y_x^1\) and \(Y_x^3\)
for the soft VRPSPDTW Split mirrors that of the preceding Splits. The fully penalized region uses
the maximum-load maintenance of the soft VRPSPD together with the
fixed time-warp difference from Condition~\eqref{cond:P4}, while the
unpenalized region uses the base-cost dominance rule.

The singly-penalized region \(Y_x^2\) requires two representations depending on
which penalty is active. When it is time-warp-only, it is represented
by \(\Lambda^2\), and the same constant-difference dominance
maintenance used for the single-penalty middle region in the soft VRPTW
Split applies. We denote
\(\operatorname{dominated}_{\beta a}(i,j,x)\) for the same
constant-difference dominance test from Equation~\eqref{eq:dominated-penalized} used for the time-warp region of
the soft VRPTW Split (excluding the capacity penalty term).
When \(Y_x^2\) is capacity-only, it is represented by \(\Gamma^2\),
whose additional operations are defined below.

We denote the indicator function that determines the type of penalty applied to the current $Y^2_x$ as
\begin{equation}\label{eq:spdtw-capacity-only}
\operatorname{capacityOnly}(x)
\Longleftrightarrow
\exists i\in Y_x^2:\rho_{\alpha h}(i)\leq x.
\end{equation}
Thus, \(\operatorname{capacityOnly}(x)\) is true when \(Y_x^2\) is
capacity-only and false when it is time-warp-only or empty. Let
\(\operatorname{capacityOnly}(0)=\mathrm{false}\).

Algorithm~\ref{alg:soft-vrpspdtw-split} presents the resulting Split. For algorithm readability, predecessors entering \(Y_x^1\) and \(Y_x^2\) are temporarily stored
in the buffers \(\mathcal T^1\) and \(\mathcal T^2\), respectively.
The first buffer feeds \(\Lambda^1\). The second feeds \(\Gamma^2\)
when \(\operatorname{capacityOnly}(x)\) is true and \(\Lambda^2\)
otherwise.

\begin{algorithm}[!t]
\caption{Linear Split for the soft VRPSPDTW}
\label{alg:soft-vrpspdtw-split}\label{alg:soft-vrpspdtw}
\scriptsize
\DontPrintSemicolon
\LinesNumbered
\SetKwInOut{Input}{Input}
\SetKwInOut{Output}{Output}

\Input{Tour length \(n\); route-cost data and multipliers
\(\alpha,\beta\); limits \(\rho_{\alpha h},\rho_{\beta a}\); and
preprocessed data for constant-time maximum-load and penalty
evaluation}
\Output{Shortest-path costs
\(\operatorname{pot}(1),\ldots,\operatorname{pot}(n)\) and
predecessors
\(\operatorname{pred}(1),\ldots,\operatorname{pred}(n)\)}

\(\operatorname{pot}(0)\leftarrow0\);
\(\Lambda^1,\Lambda^2,\Lambda^3,\Gamma^2\leftarrow()\)
\nllabel{line:spdtw-init}\;

\For{\(x\leftarrow1\) \KwTo \(n\)}{
  \(\mathcal T^1,\mathcal T^2\leftarrow()\)
  \nllabel{line:spdtw-buffer-init}\;

  \eIf{\(\operatorname{capacityOnly}(x-1)
    \nllabel{line:spdtw-middle-exit-start}\)}{
    \While{\(\Gamma^2\neq()\) \textnormal{and}
           \(\Gamma^2.\mathrm{front}\in Y_x^1\)}{
      \(i\leftarrow\Gamma^2.\mathrm{popFront}()\)\;
      \(\mathcal T^1.\mathrm{pushBack}(i)\)\;
    }
  }{
    \While{\(\Lambda^2\neq()\) \textnormal{and}
           \(\Lambda^2.\mathrm{front}\in Y_x^1\)}{
      \(i\leftarrow\Lambda^2.\mathrm{popFront}()\)\;
      \(\mathcal T^1.\mathrm{pushBack}(i)\)
      \nllabel{line:spdtw-middle-exit-end}\;
    }
  }

  \While{\(\Lambda^3\neq()\) \textnormal{and}
         \(\Lambda^3.\mathrm{front}\notin Y_x^3
         \nllabel{line:spdtw-y3-exit-start}\)}{
    \(i\leftarrow\Lambda^3.\mathrm{popFront}()\)\;
    \eIf{\(i\in Y_x^1\)}{
      \(\mathcal T^1.\mathrm{pushBack}(i)\)\;
    }{
      \(\mathcal T^2.\mathrm{pushBack}(i)\)
      \nllabel{line:spdtw-y3-exit-end}\;
    }
  }

  \(\operatorname{maintain}_{Y_x^1}(\mathcal T^1,x)\)
  \nllabel{line:spdtw-maintain-y1}\;

  \eIf{\(\operatorname{capacityOnly}(x)\)}{
    \(\Gamma^2.\mathrm{updateLoads}(x)\)
    \nllabel{line:spdtw-middle-capacity-start}\;

    \While{\(\mathcal T^2\neq()\)}{
      \(i\leftarrow\mathcal T^2.\mathrm{popFront}()\)\;
      \(\Gamma^2.\mathrm{pushBack}(i,x)\)
      \nllabel{line:spdtw-middle-capacity-end}\;
    }
  }{
    \While{\(\mathcal T^2\neq()
      \nllabel{line:spdtw-middle-time-start}\)}{
      \(i\leftarrow\mathcal T^2.\mathrm{popFront}()\)\;

      \While{\(\Lambda^2\neq()\) \textnormal{and}
             \(\operatorname{dominated}_{\beta a}
               (\Lambda^2.\mathrm{back},i,x)\)}{
        \(\Lambda^2.\mathrm{removeBack}()\)\;
      }

      \(\Lambda^2.\mathrm{pushBack}(i)\)
      \nllabel{line:spdtw-middle-time-end}\;
    }
  }

  \(\operatorname{maintain}_{Y_x^3}(x-1,x)\)
  \nllabel{line:spdtw-maintain-y3}\;

  \(\operatorname{pred}(x)\leftarrow\) the newest minimizer of
  \(\operatorname{pot}(i)+c_{\alpha h,\beta a}(i,x)\) among the
  nonempty candidates
  \(\Lambda^1.\mathrm{back}\),
  \(\Gamma^2.\mathrm{minimum}\),
  \(\Lambda^2.\mathrm{front}\), and
  \(\Lambda^3.\mathrm{front}\)
  \nllabel{line:spdtw-representatives}\;

  \(\operatorname{pot}(x)\leftarrow
    \operatorname{pot}(\operatorname{pred}(x))
    +c_{\alpha h,\beta a}(\operatorname{pred}(x),x)\)
  \nllabel{line:spdtw-record}\;
}
\end{algorithm}

Lines~\ref{line:spdtw-middle-exit-start}--\ref{line:spdtw-middle-exit-end} transfer predecessors leaving \(Y_x^2\) to \(\mathcal T^1\), while
Lines~\ref{line:spdtw-y3-exit-start}--\ref{line:spdtw-y3-exit-end} transfer predecessors leaving \(Y_x^3\)
to \(\mathcal T^1\) or \(\mathcal T^2\), depending on whether or not the predecessor is immediately doubly-penalized starting at iteration $x$.

Line~\ref{line:spdtw-maintain-y1} maintains the fully penalized region using logic similar to maintenance of the penalized region of Algorithm \ref{alg:soft-vrpspd-split}.
Lines~\ref{line:spdtw-middle-capacity-start}--
\ref{line:spdtw-middle-capacity-end} maintain \(\Gamma^2\) when only
the capacity penalty is active, while
Lines~\ref{line:spdtw-middle-time-start}--
\ref{line:spdtw-middle-time-end} maintain \(\Lambda^2\) as a monotonically increasing deque using the
time-warp dominance rule from the soft VRPTW Split.
Line~\ref{line:spdtw-maintain-y3} maintains the unpenalized region.
Finally, Lines~\ref{line:spdtw-representatives}--\ref{line:spdtw-record} compare the minimum-cost retained predecessor
reported by each nonempty region and record the resulting shortest-path
label.

\subsection{Preprocessing}

Most route preprocessing required has been performed in the preceding
Splits. Algorithms~\ref{alg:hard-spd-limits} and
\ref{alg:hard-tw-limits} provide
\(\rho_{\alpha h}\), \(\rho_{\beta a}\), and \(r_a\);
\ref{app:time-warp-preprocessing} provides \(W\) and
constant-time evaluation of the time-warp penalty; and the range-maximum
structure of \ref{app:soft-spd-rmq} provides \(h(i,x)\) in
constant time.

The values \(\operatorname{capacityOnly}(x)\) can be preprocessed in linear time.
Since \(\rho_{\alpha h}\) and \(\rho_{\beta a}\) are nondecreasing,
the predecessors penalized by each penalty form a prefix at every
customer \(x\). Two forward scans over the penalty limits therefore
identify the ends of these prefixes for all \(x\) in \(O(n)\) time.
The middle region is capacity-only when the capacity-penalized
prefix is longer than the time-warp-penalized prefix, and false otherwise.

\subsection{Maintenance of \texorpdfstring{\(Y_x^1\)}{Y1}}

Algorithm \ref{alg:spdtw-maintain-y1} presents the $\operatorname{maintain}_{Y^1_x}$ routine. In \(Y_x^1\), both penalties are active. By
Condition~\eqref{cond:P4}, the difference between the time-warp
penalties of two predecessors is fixed. The capacity maintenance from
the soft VRPSPD Split therefore applies after including this fixed
time-warp difference.

Analogous to the comparison functions of the soft VRPSPD Split, we first denote the permanent comparison
\begin{equation}\label{eq:spdtw-full-permanent-dominance}
\begin{aligned}
\operatorname{dominated}_{\alpha p,\beta a}(i,j)
\Longleftrightarrow{}&
\operatorname{pot}(i)+c_{0,i+1}-C[i+1]
+\alpha(P[j]-P[i])\\
&+\beta\Bigl(
r_a(i)-W[\rho_{\beta a}(i)]
-r_a(j)+W[\rho_{\beta a}(j)]
\Bigr)\\
&\geq
\operatorname{pot}(j)+c_{0,j+1}-C[j+1],
\qquad i<j.
\end{aligned}
\end{equation}
This is the same comparison as defined in
Equation~\eqref{eq:spd-permanent-comparison}, with the fixed
difference between the two time-warp penalties added. For every
retained adjacent pair \(i<j\) in \(\Lambda^1\),
\(\operatorname{dominated}_{\alpha p,\beta a}(i,j)\) is false after
maintenance. If the two predecessors later acquire the same
maximum-load position, the lower-bound capacity difference used by
this comparison becomes exact. The older predecessor is then strictly
cheaper, which justifies the suffix removals on Lines~3--4 of
Algorithm~\ref{alg:spdtw-maintain-y1}. Lines~6--7 then compare the
remaining suffix representative with its preceding neighbor and remove
it if that neighbor is strictly cheaper.

For each incoming predecessor \(i\), Line~12 retains it for further
consideration only when \(\Lambda^1\) is empty or the current back is
not strictly cheaper. Otherwise, \(i\) is discarded; by
Condition~\eqref{cond:S3} and the fixed time-warp difference, it
cannot become cheaper later. If \(i\) is retained,
Lines~13--14 remove older predecessors permanently dominated by \(i\),
and Line~16 appends \(i\) to \(\Lambda^1\).

For \(i<j\),  the current-cost comparison used on 
Lines~6 and~12 is 
\begin{equation}\label{eq:spdtw-full-cheaper}
\begin{aligned}
\operatorname{cheaper}_{\alpha h,\beta a}(i,j,x)
\Longleftrightarrow{}&
\operatorname{pot}(i)+c(i,x)
+\operatorname{pen}_{\alpha h}(i,x)
+\operatorname{pen}_{\beta a}(i,x)\\
<&\operatorname{pot}(j)+c(j,x)
+\operatorname{pen}_{\alpha h}(j,x)
+\operatorname{pen}_{\beta a}(j,x).
\end{aligned}
\end{equation}
This is likewise the same comparison as defined in Equation \eqref{eq:spd-current-comparison}, now with time warp included.

After maintenance, the deque \(\Lambda^1\) is in increasing predecessor order, with current candidate costs nonincreasing from front to back.
Therefore, \(\Lambda^1.\mathrm{back}\) is its newest minimum-cost
predecessor.

\begin{algorithm}[t]
\caption{Maintenance of the fully penalized region}
\label{alg:spdtw-maintain-y1}
\footnotesize
\DontPrintSemicolon
\SetKwProg{Proc}{Procedure}{:}{}
\Proc{\(\operatorname{maintain}_{Y_x^1}(\mathcal T^1,x)\)}{

  \If{\(\Lambda^1\neq()\) \textnormal{and}
      \(h(\Lambda^1.\mathrm{back},x)=x\)
      \nllabel{line:spdtw-y1-sub-update}}{

    \While{\(\Lambda^1.\mathrm{size}>1\) \textnormal{and}
        \(h(\Lambda^1.\mathrm{back2},x)=x\)
        \nllabel{line:spdtw-y1-shared-start}}{
      \(\Lambda^1.\mathrm{removeBack}()\)
      \nllabel{line:spdtw-y1-shared-end}\;
    }

    \If{\(\Lambda^1.\mathrm{size}>1\) \textnormal{and}
        \(\operatorname{cheaper}_{\alpha h,\beta a}
        (\Lambda^1.\mathrm{back2},
         \Lambda^1.\mathrm{back},x)\)}{
      \(\Lambda^1.\mathrm{removeBack}()\)
      \nllabel{line:spdtw-y1-sub-update-end}\;
    }
  }

  \While{\(\mathcal T^1\neq()
    \nllabel{line:spdtw-y1-sub-incoming}\)}{
    \(i\leftarrow\mathcal T^1.\mathrm{popFront}()\)\;

    \If{\(\Lambda^1=()\) \textnormal{or}
        \(\neg\operatorname{cheaper}_{\alpha h,\beta a}
        (\Lambda^1.\mathrm{back},i,x)
        \nllabel{line:spdtw-y1-incoming-test}\)}{
      \While{\(\Lambda^1\neq()\) \textnormal{and}
          \(\operatorname{dominated}_{\alpha p,\beta a}
          (\Lambda^1.\mathrm{back},i)
          \nllabel{line:spdtw-y1-incoming-dom-start}\)}{
        \(\Lambda^1.\mathrm{removeBack}()\)
        \nllabel{line:spdtw-y1-incoming-dom-end}\;
      }
      \(\Lambda^1.\mathrm{pushBack}(i)\)
      \nllabel{line:spdtw-y1-sub-insert}\;
    }
  }
}
\end{algorithm}

Conditions~\eqref{cond:S1}--\eqref{cond:S3} give the same
maximum-load ordering used in the soft VRPSPD Split, while
Condition~\eqref{cond:P4} fixes the time-warp difference once both
predecessors are penalized. Thus,
\(\operatorname{dominated}_{\alpha p,\beta a}\) is permanent under
the same conditions as
\(\operatorname{dominated}_{\alpha p}\).
The current-cost comparison
\(\operatorname{cheaper}_{\alpha h,\beta a}\) is used where the
relative capacity penalties may still change.

Each successful loop iteration removes a predecessor, and every
predecessor enters \(\Lambda^1\) at most once. The total work over all
calls from Algorithm \ref{alg:soft-vrpspdtw-split} is therefore \(O(n)\).

\subsection{Maintenance of \texorpdfstring{\(Y_x^3\)}{Y3}}

The unpenalized region uses the same base-cost dominance rule as the
preceding Splits. The following routine extends its back-removal across
the boundaries between the three regional structures.

\begin{algorithm}[t]
\caption{Maintenance of the unpenalized region}
\label{alg:spdtw-maintain-y3}
\footnotesize
\DontPrintSemicolon
\SetKwProg{Proc}{Procedure}{:}{}
\SetKw{Break}{break}
\Proc{\(\operatorname{maintain}_{Y_x^3}(k,x)\)}{
  \While{\textnormal{at least one regional structure is nonempty}
      \nllabel{line:spdtw-y3-sub-loop}}{
    \(i\leftarrow\) \textnormal{the newest retained predecessor}
      \nllabel{line:spdtw-y3-sub-init}\;
    \If{\(\textnormal{not }\operatorname{dominated}(i,k)
        \nllabel{line:spdtw-y3-sub-test}\)}{
      \Break\;
    }
    \textnormal{Remove \(i\) from the back of its regional structure}
      \nllabel{line:spdtw-y3-sub-remove}\;
  }
  \(\Lambda^3.\mathrm{pushBack}(k)\)
    \nllabel{line:spdtw-y3-sub-insert}\;
}
\end{algorithm}

When \(\operatorname{dominated}(i,k)\) holds, the base cost through
\(i\) is no smaller than through \(k\). Condition~\eqref{cond:P2}
ensures that neither penalty can reverse this ordering at a later
customer. Each successful loop iteration removes a predecessor, and each call performs at most one terminating unsuccessful comparison. Thus, all calls to the routine take $O(n)$ time.

\subsection{Middle-region structures}

When \(Y_x^2\) is time-warp-only, \(\Lambda^2\) is maintained using
the same time-warp dominance rule as in the soft VRPTW Split. Its front
is therefore its newest minimum-cost retained predecessor.

When \(Y_x^2\) is capacity-only, \(\Gamma^2\) is used. Its retained
predecessors are partitioned into inner deques \(\Lambda_z\), where
\[
i\in\Lambda_z
\quad\Longleftrightarrow\quad
h(i,x)=z.
\]
Thus, all predecessors in an inner deque share the same current
maximum-load position. For a given iteration $x$, as the predecessor index $i$ increases, \(
\max_{i\le r\le x}\{P[r]-D[r]\}
\) for a route $R(i,x)$ can only decrease, since increasing $i$ excludes candidates. Thus, successive
inner deques from front to back have strictly decreasing values of
\(P[z]-D[z]\).

Within each inner deque \(\Lambda_z\), predecessors are maintained
using the permanent capacity dominance test
\(\operatorname{dominated}_{\alpha p}\) from Equation \eqref{eq:spd-permanent-comparison} for the soft VRPSPD Split.
Suppose \(i<j\) belong to the same \(\Lambda_z\) and
\(\operatorname{dominated}_{\alpha p}(i,j)\) holds.
By Condition~\eqref{cond:S1}, the two predecessors continue to share
a maximum-load position after every extension, so the capacity
comparison remains permanent. In addition,
Condition~\eqref{cond:P2} ensures that the time-warp penalty of \(i\)
can never be smaller than that of \(j\). Therefore, \(i\) can never
become cheaper than \(j\) and may be removed permanently.

The front of each \(\Lambda_z\) is its minimum-cost retained
predecessor. Since \(Y_x^2\) is capacity-only, fronts of different
inner deques can be compared using
\(\operatorname{cheaper}_{\alpha h}\) from
Equation~\eqref{eq:spd-current-comparison}. For \(i=\Lambda_z.\mathrm{front}\), the capacity-only candidate cost
can be written as
\[
\begin{aligned}
\operatorname{pot}(i)+c_{\alpha h}(i,x)
={}&C[x]+c_{x0}+\alpha(D[x]-Q)\\
&+\operatorname{pot}(i)+c_{0,i+1}-C[i+1]
-\alpha P[i]+\alpha(P[z]-D[z]).
\end{aligned}
\]
The first line is common to every capacity-only predecessor at customer
\(x\). The remaining terms depend only on the front predecessor \(i\)
and its maximum-load position \(z\). Consequently, for the fronts of
any two inner deques, their candidate-cost difference remains unchanged
as \(x\) advances, provided that neither front nor maximum-load position
changes.

This property allows the outer structure of \(\Gamma^2\) to maintain
its minimum without recomparing every inner deque at each customer.
The stored minimum needs to be updated only when an inner-deque front
changes or when an inner deque is created, removed, or merged.

The operations of \(\Gamma^2\) are as follows.

\begin{description}

\item[\(\Gamma^2.\mathrm{front}\), \(\Gamma^2.\mathrm{back}\):]
Return the oldest or newest predecessor from the first or last inner
deque, respectively.

\item[\(\mathrm{\Gamma^2.pushBack}(i,x)\):]
Compute \(z=h(i,x)\). If the final inner deque is not \(\Lambda_z\),
create a new inner deque. Otherwise, remove predecessors from its back
while
$\operatorname{dominated}_{\alpha p}
\bigl(\Lambda_z.\mathrm{back},i\bigr)$,
and then append \(i\) via $\Lambda_z.\operatorname{pushBack}(i)$.

\item[\(\Gamma^2.\mathrm{popFront}()\), \(\Gamma^2.\mathrm{popBack}()\):]
Remove and return the oldest or newest predecessor from the first or last deque, respectively. Delete its
inner deque if it becomes empty.

\item[\(\Gamma^2.\mathrm{updateLoads}(x)\):]
The inner deques satisfying
$
P[z]-D[z]\leq P[x]-D[x]
$ 
form a suffix of $\Gamma^2$ by Condition~\eqref{cond:S2}. For every predecessor in this suffix, customer \(x\)
becomes the new maximum-load position, so these inner deques are merged
in predecessor order into a single deque \(\Lambda_x\).

When joining an older inner deque $\Lambda_z$ to the newest one $\Lambda_x$, remove predecessors
from the back of the older deque while
$
\operatorname{dominated}_{\alpha p}
\bigl(\Lambda_{z}.\mathrm{back},
      \Lambda_{x}.\mathrm{front}\bigr)
$
holds, and then concatenate the two deques into a single $\Lambda_x$. These removals are permanent
because, after the merge, all remaining predecessors share the same
maximum-load position \(x\).

\item[\(\Gamma^2.\mathrm{minimum}\):]
Return the newest front predecessor minimizing
\[
\operatorname{pot}(i)+c_{0,i+1}-C[i+1]-\alpha P[i]
+\alpha\bigl(P[z]-D[z]\bigr),
\qquad i=\Lambda_z.\mathrm{front}.
\]

\end{description}

Following the two-stack aggregation approach of
\citet{tangwongsan2017low}, the outer deque of \(\Gamma^2\) is
represented by two aggregation stacks, one containing the older inner
deques and one containing the newer inner deques. Each stack stores,
together with every entry, the minimum-cost front among the inner
deques below it, with ties broken in favor of the newer predecessor.

As shown above, the relative costs of two inner-deque fronts remain
unchanged while neither front nor maximum-load position changes.
Consequently, an aggregate value needs to be updated only when an
inner deque is inserted or removed, when inner deques are merged, or
when the front of an inner deque changes. Inner-deque insertions, removals, and front changes can be charged to
outer-deque endpoint operations or permanent predecessor removals.
A merge permanently consumes one or more existing inner deques and
can instead be charged to their creation. Thus, the total number of
aggregate updates is \(O(n)\).

Because \(\Gamma^2\) supports removals from both ends, we additionally
rebalance the two stacks when an endpoint operation requires access to
an empty stack. The inner deques are redistributed in order so that
the two stack sizes differ by at most one, and the aggregate minima are
rebuilt. Redistributing \(k\) inner deques takes \(O(k)\) time, but
another such redistribution cannot be required until
\(\Omega(k)\) endpoint operations have occurred. The rebuilding cost
is therefore amortized over those operations, giving \(O(1)\)
amortized time per outer-deque operation.

\begin{lemma}\label{lem:capacity-region-queue}
After maintenance of $\Gamma^2$ at iteration $x$,
\(\Gamma^2.\mathrm{minimum}\) is its newest minimum-cost retained
predecessor. All operations on \(\Gamma^2\) take \(O(n)\) total time
and \(O(n)\) space over the complete Split.
\end{lemma}

\begin{proof}
Within each inner deque, the front minimizes the fixed predecessor
term above. The two aggregation stacks compare exactly the resulting
inner-deque minima. When customer \(x\) becomes the maximum-load
position for a suffix of inner deques, these deques are merged into
\(\Lambda_x\), and the permanent dominance rule above restores the
required order.

Each predecessor enters \(\Gamma^2\) at most once and is removed at
most once. Each inner deque is created once and can be consumed by at
most one later merge. Thus all inner-deque operations and merges take
\(O(n)\) total time. The two-stack representation adds amortized
constant work per outer-deque operation, and all structures together
use \(O(n)\) space.
\end{proof}

\subsection{Correctness and complexity}

\begin{theorem}
\label{thm:soft-vrpspdtw-split}\label{thm:soft-vrpspdtw}
Under the assumptions of Section~\ref{sec:split-problem}, including
singleton route feasibility and the weakened triangle
inequalities~\eqref{eq:weak-triangle-out} and
\eqref{eq:weak-triangle-in}, Algorithm \ref{alg:soft-vrpspdtw-split} returns a minimum-cost
partition for \(\mathcal G_{\alpha h,\beta a}\) in \(O(n)\) time and
\(O(n)\) space.
\end{theorem}

\begin{proof}
Assume after iteration \(x-1\) that each regional structure reports a
minimum-cost retained predecessor of its region and that every
predecessor removed so far is permanently dominated by a retained
predecessor.
Predecessors acquiring their second penalty form a prefix of the old
middle region and are transferred to \(\mathcal T^1\).
Predecessors leaving \(Y_x^3\) likewise form a prefix and are
transferred to the appropriate buffer. If the active penalty in the middle region changes between customers
\(x-1\) and \(x\), then every predecessor in the old middle region
must enter \(Y_x^1\). By Condition~\eqref{cond:P1}, each such
predecessor retains its existing penalty. If it did not acquire its
second penalty, it would remain in \(Y_x^2\) with the same sole active
penalty, and the middle-region type could not change.

Algorithm~\ref{alg:spdtw-maintain-y1} reports the minimum retained
predecessor in \(Y_x^1\). When \(Y_x^2\) is time-warp-only, the
dominance rule from the soft VRPTW Split gives its minimum retained
predecessor. When \(Y_x^2\) is capacity-only,
Lemma~\ref{lem:capacity-region-queue} gives its minimum retained
predecessor. Algorithm~\ref{alg:spdtw-maintain-y3} gives the minimum
retained predecessor in \(Y_x^3\).

All deletions are permanent. Deletions from the unpenalized region use
the base-cost dominance rule together with Condition~\eqref{cond:P2}.
Time-warp-only deletions use Condition~\eqref{cond:P4} while only
time warp is active, together with assurance from Condition~\eqref{cond:P2} 
that a later capacity penalty cannot reverse the comparison. Deletions in
the fully penalized region use Conditions~\eqref{cond:S1}--
\eqref{cond:S3} together with the fixed time-warp difference.
Finally, deletions within \(\Gamma^2\) use the permanent capacity
comparison of Equation~\eqref{eq:spd-permanent-comparison} under
Condition~\eqref{cond:S1}, together with Condition~\eqref{cond:P2}
for any later time-warp penalty. 

Lines~\ref{line:spdtw-representatives}--
\ref{line:spdtw-record} compare the minimum-cost retained predecessor
from each nonempty region and record an optimal predecessor and its
shortest-path cost.

Each predecessor passes through each region at most once and is
permanently removed at most once. The buffer transfers,
\(\Lambda\)-maintenance, and all \(\Gamma^2\) operations therefore
take \(O(n)\) total time. Only a constant number of regional
representatives are compared to determine each $\operatorname{pred}(x)$. The complete Split
therefore takes \(O(n)\) time and \(O(n)\) space.
\end{proof}

\bibliographystyle{elsarticle-harv}
\bibliography{mybibliography}

\end{document}